\documentclass[aps,twocolumn,nofootinbib,groupedaddress,superscriptaddress,longbibliography,notitlepage]{revtex4-2}

\usepackage{amsmath,amssymb}
\usepackage[normalem]{ulem}
\usepackage{graphicx}
\usepackage{longtable,multirow}
\usepackage{mathtools}
\usepackage{dsfont}
\usepackage{amsfonts}
\usepackage{xcolor}
\usepackage{url}
\usepackage{nicematrix}
\usepackage{booktabs}

\colorlet{themecolor}{teal}

\usepackage[colorlinks=true,linkcolor=themecolor,citecolor=themecolor,urlcolor=themecolor,hypertexnames=false]{hyperref}
\newcommand{\nocontentsline}[3]{}
\newcommand{\tocless}[2]{\bgroup\let\addcontentsline=\nocontentsline#1{#2}\egroup}

\def\ba#1\ea{\begin{align}#1\end{align}}
\def\bg#1\eg{\begin{gather}#1\end{gather}}
\def\bpm{\begin{pmatrix}}
\def\epm{\end{pmatrix}}
\def\bsm{\begin{psmallmatrix}}
\def\esm{\end{psmallmatrix}}
\def\bbm{\begin{bmatrix}}
\def\ebm{\end{bmatrix}}

\newcommand{\nn}{\nonumber \\ }
\newcommand{\bb}[1]{{\mathbf #1}}
\newcommand{\bs}[1]{{\boldsymbol #1}}
\newcommand{\bx}{\bb x}
\newcommand{\bk}{\bb k}

\newcommand{\bR}{\bb R}

\newcommand{\td}[1]{\widetilde{#1}}

\newcommand{\mc}[1]{\mathcal{#1}}

\newcommand{\dg}{\dagger}

\newcommand{\sg}{\sigma}

\newcommand{\Z}{\mathbb{Z}}
\newcommand{\R}{\mathbb{R}}
\newcommand{\N}{\mathbb{N}}
\newcommand{\Q}{\mathbb{Q}}

\newcommand{\afset}{{\mc H_{\rm AF}}}
\newcommand{\aiset}{{\mc H_{\rm A}}}

\newcommand{\ourtitle}{Fragile Topology is Unstable Under Translation Refinement}

\allowdisplaybreaks

\begin{document}
\title{\ourtitle}

\author{Yoonseok Hwang}
\affiliation{Department of Physics and Anthony J. Leggett Institute for Condensed Matter Theory, University of Illinois Urbana-Champaign, Urbana, IL 61801, USA}
\affiliation{Blackett Laboratory, Imperial College London, London SW7 2AZ, United Kingdom}

\author{Saavanth Velury}
\affiliation{Department of Physics and Anthony J. Leggett Institute for Condensed Matter Theory, University of Illinois Urbana-Champaign, Urbana, IL 61801, USA}
\affiliation{Department of Physics, University of Florida, Gainesville, FL 32611, USA}

\author{Taylor L. Hughes}
\affiliation{Department of Physics and Anthony J. Leggett Institute for Condensed Matter Theory, University of Illinois Urbana-Champaign, Urbana, IL 61801, USA}

\begin{abstract}
Fragile topological phases become trivial upon the addition of suitable trivial bands, distinguishing them from stable topological phases.
Nevertheless, various response phenomena and material realizations have been proposed for fragile phases.
At the same time, many of these phenomena can also occur in atomic insulators, leaving open the question of what properties are specific to fragile phases.
Enlarging the unit cell offers a natural perspective on this question.
Band folding increases the number of bands in a manner analogous to adding trivial bands.
In this work, we establish a systematic framework for determining the stability of fragile topology under unit-cell enlargement.
We first establish a systematic criterion for trivialization under enlargements compatible with space-group symmetry, grounded in a physical electron–positron picture and formulated through a Hilbert-basis analysis of momentum-space symmetry data.
We then show that, for all two-dimensional wallpaper groups, with or without spin-orbit coupling and/or time-reversal symmetry, every symmetry-indicated fragile phase is adiabatically connected to an atomic insulator in a suitable finite supercell and can therefore be trivialized by an arbitrarily small symmetry-preserving perturbation.
Our results reveal that fragile topology has only finite stability under translation-symmetry refinement.
This highlights that the fate of a fragile phase can depend on translation-symmetry-breaking perturbations, such as charge-density-wave ordering, and suggests that physical signatures insensitive to translation refinement are unlikely to uniquely characterize fragile topology.
\end{abstract}

\maketitle

\let\oldaddcontentsline\addcontentsline
\renewcommand{\addcontentsline}[3]{}

\section{Introduction}
\label{sec:intro}
Crystalline symmetry enriches the classification of symmetry-protected topological phases~\cite{shiozaki2016topology,shiozaki2017topological,shiozaki2022atiyah,kruthoff2017topological,po2017symmetry,bradlyn2017topological} and nodal structures~\cite{bradlyn2016beyond,wieder2018wallpaper,hwang2023magnetic}.
In this context, fragile topology~\cite{po2018fragile} has emerged as a distinct form of crystalline band topology characterized by the absence of exponentially localized and symmetric Wannier functions, while becoming trivial upon the addition of suitable trivial bands~\cite{cano2018topology,bradlyn2019disconnected,bouhon2019wilson,kooi2019classification,bouhon2020geometric,brouwer2023homotopic}. Fragile phases have been proposed in various material platforms, including twisted bilayer graphene, moir\'e systems~\cite{song2019all,po2019faithful,ahn2019failure}, and other stoichiometric materials~\cite{vergniory2022all}, as well as synthetic platforms~\cite{chiu2020fragile,zhang2021tunable,peri2020experimental,bird2025design}, including photonic and phononic crystals~\cite{de2019engineering,alexandradinata2020crystallographic,manes2020fragile,park2021topological}.
Moreover, fragile topology also plays an important role in our understanding of higher-order topological insulators and topological semimetals~\cite{wieder2018axion,wieder2020strong,kobayashi2021fragile}.

Beyond its band-theoretic definition, it is known that fragile topology can also arise, or be stabilized, in the presence of interactions~\cite{else2019fragile,latimer2021correlated,herzog2024interacting}. Additionally, fragile topology has been associated with a variety of physical signatures, including corner charges~\cite{wieder2018axion,hwang2019fragile}, twisted bulk-boundary correspondence~\cite{song2020twisted,peri2020experimental}, defect responses such as disclination-induced fractional charge~\cite{liu2019shift}, magnetic and optical responses~\cite{liu2019shift,herzog2020hofstadter,lian2020landau,guan2022landau,jankowski2025optical}, and nonzero lower bounds on the superfluid weight~\cite{xie2020topology,peri2021fragile,herzog2022superfluid}.
At the same time, several of these phenomena also arise in obstructed atomic limits or other symmetry-protected trivial phases~\cite{hwang2021geometric,wu2021landau,herzog2023hofstadter,li2020fractional,zhang2022fractional,manjunath2024characterization,hwang2026stable}.
For example, quantized corner charges and defect-bound fractional charge can already appear in atomic or obstructed atomic insulators~\cite{song2017d,benalcazar2017quantized,benalcazar2019quantization,schindler2019fractional,benalcazar2019quantization,lee2020fractional,takahashi2021general,rao2023effective,velury2025global}.
As a result, while fragile topology is well-defined at the level of classification, the extent to which it gives rise to response properties uniquely distinguishable from atomic or trivial insulators remains an open conceptual question.

The defining property of fragile topology is its instability to the addition of trivial bands to the ground state wavefunction~\cite{po2018fragile}.
A natural and closely related question is how fragile topology behaves under modifications of the translation symmetry that preserve a given space-group symmetry.
In particular, enlarging the unit cell refines the translation symmetry and induces band folding, thereby modifying the symmetry data of the occupied bands, namely the multiplicities of irreducible representations at high-symmetry momenta in the Brillouin zone.
This therefore provides a complementary probe of the robustness of fragile topology.
In such a setting, does a fragile phase necessarily remain nontrivial, or can it always be trivialized by a suitable enlargement?

In this work, we investigate the stability of fragile topology under unit-cell enlargement.
Focusing on two-dimensional (2D) wallpaper groups, we systematically analyze symmetry-compatible unit-cell enlargements for each group and determine whether fragile phases defined in the primitive unit cell become unstable under some finite enlargement.
More precisely, we ask whether, when described in a suitable supercell, the phase can be connected to an atomic insulator by an arbitrarily small symmetry-preserving perturbation.
We begin with a physical real-space picture based on electron–positron annihilation~\cite{else2019fragile}, which provides intuitive insight into the mechanism of trivialization.
To complement this intuition with a systematic and general criterion, we focus on symmetry-indicated fragile topology, which can be diagnosed from momentum-space symmetry data~\cite{hwang2019fragile,song2020twisted,song2020fragile}.
Within this setting, building on Ref.~\cite{song2020fragile}, we introduce a Hilbert-basis characterization of such symmetry data formulating unit-cell enlargement as an integer linear map.

We find that, in all 2D wallpaper groups, every symmetry-indicated fragile phase can be trivialized after a suitable, finite symmetry-compatible unit-cell enlargement.
Equivalently, for each symmetry class we define a minimal enlargement scale as the smallest supercell index for which there exists such a symmetry-compatible enlargement that trivializes all fragile phases.
The minimal enlargement scale depends on the wallpaper group and on the presence or absence of time-reversal symmetry and spin-orbit coupling.
For example, for spinless electrons with or without time-reversal symmetry, fragile phases in wallpaper group $p2$ are trivialized at the smallest nontrivial enlargement, whereas in higher-symmetric groups, such as $p6$, fragile phases may survive small enlargements but are annihilated at a larger one, implying the existence of explicit finite bounds for trivialization.

These results show that symmetry-indicated fragile topology has is eventually unstable under symmetry-compatible unit-cell enlargement.
This finite stability places general constraints on response phenomena that distinguish fragile phases from atomic insulators.
In particular, physical signatures that do not persist under translation-symmetry refinement in concert with the fragile phase, or those that are directly linked to fragile invariants such as the Euler class~\cite{zhao2017pt,ahn2018band,ahn2019failure}, may provide a natural setting for distinguishing fragile phases from atomic insulators.

\section{Example: fragile topology in $p2$}
\label{sec:p2}
In this section, we introduce the necessary notation for wallpaper groups and band topology in the simplest setting of the group $p2$.
Using this example, we present a representative fragile phase and analyze its behavior under symmetry-compatible unit-cell enlargement.
This example serves as a concrete illustration before turning to the general framework developed in Secs.~\ref{sec:hilbert} and \ref{sec:enlargement}.

\subsection*{Real-space and momentum-space description}
\begin{figure*}[t]
\centering
\includegraphics[width=0.98\textwidth]{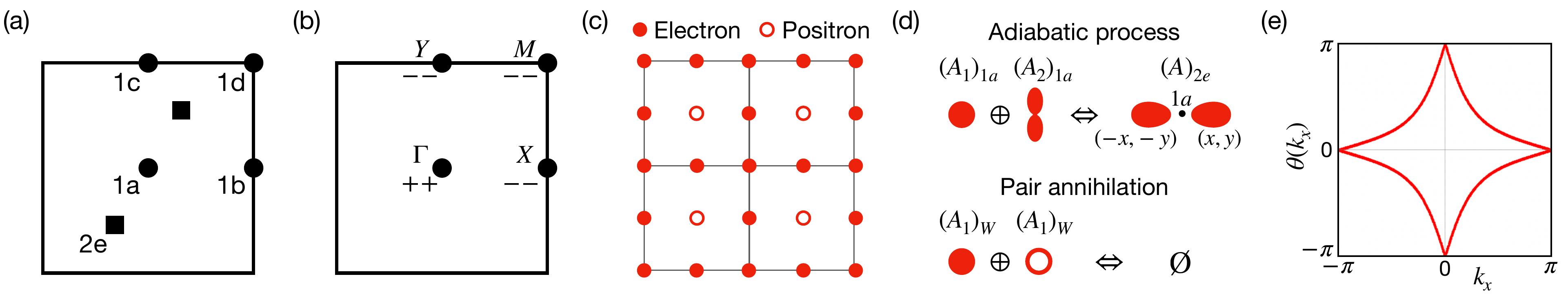}
\caption{
(a) Wyckoff positions (WPs), $1a:(0,0)$, $1b: (1/2,0)$, $1c: (0,1/2)$, $1d: (1/2,1/2)$, and $2e: \{ (x,y), (-x,-y) \}$, in a primitive unit cell.
(b) Brillouin zone and an example of symmetry data.
The high-symmetry momenta (HSM) are $\Gamma=(0,0)$, $X=(\pi,0)$, $Y=(0,\pi)$, and $M=(\pi,\pi)$.
Multiplicities of $C_2$ eigenvalues at all HSM determine the symmetry-data vector.
$\bb v = (2,0,0,2,0,2,0,2)$ according to the definition in Eq.~\eqref{eq:p2_sym_vec}.
(c) Real-space electron-positron configuration corresponding to $(A_1)_{1b} \oplus (A_1)_{1c} \oplus (A_1)_{1d} \ominus (A_1)_{1a}$, where every electron and the positron transforms according to the $A_1$ site-symmetry irrep at its respective WP.
(d) Top: symmetry-preserving adiabatic deformation, $(A_1)_W \oplus (A_2)_W \Leftrightarrow (A)_{2e}$, valid for any maximal WP $W=1a,1b,1c,1d$.
Bottom: example of electron-positron pair annihilation.
More generally, pair annihilation is allowed between an electron and a positron occupying the same WP and transforming according to the same site-symmetry irrep.
(e) Wilson-loop spectrum shows the characteristic winding of fragile topology.}
\label{fig:p2}
\end{figure*}

We begin with the simplest setting to fix notation: spinless electrons (without spin–orbit coupling) in the 2D wallpaper group $p2$ without time-reversal symmetry.
The wallpaper group consists of lattice translations $\{E|\bb v \in \Z^2\}$ and a twofold rotation $C_2 = \{c_2|\bb 0\}$, where $E$ is the trivial point-group element, and $c_2$ maps $(x,y)$ to $(-x,-y)$.
In a primitive unit cell, the Wyckoff positions (WPs) are
$1a:(0,0)$, $1b:(1/2,0)$, $1c:(0,1/2)$, $1d:(1/2,1/2)$, and the nonmaximal position $2e:\{(x,y),(-x,-y)\}$, as shown in Fig.~\ref{fig:p2}(a).

At the maximal WPs $1a$, $1b$, $1c$, and $1d$, the site-symmetry group consists of the symmetry operations that leave the position invariant.
For example, at $1a$ this group contains $\{c_2|\bb 0\}$, while at $1b$ it contains $\{c_2|1,0\}$.
At each maximal WP, the site-symmetry group has two one-dimensional irreducible representations (irreps), which we denote by $A_1$ and $A_2$, corresponding to $C_2$ eigenvalues $+1$ and $-1$, respectively.
At the nonmaximal WP $2e$, the site-symmetry group is trivial and only the trivial irrep $A$ exists.
We denote by $(\rho)_W$ the site-symmetry irrep $\rho$ defined at the WP $W$.
These site-symmetry irreps correspond physically to localized Wannier orbitals.
Each such orbital induces a band representation (BR) in momentum space, which corresponds to trivial band topology~\cite{bradlyn2017topological}.

In momentum space, the Brillouin zone (BZ) contains four high-symmetry momenta (HSM), $\Gamma=(0,0)$, $X=(\pi,0)$, $Y=(0,\pi)$, and $M=(\pi,\pi)$.
Since $C_2$ leaves the momentum at each of these points invariant up to a reciprocal lattice vector, the little group of each point contains $C_2$.
Thus, the little group has two one-dimensional irreps characterized by $C_2$ eigenvalues $\pm 1$, and each Bloch state transforms according to one of them.
We denote the little-group irreps by $\Gamma_\pm$, $X_\pm$, $Y_\pm$, and $M_\pm$, and collect their multiplicities into a symmetry-data vector,
\bg
\bb v = (n_{\Gamma_+}, n_{\Gamma_-}, n_{X_+}, n_{X_-}, n_{Y_+}, n_{Y_-}, n_{M_+}, n_{M_-})^{\rm T}.
\label{eq:p2_sym_vec}
\eg
See Fig.~\ref{fig:p2}(b) for the HSM and an example of symmetry-data vector.

We also introduce a real-space multiplicity vector $\bb m$, whose entries count the number of Wannier orbitals of each site-symmetry irrep defined within the unit cell.
We order it as
\ba
\bb m = \big[ & n_{(A_1)_{1a}}, n_{(A_2)_{1a}}, n_{(A_1)_{1b}}, n_{(A_2)_{1b}}, n_{(A_1)_{1c}},
\nn
& n_{(A_2)_{1c}}, n_{(A_1)_{1d}}, n_{(A_2)_{1d}}, n_{(A)_{2e}} \big]^{\rm T}.
\label{eq:p2_m_vec}
\ea
The relation between real-space data and momentum-space symmetry data is linear,
\bg
\bb v = BR \cdot \bb m,
\label{eq:v_from_m}
\eg
where the BR matrix $BR$ encodes how each BR induced from a Wannier orbital contributes to little-group irreps at the HSM.
Each column of $BR$ is the symmetry-data vector induced by the corresponding Wannier orbital.
In the ordering defined in Eqs.~\eqref{eq:p2_sym_vec} and \eqref{eq:p2_m_vec}, the BR matrix takes the explicit form
\bg
BR = \bpm
1 & 0 & 1 & 0 & 1 & 0 & 1 & 0 & 1 \\
0 & 1 & 0 & 1 & 0 & 1 & 0 & 1 & 1 \\
1 & 0 & 0 & 1 & 1 & 0 & 0 & 1 & 1 \\
0 & 1 & 1 & 0 & 0 & 1 & 1 & 0 & 1 \\
1 & 0 & 1 & 0 & 0 & 1 & 0 & 1 & 1 \\
0 & 1 & 0 & 1 & 1 & 0 & 1 & 0 & 1 \\
1 & 0 & 0 & 1 & 0 & 1 & 1 & 0 & 1 \\
0 & 1 & 1 & 0 & 1 & 0 & 0 & 1 & 1
\epm.
\label{eq:p2_br_mat}
\eg
%

Given a symmetry-data vector $\bb v$ of a set of bands, we ask whether it can be realized by Wannier orbitals in real space.
Formally, this amounts to solving Eq.~\eqref{eq:v_from_m} for $\bb m$.
It is convenient to organize the possibilities as follows~\cite{po2017symmetry,bradlyn2017topological,po2018fragile,bradlyn2019disconnected}.
(i) If no integer solution $\bb m \in \Z$ exists, then the set of bands represents symmetry-indicated stable topology, diagnosed by nontrivial symmetry indicators~\cite{fu2007topological,po2017symmetry,watanabe2018structure,khalaf2018symmetry,song2018quantitative}.
(ii) If an integer solution exists but necessarily contains at least one negative entry, then the set of bands possesses symmetry-indicated fragile topology.
Since negative multiplicities correspond to subtracting the associated BRs, the solution naturally takes the form of a formal difference of BRs.
(iii) Finally, if a solution $\bb m$ exists with all entries nonnegative, the phase corresponds to an atomic (trivial) insulator.

In practice, one may compute $\bb m$ from $\bb v$ using a pseudoinverse method,
\bg
\bb m = BR^\ddagger \cdot \bb v + \bb m_{\rm ker},
\label{eq:m_from_v}
\eg
where $BR^\ddagger$ is an integer pseudoinverse and $\bb m_{\rm ker}$ lies in ${\rm ker} \, BR$, the kernel of the $BR$ matrix.
In other words, $BR \cdot \bb m_{\rm ker} = 0$.
The kernel reflects redundancies in representing a given symmetry-data vector in terms of Wannier orbitals, since different integer combinations of site-symmetry irreps can induce identical symmetry data~\cite{bradlyn2017topological,cano2022topology,hwang2026stable}.
Details of the pseudoinverse construction are summarized in the Supplemental Material (SM)~\cite{supple}.
%

\subsection*{Electron-positron picture and adiabatic processes}
As explained above, fragile topology can be understood as a formal difference of BRs.
Since each BR is induced from a site-symmetry irrep, or equivalently from a Wannier orbital, this admits an electron-positron picture, in which negative multiplicities represent positrons that can annihilate with electrons of the same Wannier-orbital type.
This viewpoint has also proven useful in the discussion of interacting fragile phases~\cite{else2019fragile}.

As a concrete example, consider the two-band configuration with two $C_2$ eigenvalues $+1$ at $\Gamma$ and two $-1$ at $X$, $Y$, and $M$.
Its symmetry-data vector is
\bg
\bb v = (2,0,0,2,0,2,0,2)^{\rm T},
\label{eq:p2_v_example}
\eg
shown in Fig.~\ref{fig:p2}(b).
Solving $\bb v = BR \cdot \bb m$ yields, up to kernel contributions of the BR matrix,
\bg
\bb m = (-1,0,1,0,1,0,1,0,0)^{\rm T}.
\label{eq:p2_m_example}
\eg
Substituting Eq.~\eqref{eq:p2_m_example} into Eq.~\eqref{eq:v_from_m} reproduces the symmetry-data vector $\bb v$ in Eq.~\eqref{eq:p2_v_example}.
Note that Eq.~\eqref{eq:p2_m_example} is not the unique solution.
For example,
\bg
(1,1,-1,-1,0,0,0,0,0)^{\rm T} \in {\rm ker} \, BR,
\eg
so that $(0,1,0,-1,1,0,1,0,0)^{\rm T}$ is another solution corresponding to the same symmetry-data vector $\bb v$.
More generally, adding any kernel vector $\bb m_{\rm ker}$ leaves $\bb v$ unchanged.
However, every representative obtained by adding a kernel vector still contains at least one negative multiplicity.
Thus, $\bb v$ in Eq.~\eqref{eq:p2_v_example} corresponds to a symmetry-indicated fragile phase.
Since the discussions below apply to any such representative, we use the solution in Eq.~\eqref{eq:p2_m_example} throughout the remainder of this section.

In terms of site-symmetry irreps, this corresponds to the real-space configuration
\bg
\rho = (A_1)_{1b} \oplus (A_1)_{1c} \oplus (A_1)_{1d} \ominus (A_1)_{1a},
\label{eq:p2_rho}
\eg
illustrated in Fig.~\ref{fig:p2}(c).
This provides a real-space interpretation of the electron-positron picture.
The positive summands in $\rho$ correspond to electrons, whereas the negative summand corresponds to a positron.
In this picture, the electron number or filling associated with a positron is counted with the opposite sign~\cite{else2019fragile}.
Thus, there are two electrons per unit cell, as expected for the two-band system.
Also, pair annihilation is allowed only between an electron and a positron at the same WP transforming under the same site-symmetry irrep [See Fig.~\ref{fig:p2}(d)].
However, fragile topology requires more than the existence of a positron: the positron must not be removable through any symmetry-preserving adiabatic deformation while keeping the bulk energy gap open.

In the present example, every electron and the positron are localized at maximal WPs, with exactly one site-symmetry irrep assigned to each WP.
In this situation, moving any electron or the positron away from its WP necessarily breaks the $p2$ symmetry.
More precisely, there is no combination of site-symmetry irreps that can be combined and continuously moved to another WP while preserving the wallpaper-group symmetry.
Therefore, the positron $(A_1)_{1a}$ cannot annihilate with the remaining electrons while preserving both the symmetry and the energy gap.
At the same time, the electron-positron picture immediately identifies which BR should be added to trivialize the fragile topology.
Adding the BR induced by $(A_1)_{1a}$ annihilates the positron, leaving the atomic configuration $(A_1)_{1b} \oplus (A_1)_{1c} \oplus (A_1)_{1d}$.

The above argument can be formulated more systematically in terms of symmetry-preserving adiabatic deformations.
Such deformations allow site-symmetry irreps to move between WPs while preserving both the wallpaper-group symmetry and the bulk energy gap~\cite{huang2017building,van2018higher,hwang2019fragile,song2020twisted,xu2021three,hwang2026stable}.
For instance, continuously deforming the trivial irrep $(A)_{2e}$ at the $2e$ WP to the maximal WP $1a$ at $(0,0)$ yields $(A_1)_{1a} \oplus (A_2)_{1a}$.
Conversely, the reverse deformation recombines them into $(A)_{2e}$, as illustrated in Fig.~\ref{fig:p2}(d).
Physically, the two $A$ irreps at $(x,y)$ and $(-x,-y)$ are brought to the $1a$ WP and reorganized into symmetric and antisymmetric linear combinations.
The symmetric combination transforms as $A_1$, whereas the antisymmetric combination transforms as $A_2$ under the site-symmetry group at $1a$.
The reverse deformation is equally allowed.
In terms of the multiplicity vector, this process shifts $\bb m$ by $\Delta \bb m = \pm (1, 1, 0, 0, 0, 0, 0, 0, -1)^{\rm T}$.
The same analysis applies to the WPs $1b$, $1c$, and $1d$, giving the adiabatic equivalence
\bg
(A)_{2e} \Leftrightarrow (A_1)_W \oplus (A_2)_W
\label{eq:p2_adia}
\eg
for $W=1a,1b,1c,1d$.

Any adiabatic deformation process changes the multiplicity vector $\bb m$ by an integer vector while leaving the symmetry-data vector $\bb v$ invariant~\cite{hwang2026stable}.
The independent generators are assembled into the adiabatic-process matrix $q_{\rm adia}$:
\bg
q_{\rm adia}=
\begin{pNiceArray}[last-col,code-for-last-col=\scriptstyle]{cccc}
1 & 0 & 0 & 0 & (A_1)_{1a} \\
1 & 0 & 0 & 0 & (A_2)_{1a} \\
0 & 1 & 0 & 0 & (A_1)_{1b} \\
0 & 1 & 0 & 0 & (A_2)_{1b} \\
0 & 0 & 1 & 0 & (A_1)_{1c} \\
0 & 0 & 1 & 0 & (A_2)_{1c} \\
0 & 0 & 0 & 1 & (A_1)_{1d} \\
0 & 0 & 0 & 1 & (A_2)_{1d} \\
-1 & -1 & -1 & -1 & (A)_{2e}
\end{pNiceArray}.
\label{eq:p2_qadia}
\eg
Each column of $q_{\rm adia}$ corresponds to one independent adiabatic generator, labeled by the associated site-symmetry irreps.
The general structure of adiabatic-process matrices has been systematically analyzed in Refs.~\cite{song2020twisted,xu2021three,hwang2026stable}, with Ref.~\cite{hwang2026stable} providing a systematic algorithm to generate $q_{\rm adia}$ for all space groups.

If $q_{\rm adia}$ has size $D_{\bb m} \times N_{\rm adia}$, any symmetry-preserving adiabatic deformation is represented by
\bg
\bb m_{\rm new} = \bb m + q_{\rm adia} \cdot \bb z,
\quad
\bb z \in \Z^{N_{\rm adia}}.
\label{eq:m_adia}
\eg
Each component of $\bb z$ specifies how many times the corresponding independent adiabatic generator (a column of $q_{\rm adia}$) is applied.
Whenever the inequalities $\bb m_{\rm new} \ge 0$ admit no integer solution for $\bb z$, the corresponding real-space configuration necessarily represents fragile topology.
The present example is precisely such a case.
Equivalently, this means that the positron in Eq.~\eqref{eq:p2_rho} cannot be annihilated by any such deformation.
This confirms the fragile topology of the present example.

For completeness, we also note that the same conclusion can be verified using the more conventional Wilson-loop analysis~\cite{yu2011equivalent,fidkowski2011model,alexandradinata2014wilson}.
Within an explicit tight-binding model, the $k_y$-directed Berry phase as a function of $k_x$ exhibits the characteristic winding associated with fragile topology~\cite{alexandradinata2014wilson,wieder2018axion,hwang2019fragile}, as shown in Fig.~\ref{fig:p2}(e).
Details of the tight-binding model are provided in the SM~\cite{supple}.

\begin{figure*}[t]
\centering
\includegraphics[width=0.98\textwidth]{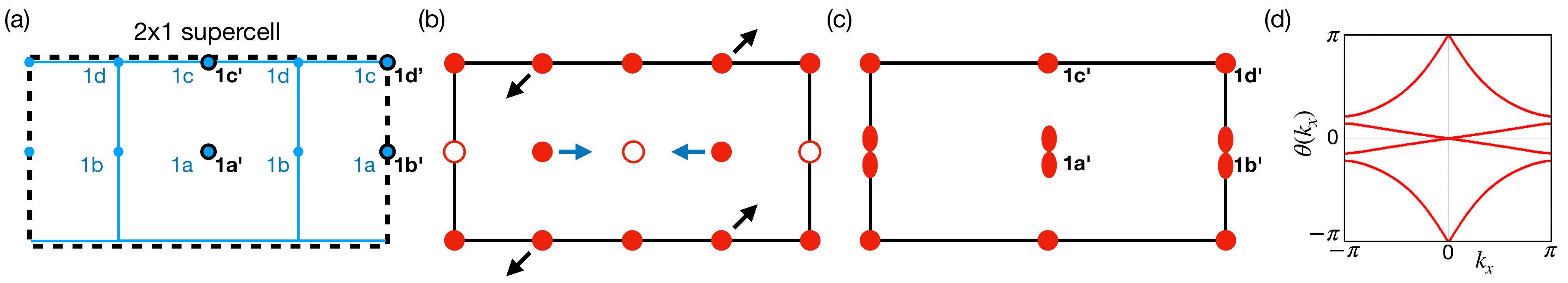}
\caption{
(a) $2 \times 1$ supercell.
The dashed black rectangle (black circles) denotes the supercell (its maximal WPs $1a',1b',1c',1d'$), while the blue solid rectangles (blue dots) denote the primitive unit cells (their maximal WPs $1a,1b,1c,1d$).
(b) Real-space electron-positron configuration corresponding to Eq.~\eqref{eq:p2_rho_sc1}.
The blue arrows illustrate the adiabatic deformation $(A)_{2e'} \Leftrightarrow (A_1)_{1a'}\oplus(A_2)_{1a'}$, depicted schematically in Fig.~\ref{fig:p2}(d).
The black arrows indicate analogous deformations of the remaining $(A)_{2e'}$ orbitals into the $1b'$ WP.
All arrows denote symmetry-preserving adiabatic deformations consistent with the $p2$ wallpaper group.
(c) Real-space configuration corresponding to Eq.~\eqref{eq:p2_rho_sc2} after the adiabatic deformations shown in (b).
The $(A_1)$ positrons at $1a'$ and $1b'$ are annihilated, leaving the $(A_2)$ electrons at those WPs.
(d) Wilson-loop spectrum of the same tight-binding model as in Fig.~\ref{fig:p2}(e), after the supercell enlargement.
After adding a symmetry-preserving perturbation, the accidental winding disappears, confirming the trivialization of the fragile topology.}
\label{fig:p2_sc}
\end{figure*}

\subsection*{Unit-cell enlargement and trivialization}
We now introduce the concept of unit-cell enlargement.
Starting from primitive lattice vectors $\bb a_1$ and $\bb a_2$, a supercell is specified by an integer matrix
\bg
\mc S_{\rm uc} = \bpm n_1 & n_2 \\ n_3 & n_4 \epm,
\quad
(\bb a_1', \bb a_2') = (\bb a_1, \bb a_2) \cdot \mc S_{\rm uc},
\label{eq:p2_sc}
\eg
with nonzero area index $N_{\rm sc} = |\det \mc S_{\rm uc}| = |n_1 n_4 - n_2 n_3|$ (in general, not every integer matrix $\mc S_{\rm uc}$ preserves the wallpaper-group symmetry.
The corresponding compatibility conditions are discussed in SM~\cite{supple}).
In $p2$, any positive integer $N_{\rm sc}$ can be realized by a suitable choice of $\mc S_{\rm uc}$.
The smallest nontrivial enlargement has $N_{\rm sc}=2$, which includes geometrically distinct choices such as $2 \times 1$, $1 \times 2$, and $\sqrt{2} \times \sqrt{2}$, as well as sheared variants.
In the main text, we focus on the representative $2 \times 1$ enlargement for concreteness.
Other choices of supercells, whether of the same or larger index, can be treated in exactly the same manner within our framework.
Throughout this section, we keep the origin fixed under enlargement, since changing the origin merely relabels WPs and site-symmetry irreps without affecting the topological classification.

In the $2 \times 1$ supercell, band folding doubles the number of bands.
At the same time, the WPs and the site-symmetry irreps must be relabeled according to the enlarged unit cell.
Consequently, the real-space configuration $\rho$ in Eq.~\eqref{eq:p2_rho} is rewritten in terms of the WPs and site-symmetry irreps of the supercell.
We then analyze the resulting real-space configuration $\rho'$ and its corresponding site-symmetry irrep multiplicity vector $\bb m'$ using the adiabatic deformations introduced previously.
Specifically, we examine whether there exists an integer vector $\bb z$ in Eq.~\eqref{eq:m_adia} such that all entries of the resulting multiplicity vector become nonnegative.
If such a vector $\bb z$ exists, the positron can be completely annihilated adiabatically.
Consequently, after unit-cell enlargement the system is adiabatically connected to an atomic insulator and can therefore be trivialized by an arbitrarily small symmetry-preserving perturbation.

To determine the supercell configuration $\rho'$ and its corresponding irrep multiplicity vector $\bb m'$, we compare the primitive cell with the $2 \times 1$ supercell shown in Fig.~\ref{fig:p2_sc}(a).
The WPs of the primitive cell are first reorganized into those of the supercell.
For convenience, we denote supercell WPs by a prime.
Since the $2 \times 1$ supercell contains two primitive unit cells, each primitive-cell WP appears twice before being reorganized into the WPs of the supercell.

For example, the two primitive-cell $1a$ WPs become the supercell WPs $1a'$ and $1b'$; see Fig.~\ref{fig:p2_sc}(a).
Since the site-symmetry groups of $1a$, $1a'$, and $1b'$ are isomorphic, the site-symmetry irreps can be identified naturally (up to relabeling) as
\bg
(A_1)_{1a} \to (A_1)_{1a'} \oplus (A_1)_{1b'},
\nn
(A_2)_{1a} \to (A_2)_{1a'} \oplus (A_2)_{1b'}.
\eg
Likewise, the two primitive-cell $1b$ WPs are reorganized into the supercell WP $2e'$.
Since the site-symmetry group of $2e'$ is trivial, both $(A_1)_{1b}$ and $(A_2)_{1b}$ map to the trivial irrep $(A)_{2e'}$.
Repeating the same analysis for the remaining WPs yields
\bg
(A_1)_{1b} \to (A)_{2e'},
\quad
(A_2)_{1b} \to (A)_{2e'},
\nn
(A_1)_{1d} \to (A)_{2e'},
\quad
(A_2)_{1d} \to (A)_{2e'},
\nn
(A_1)_{1c} \to (A_1)_{1c'} \oplus (A_1)_{1d'},
\nn
(A_2)_{1c} \to (A_2)_{1c'} \oplus (A_2)_{1d'},
\nn
(A)_{2e} \to 2(A)_{2e'}.
\eg

Applying these rules to the real-space configuration $\rho$ in Eq.~\eqref{eq:p2_rho}, we obtain the supercell configuration
\bg
\rho' = (A_1)_{1c'} \oplus (A_1)_{1d'} \oplus 2(A)_{2e'} \ominus (A_1)_{1a'} \ominus (A_1)_{1b'},
\label{eq:p2_rho_sc1}
\eg
in the $2 \times 1$ supercell, as shown in Fig.~\ref{fig:p2_sc}(b).
The corresponding site-symmetry irrep multiplicity vector is $\bb m' = (-1,0,-1,0,1,0,1,0,2)^{\rm T}$.
Using the adiabatic-process matrix introduced above and Eq.~\eqref{eq:m_adia}, we choose $\bb z = (1,1,0,0)^{\rm T}$, which gives $\bb m'_{\rm new} = (0,1,0,1,1,0,1,0,0)^{\rm T}$.
All entries of $\bb m'_{\rm new}$ are now nonnegative, indicating an atomic real-space configuration [Fig.~\ref{fig:p2_sc}(c)].
Interpreting $\bb m'_{\rm new}$ in terms of site-symmetry irreps, we obtain
\bg
\rho' \Leftrightarrow (A_2)_{1a'} \oplus (A_2)_{1b'} \oplus (A_1)_{1c'} \oplus (A_1)_{1d'}.
\label{eq:p2_rho_sc2}
\eg
Consequently, in the supercell the positron can be annihilated through an adiabatic process.
The fragile phase represented by Eq.~\eqref{eq:p2_rho} is therefore adiabatically connected to an atomic insulator in the supercell, and can be trivialized by an arbitrarily small symmetry-preserving perturbation.

The trivialization can also be confirmed by the Wilson-loop analysis in the tight-binding model.
Although the bands in the supercell may exhibit apparent winding structure, such winding structure is not symmetry-protected.
Upon adding a small symmetry-preserving perturbation, the winding disappears, as shown in Fig.~\ref{fig:p2_sc}(d), consistent with the electron–positron picture.
The explicit form of the perturbation Hamiltonian is given in the SM~\cite{supple}.

The electron–positron picture and Wilson-loop analysis provide a concrete demonstration of trivialization by unit-cell enlargement for this specific example.
In particular, the former approach, together with the adiabatic-process matrix $q_{\rm adia}$, offers valuable physical insight and can in principle be applied to arbitrary symmetry groups.
However, it requires a case-by-case analysis for each fragile phase.
To overcome this limitation, we now turn to a more systematic approach based on symmetry-data vectors.
In the next section, we formulate unit-cell enlargement as a linear map on symmetry-data vectors and introduce the Hilbert-basis method to analyze symmetry-indicated fragile topology in a unified manner.

\section{Hilbert-basis description of atomic and fragile phases}
\label{sec:hilbert}
The example in Sec.~\ref{sec:p2} illustrates explicitly how a fragile phase can be trivialized by unit-cell enlargement.
To treat all symmetry-indicated fragile phases systematically, we now develop a general framework based on symmetry data and the Hilbert basis (we will define below).
Similar to symmetry indicators for stable topology, fragile phases can be diagnosed from symmetry data~\cite{hwang2019fragile,song2020twisted,song2020fragile}.
In particular, Ref.~\cite{song2020fragile} formulated a Hilbert-basis approach to fragile topology, on which we build.

However, existing diagnosis schemes do not directly address the question of stability under unit-cell enlargement.
They either rely on specific symmetries~\cite{hwang2019fragile}, require case-by-case analysis when analyzing the effect of unit-cell enlargement~\cite{song2020twisted,song2020fragile}, or focus on restricted symmetry classes, such as spinful electrons with time-reversal symmetry~\cite{song2020fragile}.
Here, we formulate a unified procedure applicable to all wallpaper groups and symmetry classes considered in this work, both with and without spin–orbit coupling and time-reversal symmetry, tailored specifically to unit-cell enlargement.

\begin{figure}
\centering
\includegraphics[width=0.47\textwidth]{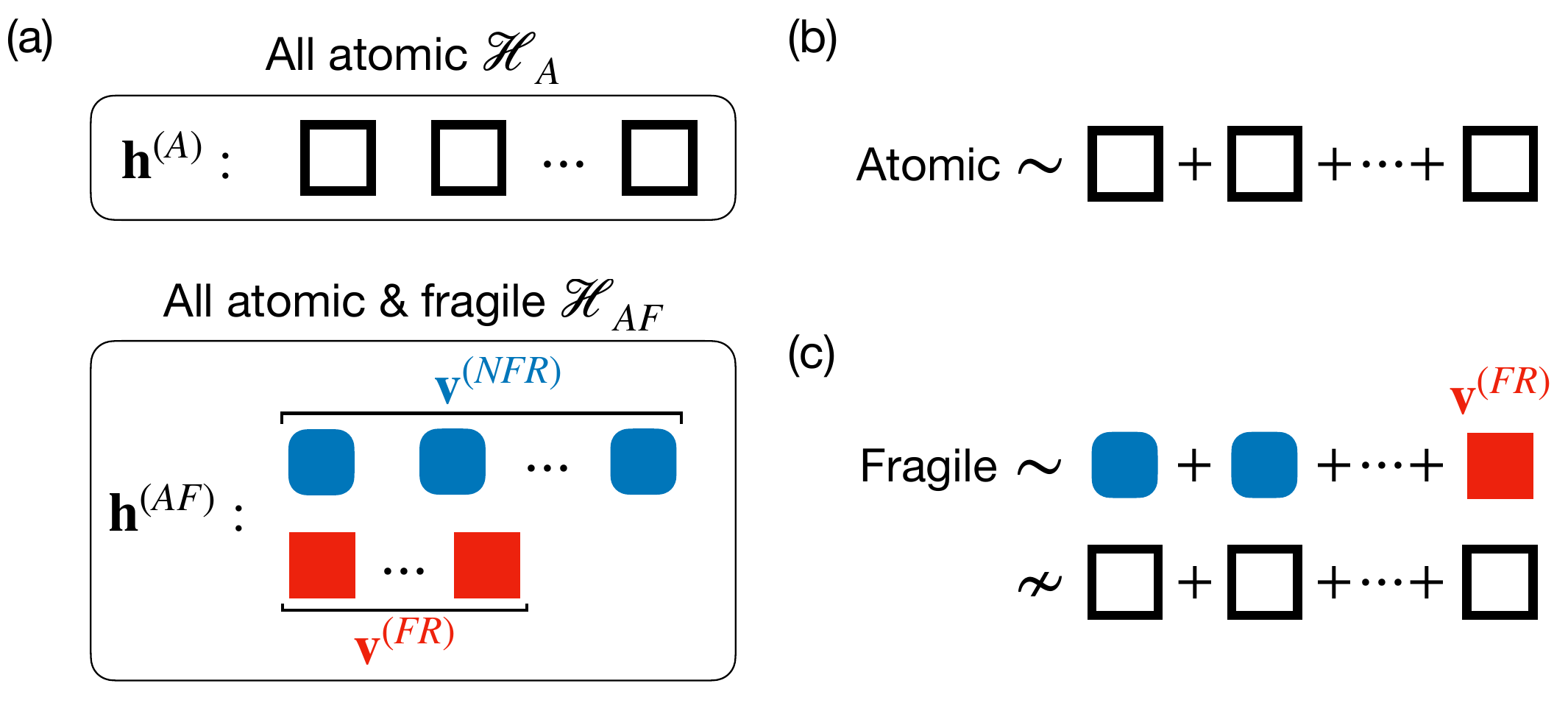}
\caption{
Schematic illustration of the Hilbert-basis construction.
(a) The Hilbert basis $\{\bb h_a^{(\rm A)}\}_{a=1}^{d_{\rm A}}$ generates all atomic symmetry-data vectors ($\aiset$).
The Hilbert basis $\{\bb h_b^{(\rm AF)}\}_{b=1}^{d_{\rm AF}}$ generates all atomic and fragile symmetry-data vectors ($\afset$).
The latter is partitioned into fragile roots
$\{\bb v_c^{(\rm FR)}\}_{c=1}^{N_{\rm FR}}$ (red squares)
and non-fragile generators
$\{\bb v_d^{(\rm NFR)}\}_{d=1}^{N_{\rm NFR}}$ (blue rounded squares).
(b) An atomic symmetry-data vector is expressed solely by $\{\bb h_a^{(\rm A)}\}$.
(c) A fragile symmetry-data vector is expressed by $\{\bb h_b^{(\rm AF)}\}$.
Every nonnegative-integer representation necessarily contains at least one fragile root, whereas no representation using only $\{\bb h_a^{(\rm A)}\}$ exists.
}
\label{fig:hilbert}
\end{figure}

We first present the general Hilbert-basis framework in an abstract setting.
For a brief overview of the framework, see Fig.~\ref{fig:hilbert}.
Once the formal structure is established, we will return to the example of $p2$ and compute the corresponding bases explicitly.
From now on, unless otherwise noted, we use the terms atomic and fragile
phases to mean symmetry-indicated atomic and fragile phases.
Accordingly, trivialization is always understood at the level of symmetry data.

\subsection*{Hilbert basis for atomic phases}
We first consider all possible symmetry-data vectors of atomic (trivial) phases and denote their set by $\aiset$.
Atomic phases are described by multiplicity vectors 
$\bb m \in \N_0^{D_{\bb m}}$, i.e., each component of $\bb m$ is a nonnegative integer $(\in \N_0)$ specifying the multiplicity of a site-symmetry irrep.
Since symmetry-data vectors are obtained through $\bb v = BR \cdot \bb m$, any element $\bb v^{(A)}$ in $\aiset$ can be expressed as a nonnegative integer combination of the columns of the BR matrix $BR$.
Thus, all atomic symmetry-data vectors can be generated from finitely many generators using nonnegative integer combinations.

The columns of $BR$ provide one generating set, but this generating set is generally redundant, since some columns can be expressed as nonnegative integer combinations of the others.
We therefore seek a minimal set of generators such that every atomic symmetry-data vector can be expressed as a nonnegative integer combination of them.
This minimal generating set is known as the Hilbert basis~\cite{bruns2009polytopes}.
Throughout this work, we compute Hilbert bases using the \texttt{Normaliz} program~\cite{bruns2010normaliz}.
For a detailed introduction to the Hilbert-basis method and its application to band topology, see also Ref.~\cite{song2020fragile}.
Similar Hilbert-basis techniques have also been applied to the study of band topology in photonic crystals~\cite{christensen2022location,hwang2026building}.

For $\aiset$, we denote the resulting Hilbert basis generators by $\{\bb h_a^{(\rm A)}\}_{a=1}^{d_{\rm A}}$.
Any atomic symmetry-data vector can therefore be expressed as a nonnegative integer combination of these generators:
\bg
\bb v^{(\rm A)} = \sum_{a=1}^{d_{\rm A}} \, n_a \bb h^{(\rm A)}_a,
\quad
n_a \in \N_0.
\label{eq:hilbert_AI}
\eg
A schematic illustration is provided in Figs.~\ref{fig:hilbert}(a) and (b).
In the next subsection, we will see that the larger set containing both atomic and fragile phases can also be formulated within the same Hilbert-basis framework.

\subsection*{Hilbert basis for atomic and fragile phases}
We now consider the set $\afset$ of all symmetry-data vectors corresponding to atomic and fragile phases.
By construction, $\aiset \subset \afset$.
Unlike atomic phases, fragile phases generally require integer multiplicity vectors with some negative entries.
Therefore, we consider symmetry-data vectors of the form $\bb v = BR \cdot \bb m \in \Z^{D_{\bb v}},$ where $\bb m \in \Z^{D_{\bb m}}$.
An example was already encountered in Sec.~\ref{sec:p2} [see Eq.~\eqref{eq:p2_m_example}].
However, physical bands must have nonnegative irrep multiplicities at every HSM.
Therefore, we impose
\bg
\bb v_i = (BR \cdot \bb m)_i \ge 0,
\label{eq:v_cond}
\eg
for all $i=1,\dots,D_{\bb v}$.

Equivalently, we consider integer vectors $\bb m \in \Z^{D_{\bb m}}$ satisfying the linear inequalities in Eq.~\eqref{eq:v_cond}.
These inequalities define a region in $\R^{D_{\bb m}}$,
\bg
\mc C = \{\bb m \in \R^{D_{\bb m}} | BR \cdot \bb m \ge 0 \},
\label{eq:cone_ineq}
\eg
where the inequality is understood componentwise.
The multiplicity vectors $\bb m$ corresponding to atomic and fragile phases are precisely the lattice points $\mc C \cap \Z^{D_{\bb m}}$, i.e., the integer points contained in $\mc C$.
Thus, the problem reduces to enumerating the lattice points of $\mc C$.
As we review below, such lattice points admit a Hilbert basis.
This allows both $\aiset$ and $\afset$ to be analyzed within a common Hilbert-basis framework.

In the language of polyhedral geometry, the region $\mc C$ is a rational cone, since the defining inequalities [Eq.~\eqref{eq:cone_ineq}] have integer coefficients~\cite{bruns2009polytopes}.
Moreover, because $\bb m$ and $-\bb m$ cannot both satisfy Eq.~\eqref{eq:v_cond} unless $\bb m=0$, the cone is pointed.
From standard results, the integer lattice points of a rational pointed cone admit a unique minimal generating set, namely its Hilbert basis~\cite{bruns2009polytopes}.

Equipped with the above result, we first compute the Hilbert basis for the multiplicity vectors $\bb m$ corresponding to atomic and fragile phases.
We denote the generators by $\{ \bb m^{(\rm AF)}_b \}_{b=1}^{d_{\rm AF}}$.
Each generator induces a symmetry-data vector
\bg
\bb h^{(\rm AF)}_b = BR \cdot \bb m^{(\rm AF)}_b.
\eg
Then, the induced symmetry-data vectors $\{\bb h^{(\rm AF)}_b\}_{b=1}^{d_{\rm AF}}$ generate all symmetry-data vectors in $\afset$ [see Fig.~\ref{fig:hilbert}(a)].
That is, any symmetry-data vector of an atomic or fragile phase can be written as
\bg
\bb v^{(\rm AF)} = \sum_{b=1}^{d_{\rm AF}} \, n_b \bb h^{(\rm AF)}_b,
\quad
n_b \in \N_0,
\label{eq:hilbert_AF}
\eg
where $\N_0$ denotes nonnegative integers.

\subsection*{Fragile roots}
The Hilbert basis generators
$\{\bb h_b^{(\rm AF)}\}_{b=1}^{d_{\rm AF}}$
generate the set $\afset$ of atomic and fragile symmetry-data vectors.
A natural question is therefore which of these generators already correspond to atomic phases.
This distinction is essential for analyzing trivialization under unit-cell enlargement.
Following Ref.~\cite{song2020fragile}, we call a generator of $\afset$ a fragile root if it does not belong to $\aiset$,
or equivalently, if it does not lie in the atomic cone.
That is, it cannot be expressed as a nonnegative integer combination
of atomic Hilbert-basis generators $\{\bb h_a^{(\rm A)}\}_{a=1}^{d_{\rm A}}$ [see Eq.~\eqref{eq:hilbert_AI}].

By construction, every atomic symmetry-data vector admits a decomposition using only $\{\bb h_a^{(\rm A)}\}_{a=1}^{d_{\rm A}}$.
In contrast, no fragile symmetry-data vector admits such a decomposition.
Instead, every fragile symmetry-data vector can be decomposed using the Hilbert-basis generators $\{\bb h_b^{(\rm AF)}\}_{b=1}^{d_{\rm AF}}$, and every such decomposition necessarily contains at least one fragile root~\cite{song2020fragile}.
Note that an atomic symmetry-data vector may also admit alternative decompositions using $\{\bb h_b^{(\rm AF)}\}_{b=1}^{d_{\rm AF}}$ that contain fragile roots.
However, the existence of a decomposition using only $\{\bb h_a^{(\rm A)}\}_{a=1}^{d_{\rm A}}$ is precisely what characterizes it as atomic.
The distinction between atomic and fragile symmetry-data vectors is illustrated in Figs.~\ref{fig:hilbert}(b) and (c).

Throughout this work, we denote fragile roots by $\bb v_c^{(\rm FR)}$, where $c = 1, \dots, N_{\rm FR}$.
The remaining Hilbert-basis generators of $\afset$ belong to $\aiset$ and are denoted by $\bb v_d^{(\rm NFR)}$, where $d=1,\dots,N_{\rm NFR}$; see Fig.~\ref{fig:hilbert}(a).
Thus, the Hilbert-basis generators of $\afset$ are partitioned as
\bg
\{\bb h_b^{(\rm AF)}\}_{b=1}^{d_{\rm AF}}
= \{\bb v_c^{(\rm FR)}\}_{c=1}^{N_{\rm FR}}
\cup
\{\bb v_d^{(\rm NFR)}\}_{d=1}^{N_{\rm NFR}},
\eg
and $N_{\rm FR}+N_{\rm NFR}=d_{\rm AF}$.
Here, FR and NFR stand for fragile and non-fragile roots, respectively.
Note that $\{\bb v_d^{(\rm NFR)}\}$ need not coincide with the atomic Hilbert basis
$\{\bb h_a^{(\rm A)}\}_{a=1}^{d_{\rm A}}$.

Because of the central role of fragile roots, for analyzing stability under unit-cell enlargement, it suffices to track the transformations of the fragile roots.
Instead of examining all possible symmetry-data vectors of fragile phases, one needs to consider only the finite set of fragile roots.

\subsection*{Example: Hilbert-basis construction for $p2$}
We now illustrate the Hilbert-basis framework explicitly in the example of $p2$ before incorporating unit-cell enlargement.
In this case, the atomic sector is particularly simple.
In symmetry-data space, the Hilbert basis for atomic phases $\aiset$, $\{\bb h_a^{(\rm A)}\}_{a=1}^{d_{\rm A}}$, was computed using \texttt{Normaliz}~\cite{bruns2010normaliz}.
Its generators coincide with the symmetry-data vectors of the elementary BRs induced from maximal WPs, namely, the first eight columns of the $BR$ matrix in Eq.~\eqref{eq:p2_br_mat}, up to ordering conventions.
Explicitly, the atomic generators $\bb h^{(\rm A)}_1, \bb h^{(\rm A)}_2, \dots, \bb h^{(\rm A)}_8$ are
\ba
& (0, 1, 0, 1, 0, 1, 0, 1)^{\rm T}, \, \,
(0, 1, 0, 1, 1, 0, 1, 0)^{\rm T}, \nn
& (0, 1, 1, 0, 0, 1, 1, 0)^{\rm T}, \, \,
(0, 1, 1, 0, 1, 0, 0, 1)^{\rm T}, \nn
& (1, 0, 0, 1, 0, 1, 1, 0)^{\rm T}, \, \,
(1, 0, 0, 1, 1, 0, 0, 1)^{\rm T}, \nn
& (1, 0, 1, 0, 0, 1, 0, 1)^{\rm T}, \, \,
(1, 0, 1, 0, 1, 0, 1, 0)^{\rm T}.
\label{eq:p2_hilbert_Ai}
\ea
We therefore have $d_{\rm A}=8$.
The Hilbert basis for $\afset$, likewise computed using \texttt{Normaliz}~\cite{bruns2010normaliz}, contains $d_{\rm AF}=16$ generators.
Eight of them coincide with the atomic generators above, while the remaining eight are fragile roots.
We list these fragile roots, $\bb v^{(\rm FR)}_1, \bb v^{(\rm FR)}_2, \dots, \bb v^{(\rm FR)}_8$, explicitly:
\ba
& (2, 0, 2, 0, 2, 0, 0, 2)^{\rm T}, \, \,
(2, 0, 0, 2, 0, 2, 0, 2)^{\rm T}, \nn
& (0, 2, 2, 0, 0, 2, 0, 2)^{\rm T}, \, \,
(0, 2, 0, 2, 2, 0, 0, 2)^{\rm T}, \nn
& (2, 0, 2, 0, 0, 2, 2, 0)^{\rm T}, \, \,
(2, 0, 0, 2, 2, 0, 2, 0)^{\rm T}, \nn
& (0, 2, 2, 0, 2, 0, 2, 0)^{\rm T}, \, \,
(0, 2, 0, 2, 0, 2, 2, 0)^{\rm T}.
\ea
Among these, $\bb v^{(\rm FR)}_2$ coincides with the two-band example in Eq.~\eqref{eq:p2_v_example} discussed in Sec.~\ref{sec:p2}.
This example will serve as a convenient benchmark when we introduce the unit-cell enlargement map below.

\section{Unit-cell enlargement in the Hilbert-basis framework}
\label{sec:enlargement}
Having introduced atomic Hilbert bases and fragile roots, we now incorporate unit-cell enlargement into the Hilbert-basis framework.
We first formulate enlargement as a linear map on symmetry data and then derive a criterion for the trivialization of fragile phases.
The procedure is illustrated explicitly for the wallpaper group $p2$.

\subsection*{Unit-cell enlargement as a linear map}
We now incorporate unit-cell enlargement into the Hilbert-basis framework.
A supercell is specified by an integer matrix 
$\mc S_{\rm uc} \in GL(2,\Z)$,
with index $N_{\rm sc}=|\det \mc S_{\rm uc}|$.
Here $GL(2,\Z)$ denotes the group of $2 \times 2$ integer matrices with nonzero determinant, ensuring that the new lattice vectors define a valid two-dimensional lattice of finite index.
The matrix $\mc S_{\rm uc}$ is further subject to symmetry-compatibility conditions with a given wallpaper group: the enlargement must leave the given wallpaper group unchanged, meaning that both the point-group operations and, in nonsymmorphic cases, the associated fractional translations remain symmetries of the enlarged lattice.
For example, in the symmorphic group $p2$, the only nontrivial point-group element is $C_2 = - \mathds{1}_{2 \times 2}$, which commutes with any $S_{\rm uc} \in GL(2,\Z)$.
Therefore, every integer supercell transformation is symmetry compatible in this case.
The detailed discussion on symmetry-compatibility conditions is provided in the SM~\cite{supple}.

Under enlargement, Wannier orbitals in the primitive cell are reorganized within the supercell.
This induces an integer linear map
\bg
\bb m' = \mc M_{\rm orb} \cdot \bb m,
\label{eq:rule_orb}
\eg
where $\mc M_{\rm orb}$ depends on the choice of supercell.
The matrix $\mc M_{\rm orb}$ encodes how Wannier orbitals in the primitive unit cell are reorganized in the enlarged cell.
More precisely, it tracks how each site-symmetry irrep $(\rho)_W$ in the primitive cell is mapped to site-symmetry irreps of WPs in the supercell.
Under enlargement, a primitive-cell WP may split into several WPs of the supercell, and the corresponding site-symmetry representation may decompose accordingly.

Consequently, the symmetry-data vector in the enlarged cell is given by $\bb v' = BR \cdot \bb m' = BR \cdot \mc M_{\rm orb} \cdot \bb m$.
Using Eq.~\eqref{eq:m_from_v}, we have
\bg
\bb v' = BR \cdot \mc M_{\rm orb} \cdot BR^\ddagger \cdot \bb v + BR \cdot \mc M_{\rm orb} \cdot \bb m_{\rm ker}.
\eg
We explicitly verified, for all symmetry-compatible enlargement maps $\mc M_{\rm orb}$ examined in this work, that the second term vanishes for every wallpaper group considered, with and without spin–orbit coupling and time-reversal symmetry.
This property is closely related to the analysis of the kernel of the $BR$ matrix in Ref.~\cite{hwang2026stable}.
We therefore obtain a linear transformation in symmetry-data space,
\bg
\bb v' = BR \cdot \mc M_{\rm orb} \cdot BR^\ddagger \cdot \bb v
= \mc M_{\bb v} \cdot \bb v,
\label{eq:v_transform}
\eg
where we defined $\mc M_{\bb v} = BR \cdot \mc M_{\rm orb} \cdot BR^\ddagger$, with $BR^\ddagger$ the integer pseudoinverse of the $BR$ matrix.

The matrices $\mc M_{\rm orb}$ and $\mc M_{\bb v}$ are constrained by physical requirements:
(i) The total number of Wannier orbitals must scale with the supercell index $N_{\rm sc}$.
For a site-symmetry irrep $(\rho)_W$, the number of orbitals is given by $|W| \, {\rm dim}(\rho)$, where $|W|$ is the multiplicity of the WP $W$ (for example, in $p2$, the WP $2e$ has multiplicity $|2e|=2$, and its trivial irrep $A$ is one-dimensional).
After enlargement, this number is multiplied by $N_{\rm sc}$.
(ii) $\bb m$ may change by symmetry-preserving adiabatic deformations that do not close the energy gap.
Such deformations do not change the symmetry-data vector $\bb v$.
In general, $\bb m$ can be shifted by $q_{\rm adia}\cdot \bb z$ for any integer vector $\bb z$, where $q_{\rm adia}$ denotes the adiabatic-process matrix introduced in Eq.~\eqref{eq:p2_qadia} for $p2$.
Consistency requires that
\bg
BR \cdot \mc M_{\rm orb} \cdot q_{\rm adia} = 0 ,
\label{eq:phys_cond}
\eg
so that $\bb v'$ remains unchanged under adiabatic deformation.

\subsection*{Trivialization criterion}
To determine whether a fragile phase survives a given enlargement, it suffices to apply $\mc M_{\bb v}$ to each fragile root $\bb v^{(\rm FR)}_c$ ($c=1,\dots,N_{\rm FR}$) and check whether the resulting vector can be written as
\bg
\mc M_{\bb v} \cdot \bb v^{(\rm FR)}_c
= \sum_{a=1}^{d_{\rm A}} \, n_{c,a} \bb h^{(\rm A)}_a,
\quad
n_{c,a} \in \N_0 .
\label{eq:check_fragile}
\eg
In this work, we refer to this situation as trivialization.
Since we focus on symmetry-indicated phases, we do not consider the possibility of additional topology not captured by symmetry data.
Note that Eq.~\eqref{eq:check_fragile} is an integer linear (Diophantine) problem.
Physically, the total number of bands on the left-hand side is fixed, while each atomic Hilbert basis carries a strictly positive number of bands and the coefficients $n_{a,b}$ are nonnegative integers.
Therefore, the possible values of $n_{a,b}$ are bounded.
Since both the set of fragile roots and the atomic Hilbert basis are finite, the classification problem is decidable in finite time.

\subsection*{Example: Unit-cell enlargement in $p2$}
We now illustrate our Hilbert-basis framework for the $2 \times 1$ enlargement in $p2$.
Under this enlargement, the size of the unit cell is doubled.
Accordingly, the WPs are reorganized as follows.
Two $1a$ ($1c$) WPs of the primitive cell correspond to the $1a$ and $1b$ ($1c$ and $1d$) WPs of the supercell.
Two copies of $1b$ (respectively, $1d$) in the primitive cell combine into a $2e$ WP of the supercell.
The $2e$ WPs map to two copies of $2e$.
This can be seen in Fig.~\ref{fig:p2_sc}(a).
Also, when a maximal WP ($1a,1b,1c,1d$) is mapped to a maximal WP in the supercell, the site-symmetry irreps $A_1$ and $A_2$ remain $A_1$ and $A_2$ at the corresponding new WP.
On the other hand, when a pair of $A_1$ or $A_2$ orbitals at maximal WPs in the primitive cell is relocated to the $2e$ WP of the supercell, the pair transforms as the trivial irrep $A$ at $2e$.
Taking into account both the reorganization of the WPs and the change of site-symmetry irreps, the induced linear map on multiplicity vectors, $\mc M_{\rm orb}$ in Eq.~\eqref{eq:rule_orb}, is given by
\bg
\mc M_{\rm orb} =
\begin{pNiceArray}[last-col,code-for-last-col=\scriptstyle]{ccccccccc}
1 & 0 & 0 & 0 & 0 & 0 & 0 & 0 & 0 & (A_1)_{1a} \\
0 & 1 & 0 & 0 & 0 & 0 & 0 & 0 & 0 & (A_2)_{1a} \\
1 & 0 & 0 & 0 & 0 & 0 & 0 & 0 & 0 & (A_1)_{1b} \\
0 & 1 & 0 & 0 & 0 & 0 & 0 & 0 & 0 & (A_2)_{1b} \\
0 & 0 & 0 & 0 & 1 & 0 & 0 & 0 & 0 & (A_1)_{1c} \\
0 & 0 & 0 & 0 & 0 & 1 & 0 & 0 & 0 & (A_2)_{1c} \\
0 & 0 & 0 & 0 & 1 & 0 & 0 & 0 & 0 & (A_1)_{1d} \\
0 & 0 & 0 & 0 & 0 & 1 & 0 & 0 & 0 & (A_2)_{1d} \\
0 & 0 & 1 & 1 & 0 & 0 & 1 & 1 & 2 & (A)_{2e}
\end{pNiceArray}.
\eg
Note that $\mc M_{\rm orb}$ satisfies the physical constraint in Eq.~\eqref{eq:phys_cond}.
We then obtain the symmetry-data transformation matrix $\mc M_{\bb v}$ defined in Eq.~\eqref{eq:v_transform},
\bg
\mc M_{\bb v} = \bpm
1 & 0 & 1 & 0 & 0 & 0 & 0 & 0 \\
1 & 2 & -1 & 0 & 0 & 0 & 0 & 0 \\
1 & 1 & 0 & 0 & 0 & 0 & 0 & 0 \\
1 & 1 & 0 & 0 & 0 & 0 & 0 & 0 \\
0 & 0 & 0 & 0 & 1 & 0 & 1 & 0 \\
2 & 2 & 0 & 0 & -1 & 0 & -1 & 0 \\
1 & 1 & 0 & 0 & 0 & 0 & 0 & 0 \\
1 & 1 & 0 & 0 & 0 & 0 & 0 & 0 \epm.
\eg

Applying $\mc M_{\bb v}$ to the fragile root $\bb v^{(\rm FR)}_2 = (2,0,0,2,0,2,0,2)^{\rm T}$ discussed in Sec.~\ref{sec:p2}, we obtain
\bg
\bb v' = (2,2,2,2,0,4,2,2)^{\rm T}.
\eg
One verifies that this vector decomposes as
\bg
\bb v' = \bb h^{(\rm A)}_1 + \bb h^{(\rm A)}_3 + \bb h^{(\rm A)}_5 + \bb h^{(\rm A)}_7,
\eg
namely as a nonnegative integer combination of atomic Hilbert-basis generators [Eq.~\eqref{eq:hilbert_AI}].
Therefore, this fragile root becomes atomic under the $2 \times 1$ enlargement.
Repeating the same procedure for all fragile roots in $p2$, we find that each of them maps to an atomic symmetry-data vector.
Hence, all symmetry-indicated fragile phases in $p2$ become atomic under the smallest nontrivial unit-cell enlargement with $N_{\rm sc}=2$.

{\renewcommand{\arraystretch}{1.3}
\begin{table}[b!]
\centering
\caption{Enlargement stability of symmetry-indicated fragile topology in 2D wallpaper groups for spinless electrons, without and with time-reversal symmetry (TRS).
For each wallpaper group, we list the number of fragile roots $N_{\rm FR}$ and the minimal enlargement index $N_*$ for which all fragile roots become atomic.
The second column indicates the smallest three nontrivial supercell indices $N_{\rm sc}$ compatible with the wallpaper symmetry.
Wallpaper groups not shown in the table have $N_{\rm FR}=0$, i.e. they do not admit symmetry-indicated fragile topology.}
\label{table:spinless}
\begin{tabular*}{0.95\columnwidth}{@{\extracolsep{\fill}} l|c c c c c c c}
\hline\hline
\multirow{2}{*}{
 \begin{tabular}{c}
Wallpaper \\ group \end{tabular}}
& \multirow{2}{*}{Allowed $N_{\rm sc}$} &
& \multicolumn{2}{c}{no TRS} &
& \multicolumn{2}{c}{TRS} \\
\cline{4-5} \cline{7-8}
& & & $N_{\rm FR}$ & $N_*$ & & $N_{\rm FR}$ & $N_*$
\\ \hline
2. $p2$ & $2, 3, 4$ & & 8 & 2 & & 8 & 2
\\
6. $p2mm$ & $2, 3, 4$ & & 8 & 2 & & 8 & 2
\\
9. $c2mm$ & $3, 4, 5$ & & 4 & 3 & & 4 & 3
\\
10. $p4$ & $2, 4, 5$ & & 88 & 4 & & 12 & 4
\\
11. $p4mm$ & $2, 4, 8$ & & 12 & 4 & & 12 & 4
\\
12. $p4gm$ & $9, 25, 49$ & & 4 & 9 & & 4 & 9
\\
13. $p3$ & $3, 4, 7$ & & 45 & 3 & & 3 & 3
\\
14. $p3m1$ & $4, 9, 16$ & & 3 & 4 & & 3 & 4
\\
15. $p31m$ & $4, 9, 16$ & & 3 & 4 & & 1 & 4
\\
16. $p6$ & $3, 4, 7$ & & 198 & 7 & & 12 & 7
\\
17. $p6mm$ & $3, 4, 9$ & & 12 & 9 & & 12 & 9
\\
\hline\hline
\end{tabular*}
\end{table}}

\section{Enlargement stability in {2D} wallpaper groups}
\label{sec:procedure}
In previous sections, we illustrated explicitly how symmetry-indicated fragile topology in $p2$ for spinless electrons can be analyzed and trivialized under symmetry-compatible unit-cell enlargement.
The same framework applies to all 2D wallpaper groups, with or without time-reversal symmetry (TRS) and spin-orbit coupling (SOC).
Since its nontrivial site-symmetry groups are generated solely by a two-fold rotation, the group $p2$ is particularly simple. The possible adiabatic processes between Wannier orbital configurations are correspondingly simple, and we can easily understand the immediate instability under unit cell enlargement. In contrast, wallpaper groups having more intricate symmetry structures may exhibit stronger enlargement stability.

The key quantity in our survey is the number of fragile roots, $N_{\rm FR}$.
If $N_{\rm FR}=0$, then the given symmetry class admits no symmetry-indicated fragile topology, and the enlargement analysis is trivial.
If $N_{\rm FR}\neq 0$, we proceed as follows:

(i) First, we determine all supercells compatible with the given wallpaper symmetry.
A supercell defined by $\mc S_{\rm uc} \in GL(2, \Z)$ must preserve the wallpaper-group symmetries, which imposes algebraic constraints on $\mc S_{\rm uc}$.
From the allowed $\mc S_{\rm uc}$, we extract the possible supercell indices $N_{\rm sc}=|\det \mc S_{\rm uc}|$.
For a given $N_{\rm sc}$, there may exist multiple supercells corresponding to different $\mc S_{\rm uc}$.
Two such supercells are regarded as equivalent if they are related by a primitive-cell basis change that preserves the wallpaper symmetry.
Although different choices of $\mc S_{\rm uc}$ may lead to different labeling of WPs and HSM, they correspond to physically equivalent enlargements.
In practice, we therefore examine one representative from each symmetry-inequivalent class of supercells.
%

{\renewcommand{\arraystretch}{1.3}
\begin{table}[b!]
\centering
\caption{Enlargement stability of symmetry-indicated fragile topology in 2D wallpaper groups for spinful electrons.
The notation and table structure are the same as in Table~\ref{table:spinless}.
When $N_{\rm FR}=0$, the minimal enlargement index $N_*$ is not applicable and indicated by ``--''.}
\label{table:spinful}
\begin{tabular*}{0.95\columnwidth}{@{\extracolsep{\fill}} l|c c c c c c c}
\hline\hline
\multirow{2}{*}{
 \begin{tabular}{c}
Wallpaper \\ group \end{tabular}}
& \multirow{2}{*}{Allowed $N_{\rm sc}$} &
& \multicolumn{2}{c}{no TRS} &
& \multicolumn{2}{c}{TRS} \\
\cline{4-5} \cline{7-8}
& & & $N_{\rm FR}$ & $N_*$ & & $N_{\rm FR}$ & $N_*$
\\ \hline
2. $p2$ & $2, 3, 4$ & & 8 & 2 & & 0 & --
\\
10. $p4$ & $2, 4, 5$ & & 88 & 4 & & 0 & --
\\
12. $p4gm$ & $9, 25, 49$ & & 4 & 9 & & 0 & --
\\
13. $p3$ & $3, 4, 7$ & & 45 & 3 & & 6 & 4
\\
14. $p3m1$ & $4, 9, 16$ & & 3 & 4 & & 6 & 4
\\
15. $p31m$ & $4, 9, 16$ & & 3 & 4 & & 2 & 4
\\
16. $p6$ & $3, 4, 7$ & & 198 & 7 & & 3 & 4
\\
17. $p6mm$ & $3, 4, 9$ & & 2 & 3 & & 3 & 4
\\
\hline\hline
\end{tabular*}
\end{table}}

(ii) For each allowed supercell, we construct the corresponding orbital map $\mc M_{\rm orb}$ defined in Eq.~\eqref{eq:rule_orb}.
This requires tracking how WPs in the primitive cell are reorganized in the supercell.
A given WP in the primitive cell can map only to WPs in the supercell whose site-symmetry group is a subgroup (possibly equal) of the original one.
Accordingly, site-symmetry irreps decompose under restriction to the subgroup.
Once $\mc M_{\rm orb}$ is determined, we construct the symmetry-data map $\bb v' = \mc M_{\bb v} \cdot \bb v$, according to Eq.~\eqref{eq:v_transform}.
The BR matrix $BR$, and the necessary group-theoretical information to perform this construction, can be obtained from the Bilbao Crystallographic Server~\cite{aroyo2006bilbao1,aroyo2006bilbao2,aroyo2011crystallography,vergniory2017graph,elcoro2017double} using the \texttt{WYCKPOS}, \texttt{Representations PG}, and \texttt{BANDREP} tools~\cite{bilbao_tools}.

For each fragile root $\bb v^{(\rm FR)}_a$, we then solve the Diophantine problem in Eq.~\eqref{eq:check_fragile} to test whether it becomes atomic.
If there exists a symmetry-compatible supercell of index $N_{\rm sc}$ for which all fragile roots map to atomic symmetry-data vectors, then all fragile phases trivialize at that enlargement scale.
Otherwise, we proceed to the next allowed $N_{\rm sc}$.
Because both the number of fragile roots and the atomic Hilbert basis are finite, and because each root corresponds to a fixed total band number, this procedure terminates in finite time.
For each symmetry class, we define $N_*$ as the minimal supercell index for which all fragile roots trivialize.
If $N_{\rm FR}=0$, we indicate $N_*$ as not applicable.
Note that $N_*$ is an existence statement.
For each symmetry class, $N_*$ is the smallest index at which all fragile roots become atomic.
Different supercells with larger indices need not trivialize them.
The results for spinless electrons are summarized in Table~\ref{table:spinless}, and the corresponding results for spinful electrons in Table~\ref{table:spinful}.

Several general conclusions can be drawn:
(i) In all symmetry classes admitting fragile roots, i.e. $N_{\rm FR} \neq 0$, a finite minimal enlargement index $N_*$ exists.
That is, symmetry-indicated fragile topology in 2D wallpaper groups always has finite enlargement stability and cannot persist under sufficiently large translation refinement compatible with the given wallpaper symmetry.
(ii) In all cases, each fragile root corresponds to at least two bands.
This is consistent with the absence of one-band fragile topology~\cite{alexandradinata2018no}, and reflects the fact that symmetry-indicated fragile topology necessarily involves multi-band structures.
(iii) In the spinless case, imposing TRS generally reduces (or leaves unchanged) the number of fragile roots for a given wallpaper group.
This can be understood from the formation of corepresentations at HSM, which pair complex-conjugate irreps and thereby constrain the structure of symmetry-data vectors.
However, this trend is not universal: in the presence of SOC, the interplay between TRS and double-valued representations can modify this behavior.
(iv) Higher-symmetry wallpaper groups can exhibit larger minimal enlargement indices $N_*$.
In contrast to the spinless $p2$ case, which we used as a prototypical example, for several higher-symmetric wallpaper groups the smallest nontrivial enlargement fails to trivialize all fragile roots, and larger $N_{\rm sc}$ are required.
Examples include $p4$ and $p4mm$ for spinless electrons, and $p6$ for spinful electrons.
For the complete list of such symmetry classes, we refer to Tables~\ref{table:spinless} and~\ref{table:spinful}.
A representative nontrivial example that requires a larger minimal enlargement index $N_*$ is discussed in the SM~\cite{supple}.

\section{Discussion}
\label{sec:discussion}
We have shown that all symmetry-indicated fragile phases in 2D wallpaper groups are eventually unstable under symmetry-compatible unit-cell enlargement.
Explicitly, whenever fragile roots exist, there always exists a finite supercell index $N_*$ at which all fragile roots become atomic.
In this sense, fragile topology has a finite stability depth with respect to translation-symmetry refinement.

This result clarifies the relation between fragile topology and translation symmetry.
While fragile topology is defined through the obstruction to forming symmetric and exponentially localized Wannier functions within a given unit cell, refining the translation subgroup can remove this obstruction.
Unit-cell enlargement therefore provides a complementary perspective to the addition of trivial bands: both operations may trivialize fragile topology, but they act in conceptually distinct ways.
The former modifies the translation symmetry, while the latter enlarges the Hilbert space at fixed choice of unit cell.

Our results also suggest constraints on physical signatures attributed to fragile topology.
Although our analysis does not by itself establish a strict criterion for response signatures that genuinely distinguish fragile phases from atomic ones, it indicates that responses invariant under symmetry-compatible unit-cell enlargement cannot serve as unique markers of fragile topology, since such phases can be rendered atomic by a suitable enlargement.
Quantities related to intrinsic fragile topological invariants may therefore offer more direct insight.

Furthermore, our results show how 2D symmetry-indicated fragile phases may be destabilized by translation-symmetry-breaking perturbations, including commensurate moiré or superlattice potentials~\cite{cano2021moire,scheer2023kagome,crepel2025efficient,yang2024topological,yang2025engineering,lhachemi2026efficient} and charge density wave modulations~\cite{mcmillan1976theory,gruner1988dynamics,wieder2020axionic,devescovi2024axion} that preserve the relevant wallpaper-group symmetries. 
Such a perturbation trivializes all fragile roots when it realizes a symmetry-compatible supercell with index equal to $N_*$.

A related setting is provided by Hofstadter systems where magnetic flux produces a projective magnetic-translation algebra and can generate topological phases at high-symmetry flux that are absent at zero flux~\cite{herzog2020hofstadter,herzog2023hofstadter,fang2023symmetry}. 
Although magnetic flux is not considered here, our results remain relevant to fragile phases protected by local rotational symmetries in wallpaper groups $pn$ $(n=2,3,4,6)$.
These rotations remain exact at nonzero flux, and, in a symmetric gauge, their local irrep labels may be identified with those at zero flux.
In contrast, generic flux breaks crystalline symmetries that reverse the magnetic field, so fragile topology protected solely by such symmetries loses its symmetry protection at infinitesimal flux. 
A complete extension of our analysis would require constructing the magnetic fragile roots and the corresponding enlargement maps for each wallpaper group and rational flux, which presents an interesting direction for future work.

An important open question is whether the finite enlargement stability established here for 2D wallpaper groups persists in 3D.
While our results demonstrate finite stability for all symmetry-indicated fragile topology in 2D, the situation in 3D may be more intricate due to the richer structure of space-group symmetries.
In this work, we focused on symmetry-indicated fragile topology diagnosed by symmetry data.
It is therefore natural to ask how non-symmetry-indicated fragile topology behaves under unit-cell enlargement.
A representative example is fragile topology characterized by an Euler class~\cite{zhao2017pt,ahn2018band,ahn2019failure} protected by $C_{2z}T$ symmetry.
Since the Euler class is defined for two-band systems, such phases are expected to trivialize under sufficiently large enlargement.

More generally, although our Hilbert-basis framework focuses on symmetry-indicated fragile topology, the adiabatic-deformation approach introduced in Sec.~\ref{sec:p2} with adiabatic-process matrix $q_{\rm adia}$, can in principle be applied on a case-by-case basis to non-indicated fragile phases.
However, a general understanding of the fate of arbitrary non-indicated fragile topology under translation refinement remains an open problem.

Finally, our analysis considered enlargements that preserve a given wallpaper-group symmetry.
More general enlargements may reduce the symmetry group or even relate distinct wallpaper groups.
Understanding how fragile topology transforms under such symmetry-changing enlargements presents another interesting direction for future study.

\begin{acknowledgments}
Y.H. and T.L.H thank the US Office of Naval Research (ONR) Multidisciplinary University Research Initiative (MURI) grant N00014-20-1-2325 on Robust Photonic Materials with High-Order Topological Protection.
Y.H. received additional support from the Air Force Office of Scientific Research under award number FA9550-21-1-0131 and a UKRI Future Leaders Fellowship MR/Y017331/1.
S.V. is supported by the Dirac Postdoctoral Fellowship, sponsored by the National High Magnetic Field Laboratory (NHMFL/MagLab).
NHMFL/MagLab is funded by NSF DMR-2128556 and the State of Florida.
\end{acknowledgments}

\bibliography{refs.bib}

\begin{thebibliography}{99}%
\makeatletter
\providecommand \@ifxundefined [1]{%
 \@ifx{#1\undefined}
}%
\providecommand \@ifnum [1]{%
 \ifnum #1\expandafter \@firstoftwo
 \else \expandafter \@secondoftwo
 \fi
}%
\providecommand \@ifx [1]{%
 \ifx #1\expandafter \@firstoftwo
 \else \expandafter \@secondoftwo
 \fi
}%
\providecommand \natexlab [1]{#1}%
\providecommand \enquote  [1]{``#1''}%
\providecommand \bibnamefont  [1]{#1}%
\providecommand \bibfnamefont [1]{#1}%
\providecommand \citenamefont [1]{#1}%
\providecommand \href@noop [0]{\@secondoftwo}%
\providecommand \href [0]{\begingroup \@sanitize@url \@href}%
\providecommand \@href[1]{\@@startlink{#1}\@@href}%
\providecommand \@@href[1]{\endgroup#1\@@endlink}%
\providecommand \@sanitize@url [0]{\catcode `\\12\catcode `\$12\catcode
  `\&12\catcode `\#12\catcode `\^12\catcode `\_12\catcode `\%12\relax}%
\providecommand \@@startlink[1]{}%
\providecommand \@@endlink[0]{}%
\providecommand \url  [0]{\begingroup\@sanitize@url \@url }%
\providecommand \@url [1]{\endgroup\@href {#1}{\urlprefix }}%
\providecommand \urlprefix  [0]{URL }%
\providecommand \Eprint [0]{\href }%
\providecommand \doibase [0]{https://doi.org/}%
\providecommand \selectlanguage [0]{\@gobble}%
\providecommand \bibinfo  [0]{\@secondoftwo}%
\providecommand \bibfield  [0]{\@secondoftwo}%
\providecommand \translation [1]{[#1]}%
\providecommand \BibitemOpen [0]{}%
\providecommand \bibitemStop [0]{}%
\providecommand \bibitemNoStop [0]{.\EOS\space}%
\providecommand \EOS [0]{\spacefactor3000\relax}%
\providecommand \BibitemShut  [1]{\csname bibitem#1\endcsname}%
\let\auto@bib@innerbib\@empty
\bibitem [{\citenamefont {Shiozaki}\ \emph {et~al.}(2016)\citenamefont
  {Shiozaki}, \citenamefont {Sato},\ and\ \citenamefont
  {Gomi}}]{shiozaki2016topology}%
  \BibitemOpen
  \bibfield  {author} {\bibinfo {author} {\bibfnamefont {K.}~\bibnamefont
  {Shiozaki}}, \bibinfo {author} {\bibfnamefont {M.}~\bibnamefont {Sato}},\
  and\ \bibinfo {author} {\bibfnamefont {K.}~\bibnamefont {Gomi}},\ }\bibfield
  {title} {\bibinfo {title} {Topology of nonsymmorphic crystalline insulators
  and superconductors},\ }\href {https://doi.org/10.1103/PhysRevB.93.195413}
  {\bibfield  {journal} {\bibinfo  {journal} {Physical Review B}\ }\textbf
  {\bibinfo {volume} {93}},\ \bibinfo {pages} {195413} (\bibinfo {year}
  {2016})}\BibitemShut {NoStop}%
\bibitem [{\citenamefont {Shiozaki}\ \emph {et~al.}(2017)\citenamefont
  {Shiozaki}, \citenamefont {Sato},\ and\ \citenamefont
  {Gomi}}]{shiozaki2017topological}%
  \BibitemOpen
  \bibfield  {author} {\bibinfo {author} {\bibfnamefont {K.}~\bibnamefont
  {Shiozaki}}, \bibinfo {author} {\bibfnamefont {M.}~\bibnamefont {Sato}},\
  and\ \bibinfo {author} {\bibfnamefont {K.}~\bibnamefont {Gomi}},\ }\bibfield
  {title} {\bibinfo {title} {Topological crystalline materials: {General}
  formulation, module structure, and wallpaper groups},\ }\href
  {https://doi.org/10.1103/PhysRevB.95.235425} {\bibfield  {journal} {\bibinfo
  {journal} {Physical Review B}\ }\textbf {\bibinfo {volume} {95}},\ \bibinfo
  {pages} {235425} (\bibinfo {year} {2017})}\BibitemShut {NoStop}%
\bibitem [{\citenamefont {Shiozaki}\ \emph {et~al.}(2022)\citenamefont
  {Shiozaki}, \citenamefont {Sato},\ and\ \citenamefont
  {Gomi}}]{shiozaki2022atiyah}%
  \BibitemOpen
  \bibfield  {author} {\bibinfo {author} {\bibfnamefont {K.}~\bibnamefont
  {Shiozaki}}, \bibinfo {author} {\bibfnamefont {M.}~\bibnamefont {Sato}},\
  and\ \bibinfo {author} {\bibfnamefont {K.}~\bibnamefont {Gomi}},\ }\bibfield
  {title} {\bibinfo {title} {{Atiyah}-{Hirzebruch} spectral sequence in band
  topology: {General} formalism and topological invariants for 230 space
  groups},\ }\href {https://doi.org/10.1103/PhysRevB.106.165103} {\bibfield
  {journal} {\bibinfo  {journal} {Physical Review B}\ }\textbf {\bibinfo
  {volume} {106}},\ \bibinfo {pages} {165103} (\bibinfo {year}
  {2022})}\BibitemShut {NoStop}%
\bibitem [{\citenamefont {Kruthoff}\ \emph {et~al.}(2017)\citenamefont
  {Kruthoff}, \citenamefont {De~Boer}, \citenamefont {Van~Wezel}, \citenamefont
  {Kane},\ and\ \citenamefont {Slager}}]{kruthoff2017topological}%
  \BibitemOpen
  \bibfield  {author} {\bibinfo {author} {\bibfnamefont {J.}~\bibnamefont
  {Kruthoff}}, \bibinfo {author} {\bibfnamefont {J.}~\bibnamefont {De~Boer}},
  \bibinfo {author} {\bibfnamefont {J.}~\bibnamefont {Van~Wezel}}, \bibinfo
  {author} {\bibfnamefont {C.~L.}\ \bibnamefont {Kane}},\ and\ \bibinfo
  {author} {\bibfnamefont {R.-J.}\ \bibnamefont {Slager}},\ }\bibfield  {title}
  {\bibinfo {title} {Topological classification of crystalline insulators
  through band structure combinatorics},\ }\href
  {https://doi.org/10.1103/PhysRevX.7.041069} {\bibfield  {journal} {\bibinfo
  {journal} {Physical Review X}\ }\textbf {\bibinfo {volume} {7}},\ \bibinfo
  {pages} {041069} (\bibinfo {year} {2017})}\BibitemShut {NoStop}%
\bibitem [{\citenamefont {Po}\ \emph {et~al.}(2017)\citenamefont {Po},
  \citenamefont {Vishwanath},\ and\ \citenamefont {Watanabe}}]{po2017symmetry}%
  \BibitemOpen
  \bibfield  {author} {\bibinfo {author} {\bibfnamefont {H.~C.}\ \bibnamefont
  {Po}}, \bibinfo {author} {\bibfnamefont {A.}~\bibnamefont {Vishwanath}},\
  and\ \bibinfo {author} {\bibfnamefont {H.}~\bibnamefont {Watanabe}},\
  }\bibfield  {title} {\bibinfo {title} {Symmetry-based indicators of band
  topology in the 230 space groups},\ }\href
  {https://doi.org/10.1038/s41467-017-00133-2} {\bibfield  {journal} {\bibinfo
  {journal} {Nature Communications}\ }\textbf {\bibinfo {volume} {8}},\
  \bibinfo {pages} {50} (\bibinfo {year} {2017})}\BibitemShut {NoStop}%
\bibitem [{\citenamefont {Bradlyn}\ \emph {et~al.}(2017)\citenamefont
  {Bradlyn}, \citenamefont {Elcoro}, \citenamefont {Cano}, \citenamefont
  {Vergniory}, \citenamefont {Wang}, \citenamefont {Felser}, \citenamefont
  {Aroyo},\ and\ \citenamefont {Bernevig}}]{bradlyn2017topological}%
  \BibitemOpen
  \bibfield  {author} {\bibinfo {author} {\bibfnamefont {B.}~\bibnamefont
  {Bradlyn}}, \bibinfo {author} {\bibfnamefont {L.}~\bibnamefont {Elcoro}},
  \bibinfo {author} {\bibfnamefont {J.}~\bibnamefont {Cano}}, \bibinfo {author}
  {\bibfnamefont {M.~G.}\ \bibnamefont {Vergniory}}, \bibinfo {author}
  {\bibfnamefont {Z.}~\bibnamefont {Wang}}, \bibinfo {author} {\bibfnamefont
  {C.}~\bibnamefont {Felser}}, \bibinfo {author} {\bibfnamefont {M.~I.}\
  \bibnamefont {Aroyo}},\ and\ \bibinfo {author} {\bibfnamefont {B.~A.}\
  \bibnamefont {Bernevig}},\ }\bibfield  {title} {\bibinfo {title} {Topological
  quantum chemistry},\ }\href {https://doi.org/10.1038/nature23268} {\bibfield
  {journal} {\bibinfo  {journal} {Nature}\ }\textbf {\bibinfo {volume} {547}},\
  \bibinfo {pages} {298} (\bibinfo {year} {2017})}\BibitemShut {NoStop}%
\bibitem [{\citenamefont {Bradlyn}\ \emph {et~al.}(2016)\citenamefont
  {Bradlyn}, \citenamefont {Cano}, \citenamefont {Wang}, \citenamefont
  {Vergniory}, \citenamefont {Felser}, \citenamefont {Cava},\ and\
  \citenamefont {Bernevig}}]{bradlyn2016beyond}%
  \BibitemOpen
  \bibfield  {author} {\bibinfo {author} {\bibfnamefont {B.}~\bibnamefont
  {Bradlyn}}, \bibinfo {author} {\bibfnamefont {J.}~\bibnamefont {Cano}},
  \bibinfo {author} {\bibfnamefont {Z.}~\bibnamefont {Wang}}, \bibinfo {author}
  {\bibfnamefont {M.}~\bibnamefont {Vergniory}}, \bibinfo {author}
  {\bibfnamefont {C.}~\bibnamefont {Felser}}, \bibinfo {author} {\bibfnamefont
  {R.~J.}\ \bibnamefont {Cava}},\ and\ \bibinfo {author} {\bibfnamefont
  {B.~A.}\ \bibnamefont {Bernevig}},\ }\bibfield  {title} {\bibinfo {title}
  {Beyond {Dirac} and {Weyl} fermions: {Unconventional} quasiparticles in
  conventional crystals},\ }\href {https://doi.org/10.1126/science.aaf5037}
  {\bibfield  {journal} {\bibinfo  {journal} {Science}\ }\textbf {\bibinfo
  {volume} {353}},\ \bibinfo {pages} {aaf5037} (\bibinfo {year}
  {2016})}\BibitemShut {NoStop}%
\bibitem [{\citenamefont {Wieder}\ \emph {et~al.}(2018)\citenamefont {Wieder},
  \citenamefont {Bradlyn}, \citenamefont {Wang}, \citenamefont {Cano},
  \citenamefont {Kim}, \citenamefont {Kim}, \citenamefont {Rappe},
  \citenamefont {Kane},\ and\ \citenamefont {Bernevig}}]{wieder2018wallpaper}%
  \BibitemOpen
  \bibfield  {author} {\bibinfo {author} {\bibfnamefont {B.~J.}\ \bibnamefont
  {Wieder}}, \bibinfo {author} {\bibfnamefont {B.}~\bibnamefont {Bradlyn}},
  \bibinfo {author} {\bibfnamefont {Z.}~\bibnamefont {Wang}}, \bibinfo {author}
  {\bibfnamefont {J.}~\bibnamefont {Cano}}, \bibinfo {author} {\bibfnamefont
  {Y.}~\bibnamefont {Kim}}, \bibinfo {author} {\bibfnamefont {H.-S.~D.}\
  \bibnamefont {Kim}}, \bibinfo {author} {\bibfnamefont {A.~M.}\ \bibnamefont
  {Rappe}}, \bibinfo {author} {\bibfnamefont {C.}~\bibnamefont {Kane}},\ and\
  \bibinfo {author} {\bibfnamefont {B.~A.}\ \bibnamefont {Bernevig}},\
  }\bibfield  {title} {\bibinfo {title} {Wallpaper fermions and the
  nonsymmorphic {Dirac} insulator},\ }\href
  {https://doi.org/10.1126/science.aan2802} {\bibfield  {journal} {\bibinfo
  {journal} {Science}\ }\textbf {\bibinfo {volume} {361}},\ \bibinfo {pages}
  {246} (\bibinfo {year} {2018})}\BibitemShut {NoStop}%
\bibitem [{\citenamefont {Hwang}\ \emph {et~al.}(2023)\citenamefont {Hwang},
  \citenamefont {Qian}, \citenamefont {Kang}, \citenamefont {Lee},
  \citenamefont {Ryu}, \citenamefont {Choi},\ and\ \citenamefont
  {Yang}}]{hwang2023magnetic}%
  \BibitemOpen
  \bibfield  {author} {\bibinfo {author} {\bibfnamefont {Y.}~\bibnamefont
  {Hwang}}, \bibinfo {author} {\bibfnamefont {Y.}~\bibnamefont {Qian}},
  \bibinfo {author} {\bibfnamefont {J.}~\bibnamefont {Kang}}, \bibinfo {author}
  {\bibfnamefont {J.}~\bibnamefont {Lee}}, \bibinfo {author} {\bibfnamefont
  {D.}~\bibnamefont {Ryu}}, \bibinfo {author} {\bibfnamefont {H.~C.}\
  \bibnamefont {Choi}},\ and\ \bibinfo {author} {\bibfnamefont {B.-J.}\
  \bibnamefont {Yang}},\ }\bibfield  {title} {\bibinfo {title} {Magnetic
  wallpaper {Dirac} fermions and topological magnetic {Dirac} insulators},\
  }\href {https://doi.org/10.1038/s41524-023-01018-3} {\bibfield  {journal}
  {\bibinfo  {journal} {npj Computational Materials}\ }\textbf {\bibinfo
  {volume} {9}},\ \bibinfo {pages} {65} (\bibinfo {year} {2023})}\BibitemShut
  {NoStop}%
\bibitem [{\citenamefont {Po}\ \emph {et~al.}(2018)\citenamefont {Po},
  \citenamefont {Watanabe},\ and\ \citenamefont {Vishwanath}}]{po2018fragile}%
  \BibitemOpen
  \bibfield  {author} {\bibinfo {author} {\bibfnamefont {H.~C.}\ \bibnamefont
  {Po}}, \bibinfo {author} {\bibfnamefont {H.}~\bibnamefont {Watanabe}},\ and\
  \bibinfo {author} {\bibfnamefont {A.}~\bibnamefont {Vishwanath}},\ }\bibfield
   {title} {\bibinfo {title} {Fragile topology and {Wannier} obstructions},\
  }\href {https://doi.org/10.1103/PhysRevLett.121.126402} {\bibfield  {journal}
  {\bibinfo  {journal} {Physical Review Letters}\ }\textbf {\bibinfo {volume}
  {121}},\ \bibinfo {pages} {126402} (\bibinfo {year} {2018})}\BibitemShut
  {NoStop}%
\bibitem [{\citenamefont {Cano}\ \emph {et~al.}(2018)\citenamefont {Cano},
  \citenamefont {Bradlyn}, \citenamefont {Wang}, \citenamefont {Elcoro},
  \citenamefont {Vergniory}, \citenamefont {Felser}, \citenamefont {Aroyo},\
  and\ \citenamefont {Bernevig}}]{cano2018topology}%
  \BibitemOpen
  \bibfield  {author} {\bibinfo {author} {\bibfnamefont {J.}~\bibnamefont
  {Cano}}, \bibinfo {author} {\bibfnamefont {B.}~\bibnamefont {Bradlyn}},
  \bibinfo {author} {\bibfnamefont {Z.}~\bibnamefont {Wang}}, \bibinfo {author}
  {\bibfnamefont {L.}~\bibnamefont {Elcoro}}, \bibinfo {author} {\bibfnamefont
  {M.}~\bibnamefont {Vergniory}}, \bibinfo {author} {\bibfnamefont
  {C.}~\bibnamefont {Felser}}, \bibinfo {author} {\bibfnamefont
  {M.}~\bibnamefont {Aroyo}},\ and\ \bibinfo {author} {\bibfnamefont {B.~A.}\
  \bibnamefont {Bernevig}},\ }\bibfield  {title} {\bibinfo {title} {Topology of
  disconnected elementary band representations},\ }\href
  {https://doi.org/10.1103/PhysRevLett.120.266401} {\bibfield  {journal}
  {\bibinfo  {journal} {Physical Review Letters}\ }\textbf {\bibinfo {volume}
  {120}},\ \bibinfo {pages} {266401} (\bibinfo {year} {2018})}\BibitemShut
  {NoStop}%
\bibitem [{\citenamefont {Bradlyn}\ \emph {et~al.}(2019)\citenamefont
  {Bradlyn}, \citenamefont {Wang}, \citenamefont {Cano},\ and\ \citenamefont
  {Bernevig}}]{bradlyn2019disconnected}%
  \BibitemOpen
  \bibfield  {author} {\bibinfo {author} {\bibfnamefont {B.}~\bibnamefont
  {Bradlyn}}, \bibinfo {author} {\bibfnamefont {Z.}~\bibnamefont {Wang}},
  \bibinfo {author} {\bibfnamefont {J.}~\bibnamefont {Cano}},\ and\ \bibinfo
  {author} {\bibfnamefont {B.~A.}\ \bibnamefont {Bernevig}},\ }\bibfield
  {title} {\bibinfo {title} {Disconnected elementary band representations,
  fragile topology, and {Wilson} loops as topological indices: {An} example on
  the triangular lattice},\ }\href {https://doi.org/10.1103/PhysRevB.99.045140}
  {\bibfield  {journal} {\bibinfo  {journal} {Physical Review B}\ }\textbf
  {\bibinfo {volume} {99}},\ \bibinfo {pages} {045140} (\bibinfo {year}
  {2019})}\BibitemShut {NoStop}%
\bibitem [{\citenamefont {Bouhon}\ \emph {et~al.}(2019)\citenamefont {Bouhon},
  \citenamefont {Black-Schaffer},\ and\ \citenamefont
  {Slager}}]{bouhon2019wilson}%
  \BibitemOpen
  \bibfield  {author} {\bibinfo {author} {\bibfnamefont {A.}~\bibnamefont
  {Bouhon}}, \bibinfo {author} {\bibfnamefont {A.~M.}\ \bibnamefont
  {Black-Schaffer}},\ and\ \bibinfo {author} {\bibfnamefont {R.-J.}\
  \bibnamefont {Slager}},\ }\bibfield  {title} {\bibinfo {title} {{Wilson} loop
  approach to fragile topology of split elementary band representations and
  topological crystalline insulators with time-reversal symmetry},\ }\href
  {https://doi.org/10.1103/PhysRevB.100.195135} {\bibfield  {journal} {\bibinfo
   {journal} {Physical Review B}\ }\textbf {\bibinfo {volume} {100}},\ \bibinfo
  {pages} {195135} (\bibinfo {year} {2019})}\BibitemShut {NoStop}%
\bibitem [{\citenamefont {Kooi}\ \emph {et~al.}(2019)\citenamefont {Kooi},
  \citenamefont {Van~Miert},\ and\ \citenamefont
  {Ortix}}]{kooi2019classification}%
  \BibitemOpen
  \bibfield  {author} {\bibinfo {author} {\bibfnamefont {S.~H.}\ \bibnamefont
  {Kooi}}, \bibinfo {author} {\bibfnamefont {G.}~\bibnamefont {Van~Miert}},\
  and\ \bibinfo {author} {\bibfnamefont {C.}~\bibnamefont {Ortix}},\ }\bibfield
   {title} {\bibinfo {title} {Classification of crystalline insulators without
  symmetry indicators: {Atomic} and fragile topological phases in twofold
  rotation symmetric systems},\ }\href
  {https://doi.org/10.1103/PhysRevB.100.115160} {\bibfield  {journal} {\bibinfo
   {journal} {Physical Review B}\ }\textbf {\bibinfo {volume} {100}},\ \bibinfo
  {pages} {115160} (\bibinfo {year} {2019})}\BibitemShut {NoStop}%
\bibitem [{\citenamefont {Bouhon}\ \emph {et~al.}(2020)\citenamefont {Bouhon},
  \citenamefont {Bzdu{\v{s}}ek},\ and\ \citenamefont
  {Slager}}]{bouhon2020geometric}%
  \BibitemOpen
  \bibfield  {author} {\bibinfo {author} {\bibfnamefont {A.}~\bibnamefont
  {Bouhon}}, \bibinfo {author} {\bibfnamefont {T.}~\bibnamefont
  {Bzdu{\v{s}}ek}},\ and\ \bibinfo {author} {\bibfnamefont {R.-J.}\
  \bibnamefont {Slager}},\ }\bibfield  {title} {\bibinfo {title} {Geometric
  approach to fragile topology beyond symmetry indicators},\ }\href
  {https://doi.org/10.1103/PhysRevB.102.115135} {\bibfield  {journal} {\bibinfo
   {journal} {Physical Review B}\ }\textbf {\bibinfo {volume} {102}},\ \bibinfo
  {pages} {115135} (\bibinfo {year} {2020})}\BibitemShut {NoStop}%
\bibitem [{\citenamefont {Brouwer}\ and\ \citenamefont
  {Dwivedi}(2023)}]{brouwer2023homotopic}%
  \BibitemOpen
  \bibfield  {author} {\bibinfo {author} {\bibfnamefont {P.~W.}\ \bibnamefont
  {Brouwer}}\ and\ \bibinfo {author} {\bibfnamefont {V.}~\bibnamefont
  {Dwivedi}},\ }\bibfield  {title} {\bibinfo {title} {Homotopic classification
  of band structures: {Stable}, fragile, delicate, and stable
  representation-protected topology},\ }\href
  {https://doi.org/10.1103/PhysRevB.108.155137} {\bibfield  {journal} {\bibinfo
   {journal} {Physical Review B}\ }\textbf {\bibinfo {volume} {108}},\ \bibinfo
  {pages} {155137} (\bibinfo {year} {2023})}\BibitemShut {NoStop}%
\bibitem [{\citenamefont {Song}\ \emph {et~al.}(2019)\citenamefont {Song},
  \citenamefont {Wang}, \citenamefont {Shi}, \citenamefont {Li}, \citenamefont
  {Fang},\ and\ \citenamefont {Bernevig}}]{song2019all}%
  \BibitemOpen
  \bibfield  {author} {\bibinfo {author} {\bibfnamefont {Z.}~\bibnamefont
  {Song}}, \bibinfo {author} {\bibfnamefont {Z.}~\bibnamefont {Wang}}, \bibinfo
  {author} {\bibfnamefont {W.}~\bibnamefont {Shi}}, \bibinfo {author}
  {\bibfnamefont {G.}~\bibnamefont {Li}}, \bibinfo {author} {\bibfnamefont
  {C.}~\bibnamefont {Fang}},\ and\ \bibinfo {author} {\bibfnamefont {B.~A.}\
  \bibnamefont {Bernevig}},\ }\bibfield  {title} {\bibinfo {title} {All magic
  angles in twisted bilayer graphene are topological},\ }\href
  {https://doi.org/10.1103/PhysRevLett.123.036401} {\bibfield  {journal}
  {\bibinfo  {journal} {Physical Review Letters}\ }\textbf {\bibinfo {volume}
  {123}},\ \bibinfo {pages} {036401} (\bibinfo {year} {2019})}\BibitemShut
  {NoStop}%
\bibitem [{\citenamefont {Po}\ \emph {et~al.}(2019)\citenamefont {Po},
  \citenamefont {Zou}, \citenamefont {Senthil},\ and\ \citenamefont
  {Vishwanath}}]{po2019faithful}%
  \BibitemOpen
  \bibfield  {author} {\bibinfo {author} {\bibfnamefont {H.~C.}\ \bibnamefont
  {Po}}, \bibinfo {author} {\bibfnamefont {L.}~\bibnamefont {Zou}}, \bibinfo
  {author} {\bibfnamefont {T.}~\bibnamefont {Senthil}},\ and\ \bibinfo {author}
  {\bibfnamefont {A.}~\bibnamefont {Vishwanath}},\ }\bibfield  {title}
  {\bibinfo {title} {Faithful tight-binding models and fragile topology of
  magic-angle bilayer graphene},\ }\href
  {https://doi.org/10.1103/PhysRevB.99.195455} {\bibfield  {journal} {\bibinfo
  {journal} {Physical Review B}\ }\textbf {\bibinfo {volume} {99}},\ \bibinfo
  {pages} {195455} (\bibinfo {year} {2019})}\BibitemShut {NoStop}%
\bibitem [{\citenamefont {Ahn}\ \emph {et~al.}(2019)\citenamefont {Ahn},
  \citenamefont {Park},\ and\ \citenamefont {Yang}}]{ahn2019failure}%
  \BibitemOpen
  \bibfield  {author} {\bibinfo {author} {\bibfnamefont {J.}~\bibnamefont
  {Ahn}}, \bibinfo {author} {\bibfnamefont {S.}~\bibnamefont {Park}},\ and\
  \bibinfo {author} {\bibfnamefont {B.-J.}\ \bibnamefont {Yang}},\ }\bibfield
  {title} {\bibinfo {title} {Failure of {Nielsen}-{Ninomiya} theorem and
  fragile topology in two-dimensional systems with space-time inversion
  symmetry: {Application} to twisted bilayer graphene at magic angle},\ }\href
  {https://doi.org/10.1103/PhysRevX.9.021013} {\bibfield  {journal} {\bibinfo
  {journal} {Physical Review X}\ }\textbf {\bibinfo {volume} {9}},\ \bibinfo
  {pages} {021013} (\bibinfo {year} {2019})}\BibitemShut {NoStop}%
\bibitem [{\citenamefont {Vergniory}\ \emph {et~al.}(2022)\citenamefont
  {Vergniory}, \citenamefont {Wieder}, \citenamefont {Elcoro}, \citenamefont
  {Parkin}, \citenamefont {Felser}, \citenamefont {Bernevig},\ and\
  \citenamefont {Regnault}}]{vergniory2022all}%
  \BibitemOpen
  \bibfield  {author} {\bibinfo {author} {\bibfnamefont {M.~G.}\ \bibnamefont
  {Vergniory}}, \bibinfo {author} {\bibfnamefont {B.~J.}\ \bibnamefont
  {Wieder}}, \bibinfo {author} {\bibfnamefont {L.}~\bibnamefont {Elcoro}},
  \bibinfo {author} {\bibfnamefont {S.~S.}\ \bibnamefont {Parkin}}, \bibinfo
  {author} {\bibfnamefont {C.}~\bibnamefont {Felser}}, \bibinfo {author}
  {\bibfnamefont {B.~A.}\ \bibnamefont {Bernevig}},\ and\ \bibinfo {author}
  {\bibfnamefont {N.}~\bibnamefont {Regnault}},\ }\bibfield  {title} {\bibinfo
  {title} {All topological bands of all nonmagnetic stoichiometric materials},\
  }\href {https://doi.org/10.1126/science.abg9094} {\bibfield  {journal}
  {\bibinfo  {journal} {Science}\ }\textbf {\bibinfo {volume} {376}},\ \bibinfo
  {pages} {eabg9094} (\bibinfo {year} {2022})}\BibitemShut {NoStop}%
\bibitem [{\citenamefont {Chiu}\ \emph {et~al.}(2020)\citenamefont {Chiu},
  \citenamefont {Ma}, \citenamefont {Song}, \citenamefont {Bernevig},\ and\
  \citenamefont {Houck}}]{chiu2020fragile}%
  \BibitemOpen
  \bibfield  {author} {\bibinfo {author} {\bibfnamefont {C.~S.}\ \bibnamefont
  {Chiu}}, \bibinfo {author} {\bibfnamefont {D.-S.}\ \bibnamefont {Ma}},
  \bibinfo {author} {\bibfnamefont {Z.-D.}\ \bibnamefont {Song}}, \bibinfo
  {author} {\bibfnamefont {B.~A.}\ \bibnamefont {Bernevig}},\ and\ \bibinfo
  {author} {\bibfnamefont {A.~A.}\ \bibnamefont {Houck}},\ }\bibfield  {title}
  {\bibinfo {title} {Fragile topology in line-graph lattices with two, three,
  or four gapped flat bands},\ }\href
  {https://doi.org/10.1103/PhysRevResearch.2.043414} {\bibfield  {journal}
  {\bibinfo  {journal} {Physical Review Research}\ }\textbf {\bibinfo {volume}
  {2}},\ \bibinfo {pages} {043414} (\bibinfo {year} {2020})}\BibitemShut
  {NoStop}%
\bibitem [{\citenamefont {Zhang}\ and\ \citenamefont
  {Yang}(2021)}]{zhang2021tunable}%
  \BibitemOpen
  \bibfield  {author} {\bibinfo {author} {\bibfnamefont {R.-X.}\ \bibnamefont
  {Zhang}}\ and\ \bibinfo {author} {\bibfnamefont {Z.-C.}\ \bibnamefont
  {Yang}},\ }\bibfield  {title} {\bibinfo {title} {Tunable fragile topology in
  floquet systems},\ }\href {https://doi.org/10.1103/PhysRevB.103.L121115}
  {\bibfield  {journal} {\bibinfo  {journal} {Physical Review B}\ }\textbf
  {\bibinfo {volume} {103}},\ \bibinfo {pages} {L121115} (\bibinfo {year}
  {2021})}\BibitemShut {NoStop}%
\bibitem [{\citenamefont {Peri}\ \emph {et~al.}(2020)\citenamefont {Peri},
  \citenamefont {Song}, \citenamefont {Serra-Garcia}, \citenamefont {Engeler},
  \citenamefont {Queiroz}, \citenamefont {Huang}, \citenamefont {Deng},
  \citenamefont {Liu}, \citenamefont {Bernevig},\ and\ \citenamefont
  {Huber}}]{peri2020experimental}%
  \BibitemOpen
  \bibfield  {author} {\bibinfo {author} {\bibfnamefont {V.}~\bibnamefont
  {Peri}}, \bibinfo {author} {\bibfnamefont {Z.-D.}\ \bibnamefont {Song}},
  \bibinfo {author} {\bibfnamefont {M.}~\bibnamefont {Serra-Garcia}}, \bibinfo
  {author} {\bibfnamefont {P.}~\bibnamefont {Engeler}}, \bibinfo {author}
  {\bibfnamefont {R.}~\bibnamefont {Queiroz}}, \bibinfo {author} {\bibfnamefont
  {X.}~\bibnamefont {Huang}}, \bibinfo {author} {\bibfnamefont
  {W.}~\bibnamefont {Deng}}, \bibinfo {author} {\bibfnamefont {Z.}~\bibnamefont
  {Liu}}, \bibinfo {author} {\bibfnamefont {B.~A.}\ \bibnamefont {Bernevig}},\
  and\ \bibinfo {author} {\bibfnamefont {S.~D.}\ \bibnamefont {Huber}},\
  }\bibfield  {title} {\bibinfo {title} {Experimental characterization of
  fragile topology in an acoustic metamaterial},\ }\href
  {https://doi.org/10.1126/science.aaz7654} {\bibfield  {journal} {\bibinfo
  {journal} {Science}\ }\textbf {\bibinfo {volume} {367}},\ \bibinfo {pages}
  {797} (\bibinfo {year} {2020})}\BibitemShut {NoStop}%
\bibitem [{\citenamefont {Bird}\ \emph {et~al.}(2025)\citenamefont {Bird},
  \citenamefont {Devescovi}, \citenamefont {Engeler}, \citenamefont {Valenti},
  \citenamefont {G{\"o}kmen}, \citenamefont {Worreby}, \citenamefont {Peri},\
  and\ \citenamefont {Huber}}]{bird2025design}%
  \BibitemOpen
  \bibfield  {author} {\bibinfo {author} {\bibfnamefont {S.}~\bibnamefont
  {Bird}}, \bibinfo {author} {\bibfnamefont {C.}~\bibnamefont {Devescovi}},
  \bibinfo {author} {\bibfnamefont {P.}~\bibnamefont {Engeler}}, \bibinfo
  {author} {\bibfnamefont {A.}~\bibnamefont {Valenti}}, \bibinfo {author}
  {\bibfnamefont {D.~E.}\ \bibnamefont {G{\"o}kmen}}, \bibinfo {author}
  {\bibfnamefont {R.}~\bibnamefont {Worreby}}, \bibinfo {author} {\bibfnamefont
  {V.}~\bibnamefont {Peri}},\ and\ \bibinfo {author} {\bibfnamefont {S.~D.}\
  \bibnamefont {Huber}},\ }\bibfield  {title} {\bibinfo {title} {Design and
  characterization of all {2D} fragile topological bands},\ }\href
  {https://doi.org/10.1093/pnasnexus/pgaf285} {\bibfield  {journal} {\bibinfo
  {journal} {PNAS nexus}\ }\textbf {\bibinfo {volume} {4}},\ \bibinfo {pages}
  {pgaf285} (\bibinfo {year} {2025})}\BibitemShut {NoStop}%
\bibitem [{\citenamefont {De~Paz}\ \emph {et~al.}(2019)\citenamefont {De~Paz},
  \citenamefont {Vergniory}, \citenamefont {Bercioux}, \citenamefont
  {Garc{\'\i}a-Etxarri},\ and\ \citenamefont {Bradlyn}}]{de2019engineering}%
  \BibitemOpen
  \bibfield  {author} {\bibinfo {author} {\bibfnamefont {M.~B.}\ \bibnamefont
  {De~Paz}}, \bibinfo {author} {\bibfnamefont {M.~G.}\ \bibnamefont
  {Vergniory}}, \bibinfo {author} {\bibfnamefont {D.}~\bibnamefont {Bercioux}},
  \bibinfo {author} {\bibfnamefont {A.}~\bibnamefont {Garc{\'\i}a-Etxarri}},\
  and\ \bibinfo {author} {\bibfnamefont {B.}~\bibnamefont {Bradlyn}},\
  }\bibfield  {title} {\bibinfo {title} {Engineering fragile topology in
  photonic crystals: {Topological} quantum chemistry of light},\ }\href
  {https://doi.org/10.1103/PhysRevResearch.1.032005} {\bibfield  {journal}
  {\bibinfo  {journal} {Physical Review Research}\ }\textbf {\bibinfo {volume}
  {1}},\ \bibinfo {pages} {032005} (\bibinfo {year} {2019})}\BibitemShut
  {NoStop}%
\bibitem [{\citenamefont {Alexandradinata}\ \emph {et~al.}(2020)\citenamefont
  {Alexandradinata}, \citenamefont {H{\"o}ller}, \citenamefont {Wang},
  \citenamefont {Cheng},\ and\ \citenamefont
  {Lu}}]{alexandradinata2020crystallographic}%
  \BibitemOpen
  \bibfield  {author} {\bibinfo {author} {\bibfnamefont {A.}~\bibnamefont
  {Alexandradinata}}, \bibinfo {author} {\bibfnamefont {J.}~\bibnamefont
  {H{\"o}ller}}, \bibinfo {author} {\bibfnamefont {C.}~\bibnamefont {Wang}},
  \bibinfo {author} {\bibfnamefont {H.}~\bibnamefont {Cheng}},\ and\ \bibinfo
  {author} {\bibfnamefont {L.}~\bibnamefont {Lu}},\ }\bibfield  {title}
  {\bibinfo {title} {Crystallographic splitting theorem for band
  representations and fragile topological photonic crystals},\ }\href
  {https://doi.org/10.1103/PhysRevB.102.115117} {\bibfield  {journal} {\bibinfo
   {journal} {Physical Review B}\ }\textbf {\bibinfo {volume} {102}},\ \bibinfo
  {pages} {115117} (\bibinfo {year} {2020})}\BibitemShut {NoStop}%
\bibitem [{\citenamefont {Ma{\~n}es}(2020)}]{manes2020fragile}%
  \BibitemOpen
  \bibfield  {author} {\bibinfo {author} {\bibfnamefont {J.~L.}\ \bibnamefont
  {Ma{\~n}es}},\ }\bibfield  {title} {\bibinfo {title} {Fragile phonon topology
  on the honeycomb lattice with time-reversal symmetry},\ }\href
  {https://doi.org/10.1103/PhysRevB.102.024307} {\bibfield  {journal} {\bibinfo
   {journal} {Physical Review B}\ }\textbf {\bibinfo {volume} {102}},\ \bibinfo
  {pages} {024307} (\bibinfo {year} {2020})}\BibitemShut {NoStop}%
\bibitem [{\citenamefont {Park}\ \emph {et~al.}(2021)\citenamefont {Park},
  \citenamefont {Hwang}, \citenamefont {Choi},\ and\ \citenamefont
  {Yang}}]{park2021topological}%
  \BibitemOpen
  \bibfield  {author} {\bibinfo {author} {\bibfnamefont {S.}~\bibnamefont
  {Park}}, \bibinfo {author} {\bibfnamefont {Y.}~\bibnamefont {Hwang}},
  \bibinfo {author} {\bibfnamefont {H.~C.}\ \bibnamefont {Choi}},\ and\
  \bibinfo {author} {\bibfnamefont {B.-J.}\ \bibnamefont {Yang}},\ }\bibfield
  {title} {\bibinfo {title} {Topological acoustic triple point},\ }\href
  {https://doi.org/10.1038/s41467-021-27158-y} {\bibfield  {journal} {\bibinfo
  {journal} {Nature Communications}\ }\textbf {\bibinfo {volume} {12}},\
  \bibinfo {pages} {6781} (\bibinfo {year} {2021})}\BibitemShut {NoStop}%
\bibitem [{\citenamefont {Wieder}\ and\ \citenamefont
  {Bernevig}(2018)}]{wieder2018axion}%
  \BibitemOpen
  \bibfield  {author} {\bibinfo {author} {\bibfnamefont {B.~J.}\ \bibnamefont
  {Wieder}}\ and\ \bibinfo {author} {\bibfnamefont {B.~A.}\ \bibnamefont
  {Bernevig}},\ }\bibfield  {title} {\bibinfo {title} {The axion insulator as a
  pump of fragile topology},\ }\href {https://arxiv.org/abs/1810.02373}
  {\bibfield  {journal} {\bibinfo  {journal} {arXiv preprint arXiv:1810.02373}\
  } (\bibinfo {year} {2018})}\BibitemShut {NoStop}%
\bibitem [{\citenamefont {Wieder}\ \emph
  {et~al.}(2020{\natexlab{a}})\citenamefont {Wieder}, \citenamefont {Wang},
  \citenamefont {Cano}, \citenamefont {Dai}, \citenamefont {Schoop},
  \citenamefont {Bradlyn},\ and\ \citenamefont {Bernevig}}]{wieder2020strong}%
  \BibitemOpen
  \bibfield  {author} {\bibinfo {author} {\bibfnamefont {B.~J.}\ \bibnamefont
  {Wieder}}, \bibinfo {author} {\bibfnamefont {Z.}~\bibnamefont {Wang}},
  \bibinfo {author} {\bibfnamefont {J.}~\bibnamefont {Cano}}, \bibinfo {author}
  {\bibfnamefont {X.}~\bibnamefont {Dai}}, \bibinfo {author} {\bibfnamefont
  {L.~M.}\ \bibnamefont {Schoop}}, \bibinfo {author} {\bibfnamefont
  {B.}~\bibnamefont {Bradlyn}},\ and\ \bibinfo {author} {\bibfnamefont {B.~A.}\
  \bibnamefont {Bernevig}},\ }\bibfield  {title} {\bibinfo {title} {Strong and
  fragile topological {Dirac} semimetals with higher-order {Fermi} arcs},\
  }\href {https://doi.org/10.1038/s41467-020-14443-5} {\bibfield  {journal}
  {\bibinfo  {journal} {Nature Communications}\ }\textbf {\bibinfo {volume}
  {11}},\ \bibinfo {pages} {627} (\bibinfo {year}
  {2020}{\natexlab{a}})}\BibitemShut {NoStop}%
\bibitem [{\citenamefont {Kobayashi}\ and\ \citenamefont
  {Furusaki}(2021)}]{kobayashi2021fragile}%
  \BibitemOpen
  \bibfield  {author} {\bibinfo {author} {\bibfnamefont {S.}~\bibnamefont
  {Kobayashi}}\ and\ \bibinfo {author} {\bibfnamefont {A.}~\bibnamefont
  {Furusaki}},\ }\bibfield  {title} {\bibinfo {title} {Fragile topological
  insulators protected by rotation symmetry without spin-orbit coupling},\
  }\href {https://doi.org/10.1103/PhysRevB.104.195114} {\bibfield  {journal}
  {\bibinfo  {journal} {Physical Review B}\ }\textbf {\bibinfo {volume}
  {104}},\ \bibinfo {pages} {195114} (\bibinfo {year} {2021})}\BibitemShut
  {NoStop}%
\bibitem [{\citenamefont {Else}\ \emph {et~al.}(2019)\citenamefont {Else},
  \citenamefont {Po},\ and\ \citenamefont {Watanabe}}]{else2019fragile}%
  \BibitemOpen
  \bibfield  {author} {\bibinfo {author} {\bibfnamefont {D.~V.}\ \bibnamefont
  {Else}}, \bibinfo {author} {\bibfnamefont {H.~C.}\ \bibnamefont {Po}},\ and\
  \bibinfo {author} {\bibfnamefont {H.}~\bibnamefont {Watanabe}},\ }\bibfield
  {title} {\bibinfo {title} {Fragile topological phases in interacting
  systems},\ }\href {https://doi.org/10.1103/PhysRevB.99.125122} {\bibfield
  {journal} {\bibinfo  {journal} {Physical Review B}\ }\textbf {\bibinfo
  {volume} {99}},\ \bibinfo {pages} {125122} (\bibinfo {year}
  {2019})}\BibitemShut {NoStop}%
\bibitem [{\citenamefont {Latimer}\ and\ \citenamefont
  {Wang}(2021)}]{latimer2021correlated}%
  \BibitemOpen
  \bibfield  {author} {\bibinfo {author} {\bibfnamefont {K.}~\bibnamefont
  {Latimer}}\ and\ \bibinfo {author} {\bibfnamefont {C.}~\bibnamefont {Wang}},\
  }\bibfield  {title} {\bibinfo {title} {Correlated fragile topology: {A}
  parton approach},\ }\href {https://doi.org/10.1103/PhysRevB.103.045128}
  {\bibfield  {journal} {\bibinfo  {journal} {Physical Review B}\ }\textbf
  {\bibinfo {volume} {103}},\ \bibinfo {pages} {045128} (\bibinfo {year}
  {2021})}\BibitemShut {NoStop}%
\bibitem [{\citenamefont {Herzog-Arbeitman}\ \emph {et~al.}(2024)\citenamefont
  {Herzog-Arbeitman}, \citenamefont {Bernevig},\ and\ \citenamefont
  {Song}}]{herzog2024interacting}%
  \BibitemOpen
  \bibfield  {author} {\bibinfo {author} {\bibfnamefont {J.}~\bibnamefont
  {Herzog-Arbeitman}}, \bibinfo {author} {\bibfnamefont {B.~A.}\ \bibnamefont
  {Bernevig}},\ and\ \bibinfo {author} {\bibfnamefont {Z.-D.}\ \bibnamefont
  {Song}},\ }\bibfield  {title} {\bibinfo {title} {Interacting topological
  quantum chemistry in {2D} with many-body real space invariants},\ }\href
  {https://doi.org/10.1038/s41467-024-45395-9} {\bibfield  {journal} {\bibinfo
  {journal} {Nature Communications}\ }\textbf {\bibinfo {volume} {15}},\
  \bibinfo {pages} {1171} (\bibinfo {year} {2024})}\BibitemShut {NoStop}%
\bibitem [{\citenamefont {Hwang}\ \emph {et~al.}(2019)\citenamefont {Hwang},
  \citenamefont {Ahn},\ and\ \citenamefont {Yang}}]{hwang2019fragile}%
  \BibitemOpen
  \bibfield  {author} {\bibinfo {author} {\bibfnamefont {Y.}~\bibnamefont
  {Hwang}}, \bibinfo {author} {\bibfnamefont {J.}~\bibnamefont {Ahn}},\ and\
  \bibinfo {author} {\bibfnamefont {B.-J.}\ \bibnamefont {Yang}},\ }\bibfield
  {title} {\bibinfo {title} {Fragile topology protected by inversion symmetry:
  {Diagnosis}, bulk-boundary correspondence, and {Wilson} loop},\ }\href
  {https://doi.org/10.1103/PhysRevB.100.205126} {\bibfield  {journal} {\bibinfo
   {journal} {Physical Review B}\ }\textbf {\bibinfo {volume} {100}},\ \bibinfo
  {pages} {205126} (\bibinfo {year} {2019})}\BibitemShut {NoStop}%
\bibitem [{\citenamefont {Song}\ \emph
  {et~al.}(2020{\natexlab{a}})\citenamefont {Song}, \citenamefont {Elcoro},\
  and\ \citenamefont {Bernevig}}]{song2020twisted}%
  \BibitemOpen
  \bibfield  {author} {\bibinfo {author} {\bibfnamefont {Z.-D.}\ \bibnamefont
  {Song}}, \bibinfo {author} {\bibfnamefont {L.}~\bibnamefont {Elcoro}},\ and\
  \bibinfo {author} {\bibfnamefont {B.~A.}\ \bibnamefont {Bernevig}},\
  }\bibfield  {title} {\bibinfo {title} {Twisted bulk-boundary correspondence
  of fragile topology},\ }\href {https://doi.org/10.1126/science.aaz7650}
  {\bibfield  {journal} {\bibinfo  {journal} {Science}\ }\textbf {\bibinfo
  {volume} {367}},\ \bibinfo {pages} {794} (\bibinfo {year}
  {2020}{\natexlab{a}})}\BibitemShut {NoStop}%
\bibitem [{\citenamefont {Liu}\ \emph {et~al.}(2019)\citenamefont {Liu},
  \citenamefont {Vishwanath},\ and\ \citenamefont {Khalaf}}]{liu2019shift}%
  \BibitemOpen
  \bibfield  {author} {\bibinfo {author} {\bibfnamefont {S.}~\bibnamefont
  {Liu}}, \bibinfo {author} {\bibfnamefont {A.}~\bibnamefont {Vishwanath}},\
  and\ \bibinfo {author} {\bibfnamefont {E.}~\bibnamefont {Khalaf}},\
  }\bibfield  {title} {\bibinfo {title} {Shift insulators: {Rotation}-protected
  two-dimensional topological crystalline insulators},\ }\href
  {https://doi.org/10.1103/PhysRevX.9.031003} {\bibfield  {journal} {\bibinfo
  {journal} {Physical Review X}\ }\textbf {\bibinfo {volume} {9}},\ \bibinfo
  {pages} {031003} (\bibinfo {year} {2019})}\BibitemShut {NoStop}%
\bibitem [{\citenamefont {Herzog-Arbeitman}\ \emph {et~al.}(2020)\citenamefont
  {Herzog-Arbeitman}, \citenamefont {Song}, \citenamefont {Regnault},\ and\
  \citenamefont {Bernevig}}]{herzog2020hofstadter}%
  \BibitemOpen
  \bibfield  {author} {\bibinfo {author} {\bibfnamefont {J.}~\bibnamefont
  {Herzog-Arbeitman}}, \bibinfo {author} {\bibfnamefont {Z.-D.}\ \bibnamefont
  {Song}}, \bibinfo {author} {\bibfnamefont {N.}~\bibnamefont {Regnault}},\
  and\ \bibinfo {author} {\bibfnamefont {B.~A.}\ \bibnamefont {Bernevig}},\
  }\bibfield  {title} {\bibinfo {title} {{Hofstadter} topology: noncrystalline
  topological materials at high flux},\ }\href
  {https://doi.org/10.1103/PhysRevLett.125.236804} {\bibfield  {journal}
  {\bibinfo  {journal} {Physical Review Letters}\ }\textbf {\bibinfo {volume}
  {125}},\ \bibinfo {pages} {236804} (\bibinfo {year} {2020})}\BibitemShut
  {NoStop}%
\bibitem [{\citenamefont {Lian}\ \emph {et~al.}(2020)\citenamefont {Lian},
  \citenamefont {Xie},\ and\ \citenamefont {Bernevig}}]{lian2020landau}%
  \BibitemOpen
  \bibfield  {author} {\bibinfo {author} {\bibfnamefont {B.}~\bibnamefont
  {Lian}}, \bibinfo {author} {\bibfnamefont {F.}~\bibnamefont {Xie}},\ and\
  \bibinfo {author} {\bibfnamefont {B.~A.}\ \bibnamefont {Bernevig}},\
  }\bibfield  {title} {\bibinfo {title} {{Landau} level of fragile topology},\
  }\href {https://doi.org/10.1103/PhysRevB.102.041402} {\bibfield  {journal}
  {\bibinfo  {journal} {Physical Review B}\ }\textbf {\bibinfo {volume}
  {102}},\ \bibinfo {pages} {041402} (\bibinfo {year} {2020})}\BibitemShut
  {NoStop}%
\bibitem [{\citenamefont {Guan}\ \emph {et~al.}(2022)\citenamefont {Guan},
  \citenamefont {Bouhon},\ and\ \citenamefont {Yazyev}}]{guan2022landau}%
  \BibitemOpen
  \bibfield  {author} {\bibinfo {author} {\bibfnamefont {Y.}~\bibnamefont
  {Guan}}, \bibinfo {author} {\bibfnamefont {A.}~\bibnamefont {Bouhon}},\ and\
  \bibinfo {author} {\bibfnamefont {O.~V.}\ \bibnamefont {Yazyev}},\ }\bibfield
   {title} {\bibinfo {title} {Landau levels of the {Euler} class topology},\
  }\href {https://doi.org/10.1103/PhysRevResearch.4.023188} {\bibfield
  {journal} {\bibinfo  {journal} {Physical Review Research}\ }\textbf {\bibinfo
  {volume} {4}},\ \bibinfo {pages} {023188} (\bibinfo {year}
  {2022})}\BibitemShut {NoStop}%
\bibitem [{\citenamefont {Jankowski}\ \emph {et~al.}(2025)\citenamefont
  {Jankowski}, \citenamefont {Morris}, \citenamefont {Bouhon}, \citenamefont
  {{\"U}nal},\ and\ \citenamefont {Slager}}]{jankowski2025optical}%
  \BibitemOpen
  \bibfield  {author} {\bibinfo {author} {\bibfnamefont {W.~J.}\ \bibnamefont
  {Jankowski}}, \bibinfo {author} {\bibfnamefont {A.~S.}\ \bibnamefont
  {Morris}}, \bibinfo {author} {\bibfnamefont {A.}~\bibnamefont {Bouhon}},
  \bibinfo {author} {\bibfnamefont {F.~N.}\ \bibnamefont {{\"U}nal}},\ and\
  \bibinfo {author} {\bibfnamefont {R.-J.}\ \bibnamefont {Slager}},\ }\bibfield
   {title} {\bibinfo {title} {Optical manifestations and bounds of topological
  {Euler} class},\ }\href {https://doi.org/10.1103/PhysRevB.111.L081103}
  {\bibfield  {journal} {\bibinfo  {journal} {Physical Review B}\ }\textbf
  {\bibinfo {volume} {111}},\ \bibinfo {pages} {L081103} (\bibinfo {year}
  {2025})}\BibitemShut {NoStop}%
\bibitem [{\citenamefont {Xie}\ \emph {et~al.}(2020)\citenamefont {Xie},
  \citenamefont {Song}, \citenamefont {Lian},\ and\ \citenamefont
  {Bernevig}}]{xie2020topology}%
  \BibitemOpen
  \bibfield  {author} {\bibinfo {author} {\bibfnamefont {F.}~\bibnamefont
  {Xie}}, \bibinfo {author} {\bibfnamefont {Z.}~\bibnamefont {Song}}, \bibinfo
  {author} {\bibfnamefont {B.}~\bibnamefont {Lian}},\ and\ \bibinfo {author}
  {\bibfnamefont {B.~A.}\ \bibnamefont {Bernevig}},\ }\bibfield  {title}
  {\bibinfo {title} {Topology-bounded superfluid weight in twisted bilayer
  graphene},\ }\href {https://doi.org/10.1103/PhysRevLett.124.167002}
  {\bibfield  {journal} {\bibinfo  {journal} {Physical Review Letters}\
  }\textbf {\bibinfo {volume} {124}},\ \bibinfo {pages} {167002} (\bibinfo
  {year} {2020})}\BibitemShut {NoStop}%
\bibitem [{\citenamefont {Peri}\ \emph {et~al.}(2021)\citenamefont {Peri},
  \citenamefont {Song}, \citenamefont {Bernevig},\ and\ \citenamefont
  {Huber}}]{peri2021fragile}%
  \BibitemOpen
  \bibfield  {author} {\bibinfo {author} {\bibfnamefont {V.}~\bibnamefont
  {Peri}}, \bibinfo {author} {\bibfnamefont {Z.-D.}\ \bibnamefont {Song}},
  \bibinfo {author} {\bibfnamefont {B.~A.}\ \bibnamefont {Bernevig}},\ and\
  \bibinfo {author} {\bibfnamefont {S.~D.}\ \bibnamefont {Huber}},\ }\bibfield
  {title} {\bibinfo {title} {Fragile topology and flat-band superconductivity
  in the strong-coupling regime},\ }\href
  {https://doi.org/10.1103/PhysRevLett.126.027002} {\bibfield  {journal}
  {\bibinfo  {journal} {Physical Review Letters}\ }\textbf {\bibinfo {volume}
  {126}},\ \bibinfo {pages} {027002} (\bibinfo {year} {2021})}\BibitemShut
  {NoStop}%
\bibitem [{\citenamefont {Herzog-Arbeitman}\ \emph {et~al.}(2022)\citenamefont
  {Herzog-Arbeitman}, \citenamefont {Peri}, \citenamefont {Schindler},
  \citenamefont {Huber},\ and\ \citenamefont
  {Bernevig}}]{herzog2022superfluid}%
  \BibitemOpen
  \bibfield  {author} {\bibinfo {author} {\bibfnamefont {J.}~\bibnamefont
  {Herzog-Arbeitman}}, \bibinfo {author} {\bibfnamefont {V.}~\bibnamefont
  {Peri}}, \bibinfo {author} {\bibfnamefont {F.}~\bibnamefont {Schindler}},
  \bibinfo {author} {\bibfnamefont {S.~D.}\ \bibnamefont {Huber}},\ and\
  \bibinfo {author} {\bibfnamefont {B.~A.}\ \bibnamefont {Bernevig}},\
  }\bibfield  {title} {\bibinfo {title} {Superfluid weight bounds from symmetry
  and quantum geometry in flat bands},\ }\href
  {https://doi.org/10.1103/PhysRevLett.128.087002} {\bibfield  {journal}
  {\bibinfo  {journal} {Physical Review Letters}\ }\textbf {\bibinfo {volume}
  {128}},\ \bibinfo {pages} {087002} (\bibinfo {year} {2022})}\BibitemShut
  {NoStop}%
\bibitem [{\citenamefont {Hwang}\ \emph {et~al.}(2021)\citenamefont {Hwang},
  \citenamefont {Rhim},\ and\ \citenamefont {Yang}}]{hwang2021geometric}%
  \BibitemOpen
  \bibfield  {author} {\bibinfo {author} {\bibfnamefont {Y.}~\bibnamefont
  {Hwang}}, \bibinfo {author} {\bibfnamefont {J.-W.}\ \bibnamefont {Rhim}},\
  and\ \bibinfo {author} {\bibfnamefont {B.-J.}\ \bibnamefont {Yang}},\
  }\bibfield  {title} {\bibinfo {title} {Geometric characterization of
  anomalous {Landau} levels of isolated flat bands},\ }\href
  {https://doi.org/10.1038/s41467-021-26765-z} {\bibfield  {journal} {\bibinfo
  {journal} {Nature Communications}\ }\textbf {\bibinfo {volume} {12}},\
  \bibinfo {pages} {6433} (\bibinfo {year} {2021})}\BibitemShut {NoStop}%
\bibitem [{\citenamefont {Wu}\ \emph {et~al.}(2021)\citenamefont {Wu},
  \citenamefont {Liu}, \citenamefont {Guan},\ and\ \citenamefont
  {Yazyev}}]{wu2021landau}%
  \BibitemOpen
  \bibfield  {author} {\bibinfo {author} {\bibfnamefont {Q.}~\bibnamefont
  {Wu}}, \bibinfo {author} {\bibfnamefont {J.}~\bibnamefont {Liu}}, \bibinfo
  {author} {\bibfnamefont {Y.}~\bibnamefont {Guan}},\ and\ \bibinfo {author}
  {\bibfnamefont {O.~V.}\ \bibnamefont {Yazyev}},\ }\bibfield  {title}
  {\bibinfo {title} {{Landau} levels as a probe for band topology in graphene
  moir{\'e} superlattices},\ }\href
  {https://doi.org/10.1103/PhysRevLett.126.056401} {\bibfield  {journal}
  {\bibinfo  {journal} {Physical Review Letters}\ }\textbf {\bibinfo {volume}
  {126}},\ \bibinfo {pages} {056401} (\bibinfo {year} {2021})}\BibitemShut
  {NoStop}%
\bibitem [{\citenamefont {Herzog-Arbeitman}\ \emph {et~al.}(2023)\citenamefont
  {Herzog-Arbeitman}, \citenamefont {Song}, \citenamefont {Elcoro},\ and\
  \citenamefont {Bernevig}}]{herzog2023hofstadter}%
  \BibitemOpen
  \bibfield  {author} {\bibinfo {author} {\bibfnamefont {J.}~\bibnamefont
  {Herzog-Arbeitman}}, \bibinfo {author} {\bibfnamefont {Z.-D.}\ \bibnamefont
  {Song}}, \bibinfo {author} {\bibfnamefont {L.}~\bibnamefont {Elcoro}},\ and\
  \bibinfo {author} {\bibfnamefont {B.~A.}\ \bibnamefont {Bernevig}},\
  }\bibfield  {title} {\bibinfo {title} {{Hofstadter} topology with real space
  invariants and reentrant projective symmetries},\ }\href
  {https://doi.org/10.1103/PhysRevLett.130.236601} {\bibfield  {journal}
  {\bibinfo  {journal} {Physical Review Letters}\ }\textbf {\bibinfo {volume}
  {130}},\ \bibinfo {pages} {236601} (\bibinfo {year} {2023})}\BibitemShut
  {NoStop}%
\bibitem [{\citenamefont {Li}\ \emph {et~al.}(2020)\citenamefont {Li},
  \citenamefont {Zhu}, \citenamefont {Benalcazar},\ and\ \citenamefont
  {Hughes}}]{li2020fractional}%
  \BibitemOpen
  \bibfield  {author} {\bibinfo {author} {\bibfnamefont {T.}~\bibnamefont
  {Li}}, \bibinfo {author} {\bibfnamefont {P.}~\bibnamefont {Zhu}}, \bibinfo
  {author} {\bibfnamefont {W.~A.}\ \bibnamefont {Benalcazar}},\ and\ \bibinfo
  {author} {\bibfnamefont {T.~L.}\ \bibnamefont {Hughes}},\ }\bibfield  {title}
  {\bibinfo {title} {Fractional disclination charge in two-dimensional
  {$C_n$}-symmetric topological crystalline insulators},\ }\href
  {https://doi.org/10.1103/PhysRevB.101.115115} {\bibfield  {journal} {\bibinfo
   {journal} {Physical Review B}\ }\textbf {\bibinfo {volume} {101}},\ \bibinfo
  {pages} {115115} (\bibinfo {year} {2020})}\BibitemShut {NoStop}%
\bibitem [{\citenamefont {Zhang}\ \emph {et~al.}(2022)\citenamefont {Zhang},
  \citenamefont {Manjunath}, \citenamefont {Nambiar},\ and\ \citenamefont
  {Barkeshli}}]{zhang2022fractional}%
  \BibitemOpen
  \bibfield  {author} {\bibinfo {author} {\bibfnamefont {Y.}~\bibnamefont
  {Zhang}}, \bibinfo {author} {\bibfnamefont {N.}~\bibnamefont {Manjunath}},
  \bibinfo {author} {\bibfnamefont {G.}~\bibnamefont {Nambiar}},\ and\ \bibinfo
  {author} {\bibfnamefont {M.}~\bibnamefont {Barkeshli}},\ }\bibfield  {title}
  {\bibinfo {title} {Fractional disclination charge and discrete shift in the
  {Hofstadter} butterfly},\ }\href
  {https://doi.org/10.1103/PhysRevLett.129.275301} {\bibfield  {journal}
  {\bibinfo  {journal} {Physical Review Letters}\ }\textbf {\bibinfo {volume}
  {129}},\ \bibinfo {pages} {275301} (\bibinfo {year} {2022})}\BibitemShut
  {NoStop}%
\bibitem [{\citenamefont {Manjunath}\ \emph {et~al.}(2024)\citenamefont
  {Manjunath}, \citenamefont {Calvera},\ and\ \citenamefont
  {Barkeshli}}]{manjunath2024characterization}%
  \BibitemOpen
  \bibfield  {author} {\bibinfo {author} {\bibfnamefont {N.}~\bibnamefont
  {Manjunath}}, \bibinfo {author} {\bibfnamefont {V.}~\bibnamefont {Calvera}},\
  and\ \bibinfo {author} {\bibfnamefont {M.}~\bibnamefont {Barkeshli}},\
  }\bibfield  {title} {\bibinfo {title} {Characterization and classification of
  interacting {(2+1)}-dimensional topological crystalline insulators with
  orientation-preserving wallpaper groups},\ }\href
  {https://doi.org/10.1103/PhysRevB.109.035168} {\bibfield  {journal} {\bibinfo
   {journal} {Physical Review B}\ }\textbf {\bibinfo {volume} {109}},\ \bibinfo
  {pages} {035168} (\bibinfo {year} {2024})}\BibitemShut {NoStop}%
\bibitem [{\citenamefont {Hwang}\ \emph {et~al.}(2025)\citenamefont {Hwang},
  \citenamefont {Gupta}, \citenamefont {Schindler}, \citenamefont {Elcoro},
  \citenamefont {Song}, \citenamefont {Bernevig},\ and\ \citenamefont
  {Bradlyn}}]{hwang2026stable}%
  \BibitemOpen
  \bibfield  {author} {\bibinfo {author} {\bibfnamefont {Y.}~\bibnamefont
  {Hwang}}, \bibinfo {author} {\bibfnamefont {V.}~\bibnamefont {Gupta}},
  \bibinfo {author} {\bibfnamefont {F.}~\bibnamefont {Schindler}}, \bibinfo
  {author} {\bibfnamefont {L.}~\bibnamefont {Elcoro}}, \bibinfo {author}
  {\bibfnamefont {Z.}~\bibnamefont {Song}}, \bibinfo {author} {\bibfnamefont
  {B.~A.}\ \bibnamefont {Bernevig}},\ and\ \bibinfo {author} {\bibfnamefont
  {B.}~\bibnamefont {Bradlyn}},\ }\bibfield  {title} {\bibinfo {title} {Stable
  real-space invariants and topology beyond symmetry indicators},\ }\href
  {https://arxiv.org/abs/2505.09697} {\bibfield  {journal} {\bibinfo  {journal}
  {arXiv preprint arXiv:2505.09697}\ } (\bibinfo {year} {2025})}\BibitemShut
  {NoStop}%
\bibitem [{\citenamefont {Song}\ \emph {et~al.}(2017)\citenamefont {Song},
  \citenamefont {Fang},\ and\ \citenamefont {Fang}}]{song2017d}%
  \BibitemOpen
  \bibfield  {author} {\bibinfo {author} {\bibfnamefont {Z.}~\bibnamefont
  {Song}}, \bibinfo {author} {\bibfnamefont {Z.}~\bibnamefont {Fang}},\ and\
  \bibinfo {author} {\bibfnamefont {C.}~\bibnamefont {Fang}},\ }\bibfield
  {title} {\bibinfo {title} {{($d-2$)}-dimensional edge states of rotation
  symmetry protected topological states},\ }\href
  {https://doi.org/10.1103/PhysRevLett.119.246402} {\bibfield  {journal}
  {\bibinfo  {journal} {Physical Review Letters}\ }\textbf {\bibinfo {volume}
  {119}},\ \bibinfo {pages} {246402} (\bibinfo {year} {2017})}\BibitemShut
  {NoStop}%
\bibitem [{\citenamefont {Benalcazar}\ \emph {et~al.}(2017)\citenamefont
  {Benalcazar}, \citenamefont {Bernevig},\ and\ \citenamefont
  {Hughes}}]{benalcazar2017quantized}%
  \BibitemOpen
  \bibfield  {author} {\bibinfo {author} {\bibfnamefont {W.~A.}\ \bibnamefont
  {Benalcazar}}, \bibinfo {author} {\bibfnamefont {B.~A.}\ \bibnamefont
  {Bernevig}},\ and\ \bibinfo {author} {\bibfnamefont {T.~L.}\ \bibnamefont
  {Hughes}},\ }\bibfield  {title} {\bibinfo {title} {Quantized electric
  multipole insulators},\ }\href {https://doi.org/10.1126/science.aah6442}
  {\bibfield  {journal} {\bibinfo  {journal} {Science}\ }\textbf {\bibinfo
  {volume} {357}},\ \bibinfo {pages} {61} (\bibinfo {year} {2017})}\BibitemShut
  {NoStop}%
\bibitem [{\citenamefont {Benalcazar}\ \emph {et~al.}(2019)\citenamefont
  {Benalcazar}, \citenamefont {Li},\ and\ \citenamefont
  {Hughes}}]{benalcazar2019quantization}%
  \BibitemOpen
  \bibfield  {author} {\bibinfo {author} {\bibfnamefont {W.~A.}\ \bibnamefont
  {Benalcazar}}, \bibinfo {author} {\bibfnamefont {T.}~\bibnamefont {Li}},\
  and\ \bibinfo {author} {\bibfnamefont {T.~L.}\ \bibnamefont {Hughes}},\
  }\bibfield  {title} {\bibinfo {title} {Quantization of fractional corner
  charge in {$C_n$}-symmetric higher-order topological crystalline
  insulators},\ }\href {https://doi.org/10.1103/PhysRevB.99.245151} {\bibfield
  {journal} {\bibinfo  {journal} {Physical Review B}\ }\textbf {\bibinfo
  {volume} {99}},\ \bibinfo {pages} {245151} (\bibinfo {year}
  {2019})}\BibitemShut {NoStop}%
\bibitem [{\citenamefont {Schindler}\ \emph {et~al.}(2019)\citenamefont
  {Schindler}, \citenamefont {Brzezi{\'n}ska}, \citenamefont {Benalcazar},
  \citenamefont {Iraola}, \citenamefont {Bouhon}, \citenamefont {Tsirkin},
  \citenamefont {Vergniory},\ and\ \citenamefont
  {Neupert}}]{schindler2019fractional}%
  \BibitemOpen
  \bibfield  {author} {\bibinfo {author} {\bibfnamefont {F.}~\bibnamefont
  {Schindler}}, \bibinfo {author} {\bibfnamefont {M.}~\bibnamefont
  {Brzezi{\'n}ska}}, \bibinfo {author} {\bibfnamefont {W.~A.}\ \bibnamefont
  {Benalcazar}}, \bibinfo {author} {\bibfnamefont {M.}~\bibnamefont {Iraola}},
  \bibinfo {author} {\bibfnamefont {A.}~\bibnamefont {Bouhon}}, \bibinfo
  {author} {\bibfnamefont {S.~S.}\ \bibnamefont {Tsirkin}}, \bibinfo {author}
  {\bibfnamefont {M.~G.}\ \bibnamefont {Vergniory}},\ and\ \bibinfo {author}
  {\bibfnamefont {T.}~\bibnamefont {Neupert}},\ }\bibfield  {title} {\bibinfo
  {title} {Fractional corner charges in spin-orbit coupled crystals},\ }\href
  {https://doi.org/10.1103/PhysRevResearch.1.033074} {\bibfield  {journal}
  {\bibinfo  {journal} {Physical Review Research}\ }\textbf {\bibinfo {volume}
  {1}},\ \bibinfo {pages} {033074} (\bibinfo {year} {2019})}\BibitemShut
  {NoStop}%
\bibitem [{\citenamefont {Lee}\ \emph {et~al.}(2020)\citenamefont {Lee},
  \citenamefont {Furusaki},\ and\ \citenamefont {Yang}}]{lee2020fractional}%
  \BibitemOpen
  \bibfield  {author} {\bibinfo {author} {\bibfnamefont {E.}~\bibnamefont
  {Lee}}, \bibinfo {author} {\bibfnamefont {A.}~\bibnamefont {Furusaki}},\ and\
  \bibinfo {author} {\bibfnamefont {B.-J.}\ \bibnamefont {Yang}},\ }\bibfield
  {title} {\bibinfo {title} {Fractional charge bound to a vortex in
  two-dimensional topological crystalline insulators},\ }\href
  {https://doi.org/10.1103/PhysRevB.101.241109} {\bibfield  {journal} {\bibinfo
   {journal} {Physical Review B}\ }\textbf {\bibinfo {volume} {101}},\ \bibinfo
  {pages} {241109} (\bibinfo {year} {2020})}\BibitemShut {NoStop}%
\bibitem [{\citenamefont {Takahashi}\ \emph {et~al.}(2021)\citenamefont
  {Takahashi}, \citenamefont {Zhang},\ and\ \citenamefont
  {Murakami}}]{takahashi2021general}%
  \BibitemOpen
  \bibfield  {author} {\bibinfo {author} {\bibfnamefont {R.}~\bibnamefont
  {Takahashi}}, \bibinfo {author} {\bibfnamefont {T.}~\bibnamefont {Zhang}},\
  and\ \bibinfo {author} {\bibfnamefont {S.}~\bibnamefont {Murakami}},\
  }\bibfield  {title} {\bibinfo {title} {General corner charge formula in
  two-dimensional {$C_n$}-symmetric higher-order topological insulators},\
  }\href {https://doi.org/10.1103/PhysRevB.103.205123} {\bibfield  {journal}
  {\bibinfo  {journal} {Physical Review B}\ }\textbf {\bibinfo {volume}
  {103}},\ \bibinfo {pages} {205123} (\bibinfo {year} {2021})}\BibitemShut
  {NoStop}%
\bibitem [{\citenamefont {Rao}\ and\ \citenamefont
  {Bradlyn}(2023)}]{rao2023effective}%
  \BibitemOpen
  \bibfield  {author} {\bibinfo {author} {\bibfnamefont {P.}~\bibnamefont
  {Rao}}\ and\ \bibinfo {author} {\bibfnamefont {B.}~\bibnamefont {Bradlyn}},\
  }\bibfield  {title} {\bibinfo {title} {Effective action approach to the
  filling anomaly in crystalline topological matter},\ }\href
  {https://doi.org/10.1103/PhysRevB.107.195153} {\bibfield  {journal} {\bibinfo
   {journal} {Physical Review B}\ }\textbf {\bibinfo {volume} {107}},\ \bibinfo
  {pages} {195153} (\bibinfo {year} {2023})}\BibitemShut {NoStop}%
\bibitem [{\citenamefont {Velury}\ \emph {et~al.}(2025)\citenamefont {Velury},
  \citenamefont {Hwang},\ and\ \citenamefont {Hughes}}]{velury2025global}%
  \BibitemOpen
  \bibfield  {author} {\bibinfo {author} {\bibfnamefont {S.}~\bibnamefont
  {Velury}}, \bibinfo {author} {\bibfnamefont {Y.}~\bibnamefont {Hwang}},\ and\
  \bibinfo {author} {\bibfnamefont {T.~L.}\ \bibnamefont {Hughes}},\ }\bibfield
   {title} {\bibinfo {title} {Global and local topological crystalline markers
  for rotation-symmetric insulators},\ }\href
  {https://doi.org/10.1103/1sw2-qkpw} {\bibfield  {journal} {\bibinfo
  {journal} {Physical Review B}\ }\textbf {\bibinfo {volume} {112}},\ \bibinfo
  {pages} {094204} (\bibinfo {year} {2025})}\BibitemShut {NoStop}%
\bibitem [{\citenamefont {Song}\ \emph
  {et~al.}(2020{\natexlab{b}})\citenamefont {Song}, \citenamefont {Elcoro},
  \citenamefont {Xu}, \citenamefont {Regnault},\ and\ \citenamefont
  {Bernevig}}]{song2020fragile}%
  \BibitemOpen
  \bibfield  {author} {\bibinfo {author} {\bibfnamefont {Z.-D.}\ \bibnamefont
  {Song}}, \bibinfo {author} {\bibfnamefont {L.}~\bibnamefont {Elcoro}},
  \bibinfo {author} {\bibfnamefont {Y.-F.}\ \bibnamefont {Xu}}, \bibinfo
  {author} {\bibfnamefont {N.}~\bibnamefont {Regnault}},\ and\ \bibinfo
  {author} {\bibfnamefont {B.~A.}\ \bibnamefont {Bernevig}},\ }\bibfield
  {title} {\bibinfo {title} {Fragile phases as affine monoids: {Classification}
  and material examples},\ }\href {https://doi.org/10.1103/PhysRevX.10.031001}
  {\bibfield  {journal} {\bibinfo  {journal} {Physical Review X}\ }\textbf
  {\bibinfo {volume} {10}},\ \bibinfo {pages} {031001} (\bibinfo {year}
  {2020}{\natexlab{b}})}\BibitemShut {NoStop}%
\bibitem [{\citenamefont {Zhao}\ and\ \citenamefont {Lu}(2017)}]{zhao2017pt}%
  \BibitemOpen
  \bibfield  {author} {\bibinfo {author} {\bibfnamefont {Y.}~\bibnamefont
  {Zhao}}\ and\ \bibinfo {author} {\bibfnamefont {Y.}~\bibnamefont {Lu}},\
  }\bibfield  {title} {\bibinfo {title} {{$PT$}-symmetric real {Dirac} fermions
  and semimetals},\ }\href {https://doi.org/10.1103/PhysRevLett.118.056401}
  {\bibfield  {journal} {\bibinfo  {journal} {Physical Review Letters}\
  }\textbf {\bibinfo {volume} {118}},\ \bibinfo {pages} {056401} (\bibinfo
  {year} {2017})}\BibitemShut {NoStop}%
\bibitem [{\citenamefont {Ahn}\ \emph {et~al.}(2018)\citenamefont {Ahn},
  \citenamefont {Kim}, \citenamefont {Kim},\ and\ \citenamefont
  {Yang}}]{ahn2018band}%
  \BibitemOpen
  \bibfield  {author} {\bibinfo {author} {\bibfnamefont {J.}~\bibnamefont
  {Ahn}}, \bibinfo {author} {\bibfnamefont {D.}~\bibnamefont {Kim}}, \bibinfo
  {author} {\bibfnamefont {Y.}~\bibnamefont {Kim}},\ and\ \bibinfo {author}
  {\bibfnamefont {B.-J.}\ \bibnamefont {Yang}},\ }\bibfield  {title} {\bibinfo
  {title} {Band topology and linking structure of nodal line semimetals with
  {$Z_2$} monopole charges},\ }\href
  {https://doi.org/10.1103/PhysRevLett.121.106403} {\bibfield  {journal}
  {\bibinfo  {journal} {Physical Review Letters}\ }\textbf {\bibinfo {volume}
  {121}},\ \bibinfo {pages} {106403} (\bibinfo {year} {2018})}\BibitemShut
  {NoStop}%
\bibitem [{\citenamefont {Fu}\ and\ \citenamefont
  {Kane}(2007)}]{fu2007topological}%
  \BibitemOpen
  \bibfield  {author} {\bibinfo {author} {\bibfnamefont {L.}~\bibnamefont
  {Fu}}\ and\ \bibinfo {author} {\bibfnamefont {C.~L.}\ \bibnamefont {Kane}},\
  }\bibfield  {title} {\bibinfo {title} {Topological insulators with inversion
  symmetry},\ }\href {https://doi.org/10.1103/PhysRevB.76.045302} {\bibfield
  {journal} {\bibinfo  {journal} {Physical Review B}\ }\textbf {\bibinfo
  {volume} {76}},\ \bibinfo {pages} {045302} (\bibinfo {year}
  {2007})}\BibitemShut {NoStop}%
\bibitem [{\citenamefont {Watanabe}\ \emph {et~al.}(2018)\citenamefont
  {Watanabe}, \citenamefont {Po},\ and\ \citenamefont
  {Vishwanath}}]{watanabe2018structure}%
  \BibitemOpen
  \bibfield  {author} {\bibinfo {author} {\bibfnamefont {H.}~\bibnamefont
  {Watanabe}}, \bibinfo {author} {\bibfnamefont {H.~C.}\ \bibnamefont {Po}},\
  and\ \bibinfo {author} {\bibfnamefont {A.}~\bibnamefont {Vishwanath}},\
  }\bibfield  {title} {\bibinfo {title} {Structure and topology of band
  structures in the 1651 magnetic space groups},\ }\href
  {https://doi.org/10.1126/sciadv.aat8685} {\bibfield  {journal} {\bibinfo
  {journal} {Science Advances}\ }\textbf {\bibinfo {volume} {4}},\ \bibinfo
  {pages} {eaat8685} (\bibinfo {year} {2018})}\BibitemShut {NoStop}%
\bibitem [{\citenamefont {Khalaf}\ \emph {et~al.}(2018)\citenamefont {Khalaf},
  \citenamefont {Po}, \citenamefont {Vishwanath},\ and\ \citenamefont
  {Watanabe}}]{khalaf2018symmetry}%
  \BibitemOpen
  \bibfield  {author} {\bibinfo {author} {\bibfnamefont {E.}~\bibnamefont
  {Khalaf}}, \bibinfo {author} {\bibfnamefont {H.~C.}\ \bibnamefont {Po}},
  \bibinfo {author} {\bibfnamefont {A.}~\bibnamefont {Vishwanath}},\ and\
  \bibinfo {author} {\bibfnamefont {H.}~\bibnamefont {Watanabe}},\ }\bibfield
  {title} {\bibinfo {title} {Symmetry indicators and anomalous surface states
  of topological crystalline insulators},\ }\href
  {https://doi.org/10.1103/PhysRevX.8.031070} {\bibfield  {journal} {\bibinfo
  {journal} {Physical Review X}\ }\textbf {\bibinfo {volume} {8}},\ \bibinfo
  {pages} {031070} (\bibinfo {year} {2018})}\BibitemShut {NoStop}%
\bibitem [{\citenamefont {Song}\ \emph {et~al.}(2018)\citenamefont {Song},
  \citenamefont {Zhang}, \citenamefont {Fang},\ and\ \citenamefont
  {Fang}}]{song2018quantitative}%
  \BibitemOpen
  \bibfield  {author} {\bibinfo {author} {\bibfnamefont {Z.}~\bibnamefont
  {Song}}, \bibinfo {author} {\bibfnamefont {T.}~\bibnamefont {Zhang}},
  \bibinfo {author} {\bibfnamefont {Z.}~\bibnamefont {Fang}},\ and\ \bibinfo
  {author} {\bibfnamefont {C.}~\bibnamefont {Fang}},\ }\bibfield  {title}
  {\bibinfo {title} {Quantitative mappings between symmetry and topology in
  solids},\ }\href {https://doi.org/10.1038/s41467-018-06010-w} {\bibfield
  {journal} {\bibinfo  {journal} {Nature Communications}\ }\textbf {\bibinfo
  {volume} {9}},\ \bibinfo {pages} {3530} (\bibinfo {year} {2018})}\BibitemShut
  {NoStop}%
\bibitem [{\citenamefont {Cano}\ \emph {et~al.}(2022)\citenamefont {Cano},
  \citenamefont {Elcoro}, \citenamefont {Aroyo}, \citenamefont {Bernevig},\
  and\ \citenamefont {Bradlyn}}]{cano2022topology}%
  \BibitemOpen
  \bibfield  {author} {\bibinfo {author} {\bibfnamefont {J.}~\bibnamefont
  {Cano}}, \bibinfo {author} {\bibfnamefont {L.}~\bibnamefont {Elcoro}},
  \bibinfo {author} {\bibfnamefont {M.}~\bibnamefont {Aroyo}}, \bibinfo
  {author} {\bibfnamefont {B.~A.}\ \bibnamefont {Bernevig}},\ and\ \bibinfo
  {author} {\bibfnamefont {B.}~\bibnamefont {Bradlyn}},\ }\bibfield  {title}
  {\bibinfo {title} {Topology invisible to eigenvalues in obstructed atomic
  insulators},\ }\href {https://doi.org/10.1103/PhysRevB.105.125115} {\bibfield
   {journal} {\bibinfo  {journal} {Physical Review B}\ }\textbf {\bibinfo
  {volume} {105}},\ \bibinfo {pages} {125115} (\bibinfo {year}
  {2022})}\BibitemShut {NoStop}%
\bibitem [{sup()}]{supple}%
  \BibitemOpen
  \bibinfo {note} {See the Supplemental Material for details on the
  pseudoinverse method, supercell constructions, the tight-binding model, and a
  representative nonminimal enlargement example.}\BibitemShut {Stop}%
\bibitem [{\citenamefont {Huang}\ \emph {et~al.}(2017)\citenamefont {Huang},
  \citenamefont {Song}, \citenamefont {Huang},\ and\ \citenamefont
  {Hermele}}]{huang2017building}%
  \BibitemOpen
  \bibfield  {author} {\bibinfo {author} {\bibfnamefont {S.-J.}\ \bibnamefont
  {Huang}}, \bibinfo {author} {\bibfnamefont {H.}~\bibnamefont {Song}},
  \bibinfo {author} {\bibfnamefont {Y.-P.}\ \bibnamefont {Huang}},\ and\
  \bibinfo {author} {\bibfnamefont {M.}~\bibnamefont {Hermele}},\ }\bibfield
  {title} {\bibinfo {title} {Building crystalline topological phases from
  lower-dimensional states},\ }\href
  {https://doi.org/10.1103/PhysRevB.96.205106} {\bibfield  {journal} {\bibinfo
  {journal} {Physical Review B}\ }\textbf {\bibinfo {volume} {96}},\ \bibinfo
  {pages} {205106} (\bibinfo {year} {2017})}\BibitemShut {NoStop}%
\bibitem [{\citenamefont {Van~Miert}\ and\ \citenamefont
  {Ortix}(2018)}]{van2018higher}%
  \BibitemOpen
  \bibfield  {author} {\bibinfo {author} {\bibfnamefont {G.}~\bibnamefont
  {Van~Miert}}\ and\ \bibinfo {author} {\bibfnamefont {C.}~\bibnamefont
  {Ortix}},\ }\bibfield  {title} {\bibinfo {title} {Higher-order topological
  insulators protected by inversion and rotoinversion symmetries},\ }\href
  {https://doi.org/10.1103/PhysRevB.98.081110} {\bibfield  {journal} {\bibinfo
  {journal} {Physical Review B}\ }\textbf {\bibinfo {volume} {98}},\ \bibinfo
  {pages} {081110} (\bibinfo {year} {2018})}\BibitemShut {NoStop}%
\bibitem [{\citenamefont {Xu}\ \emph {et~al.}(2021)\citenamefont {Xu},
  \citenamefont {Elcoro}, \citenamefont {Li}, \citenamefont {Song},
  \citenamefont {Regnault}, \citenamefont {Yang}, \citenamefont {Sun},
  \citenamefont {Parkin}, \citenamefont {Felser},\ and\ \citenamefont
  {Bernevig}}]{xu2021three}%
  \BibitemOpen
  \bibfield  {author} {\bibinfo {author} {\bibfnamefont {Y.}~\bibnamefont
  {Xu}}, \bibinfo {author} {\bibfnamefont {L.}~\bibnamefont {Elcoro}}, \bibinfo
  {author} {\bibfnamefont {G.}~\bibnamefont {Li}}, \bibinfo {author}
  {\bibfnamefont {Z.-D.}\ \bibnamefont {Song}}, \bibinfo {author}
  {\bibfnamefont {N.}~\bibnamefont {Regnault}}, \bibinfo {author}
  {\bibfnamefont {Q.}~\bibnamefont {Yang}}, \bibinfo {author} {\bibfnamefont
  {Y.}~\bibnamefont {Sun}}, \bibinfo {author} {\bibfnamefont {S.}~\bibnamefont
  {Parkin}}, \bibinfo {author} {\bibfnamefont {C.}~\bibnamefont {Felser}},\
  and\ \bibinfo {author} {\bibfnamefont {B.~A.}\ \bibnamefont {Bernevig}},\
  }\bibfield  {title} {\bibinfo {title} {Three-dimensional real space
  invariants, obstructed atomic insulators and a new principle for active
  catalytic sites},\ }\href {https://arxiv.org/abs/2111.02433} {\bibfield
  {journal} {\bibinfo  {journal} {arXiv preprint arXiv:2111.02433}\ } (\bibinfo
  {year} {2021})}\BibitemShut {NoStop}%
\bibitem [{\citenamefont {Yu}\ \emph {et~al.}(2011)\citenamefont {Yu},
  \citenamefont {Qi}, \citenamefont {Bernevig}, \citenamefont {Fang},\ and\
  \citenamefont {Dai}}]{yu2011equivalent}%
  \BibitemOpen
  \bibfield  {author} {\bibinfo {author} {\bibfnamefont {R.}~\bibnamefont
  {Yu}}, \bibinfo {author} {\bibfnamefont {X.~L.}\ \bibnamefont {Qi}}, \bibinfo
  {author} {\bibfnamefont {A.}~\bibnamefont {Bernevig}}, \bibinfo {author}
  {\bibfnamefont {Z.}~\bibnamefont {Fang}},\ and\ \bibinfo {author}
  {\bibfnamefont {X.}~\bibnamefont {Dai}},\ }\bibfield  {title} {\bibinfo
  {title} {Equivalent expression of {$\mathbb{Z}_2$} topological invariant for
  band insulators using the non-{Abelian} {Berry} connection},\ }\href
  {https://doi.org/10.1103/PhysRevB.84.075119} {\bibfield  {journal} {\bibinfo
  {journal} {Physical Review B}\ }\textbf {\bibinfo {volume} {84}},\ \bibinfo
  {pages} {075119} (\bibinfo {year} {2011})}\BibitemShut {NoStop}%
\bibitem [{\citenamefont {Fidkowski}\ \emph {et~al.}(2011)\citenamefont
  {Fidkowski}, \citenamefont {Jackson},\ and\ \citenamefont
  {Klich}}]{fidkowski2011model}%
  \BibitemOpen
  \bibfield  {author} {\bibinfo {author} {\bibfnamefont {L.}~\bibnamefont
  {Fidkowski}}, \bibinfo {author} {\bibfnamefont {T.}~\bibnamefont {Jackson}},\
  and\ \bibinfo {author} {\bibfnamefont {I.}~\bibnamefont {Klich}},\ }\bibfield
   {title} {\bibinfo {title} {Model characterization of gapless edge modes of
  topological insulators using intermediate {Brillouin}-zone functions},\
  }\href {https://doi.org/10.1103/PhysRevLett.107.036601} {\bibfield  {journal}
  {\bibinfo  {journal} {Physical Review Letters}\ }\textbf {\bibinfo {volume}
  {107}},\ \bibinfo {pages} {036601} (\bibinfo {year} {2011})}\BibitemShut
  {NoStop}%
\bibitem [{\citenamefont {Alexandradinata}\ \emph {et~al.}(2014)\citenamefont
  {Alexandradinata}, \citenamefont {Dai},\ and\ \citenamefont
  {Bernevig}}]{alexandradinata2014wilson}%
  \BibitemOpen
  \bibfield  {author} {\bibinfo {author} {\bibfnamefont {A.}~\bibnamefont
  {Alexandradinata}}, \bibinfo {author} {\bibfnamefont {X.}~\bibnamefont
  {Dai}},\ and\ \bibinfo {author} {\bibfnamefont {B.~A.}\ \bibnamefont
  {Bernevig}},\ }\bibfield  {title} {\bibinfo {title} {{Wilson}-loop
  characterization of inversion-symmetric topological insulators},\ }\href
  {https://doi.org/10.1103/PhysRevB.89.155114} {\bibfield  {journal} {\bibinfo
  {journal} {Physical Review B}\ }\textbf {\bibinfo {volume} {89}},\ \bibinfo
  {pages} {155114} (\bibinfo {year} {2014})}\BibitemShut {NoStop}%
\bibitem [{\citenamefont {Bruns}\ and\ \citenamefont
  {Gubeladze}(2009)}]{bruns2009polytopes}%
  \BibitemOpen
  \bibfield  {author} {\bibinfo {author} {\bibfnamefont {W.}~\bibnamefont
  {Bruns}}\ and\ \bibinfo {author} {\bibfnamefont {J.}~\bibnamefont
  {Gubeladze}},\ }\href@noop {} {\emph {\bibinfo {title} {Polytopes, rings, and
  {K}-theory}}}\ (\bibinfo  {publisher} {Springer New York, NY},\ \bibinfo
  {year} {2009})\BibitemShut {NoStop}%
\bibitem [{\citenamefont {Bruns}\ and\ \citenamefont
  {Ichim}(2010)}]{bruns2010normaliz}%
  \BibitemOpen
  \bibfield  {author} {\bibinfo {author} {\bibfnamefont {W.}~\bibnamefont
  {Bruns}}\ and\ \bibinfo {author} {\bibfnamefont {B.}~\bibnamefont {Ichim}},\
  }\bibfield  {title} {\bibinfo {title} {{Normaliz}: {Algorithms} for affine
  monoids and rational cones},\ }\href
  {https://doi.org/10.1016/j.jalgebra.2010.01.031} {\bibfield  {journal}
  {\bibinfo  {journal} {Journal of Algebra}\ }\textbf {\bibinfo {volume}
  {324}},\ \bibinfo {pages} {1098} (\bibinfo {year} {2010})}\BibitemShut
  {NoStop}%
\bibitem [{\citenamefont {Christensen}\ \emph {et~al.}(2022)\citenamefont
  {Christensen}, \citenamefont {Po}, \citenamefont {Joannopoulos},\ and\
  \citenamefont {Solja{\v{c}}i{\'c}}}]{christensen2022location}%
  \BibitemOpen
  \bibfield  {author} {\bibinfo {author} {\bibfnamefont {T.}~\bibnamefont
  {Christensen}}, \bibinfo {author} {\bibfnamefont {H.~C.}\ \bibnamefont {Po}},
  \bibinfo {author} {\bibfnamefont {J.~D.}\ \bibnamefont {Joannopoulos}},\ and\
  \bibinfo {author} {\bibfnamefont {M.}~\bibnamefont {Solja{\v{c}}i{\'c}}},\
  }\bibfield  {title} {\bibinfo {title} {Location and topology of the
  fundamental gap in photonic crystals},\ }\href
  {https://doi.org/10.1103/PhysRevX.12.021066} {\bibfield  {journal} {\bibinfo
  {journal} {Physical Review X}\ }\textbf {\bibinfo {volume} {12}},\ \bibinfo
  {pages} {021066} (\bibinfo {year} {2022})}\BibitemShut {NoStop}%
\bibitem [{\citenamefont {Hwang}\ \emph {et~al.}(2026)\citenamefont {Hwang},
  \citenamefont {Gupta}, \citenamefont {Morales-P{\'e}rez}, \citenamefont
  {Devescovi}, \citenamefont {Garc{\'\i}a-D{\'\i}ez}, \citenamefont
  {Ma{\~n}es}, \citenamefont {Vergniory}, \citenamefont {Garc{\'\i}a-Etxarri},\
  and\ \citenamefont {Bradlyn}}]{hwang2026building}%
  \BibitemOpen
  \bibfield  {author} {\bibinfo {author} {\bibfnamefont {Y.}~\bibnamefont
  {Hwang}}, \bibinfo {author} {\bibfnamefont {V.}~\bibnamefont {Gupta}},
  \bibinfo {author} {\bibfnamefont {A.}~\bibnamefont {Morales-P{\'e}rez}},
  \bibinfo {author} {\bibfnamefont {C.}~\bibnamefont {Devescovi}}, \bibinfo
  {author} {\bibfnamefont {M.}~\bibnamefont {Garc{\'\i}a-D{\'\i}ez}}, \bibinfo
  {author} {\bibfnamefont {J.~L.}\ \bibnamefont {Ma{\~n}es}}, \bibinfo {author}
  {\bibfnamefont {M.~G.}\ \bibnamefont {Vergniory}}, \bibinfo {author}
  {\bibfnamefont {A.}~\bibnamefont {Garc{\'\i}a-Etxarri}},\ and\ \bibinfo
  {author} {\bibfnamefont {B.}~\bibnamefont {Bradlyn}},\ }\bibfield  {title}
  {\bibinfo {title} {Building blocks of topological band theory for photonic
  crystals},\ }\href {https://arxiv.org/abs/2601.06293} {\bibfield  {journal}
  {\bibinfo  {journal} {arXiv preprint arXiv:2601.06293}\ } (\bibinfo {year}
  {2026})}\BibitemShut {NoStop}%
\bibitem [{\citenamefont {Aroyo}\ \emph
  {et~al.}(2006{\natexlab{a}})\citenamefont {Aroyo}, \citenamefont
  {Perez-Mato}, \citenamefont {Capillas}, \citenamefont {Kroumova},
  \citenamefont {Ivantchev}, \citenamefont {Madariaga}, \citenamefont {Kirov},\
  and\ \citenamefont {Wondratschek}}]{aroyo2006bilbao1}%
  \BibitemOpen
  \bibfield  {author} {\bibinfo {author} {\bibfnamefont {M.~I.}\ \bibnamefont
  {Aroyo}}, \bibinfo {author} {\bibfnamefont {J.~M.}\ \bibnamefont
  {Perez-Mato}}, \bibinfo {author} {\bibfnamefont {C.}~\bibnamefont
  {Capillas}}, \bibinfo {author} {\bibfnamefont {E.}~\bibnamefont {Kroumova}},
  \bibinfo {author} {\bibfnamefont {S.}~\bibnamefont {Ivantchev}}, \bibinfo
  {author} {\bibfnamefont {G.}~\bibnamefont {Madariaga}}, \bibinfo {author}
  {\bibfnamefont {A.}~\bibnamefont {Kirov}},\ and\ \bibinfo {author}
  {\bibfnamefont {H.}~\bibnamefont {Wondratschek}},\ }\bibfield  {title}
  {\bibinfo {title} {{Bilbao Crystallographic Server: I. Databases} and
  crystallographic computing programs},\ }\href
  {https://doi.org/10.1524/zkri.2006.221.1.15} {\bibfield  {journal} {\bibinfo
  {journal} {Zeitschrift f{\"u}r Kristallographie-Crystalline Materials}\
  }\textbf {\bibinfo {volume} {221}},\ \bibinfo {pages} {15} (\bibinfo {year}
  {2006}{\natexlab{a}})}\BibitemShut {NoStop}%
\bibitem [{\citenamefont {Aroyo}\ \emph
  {et~al.}(2006{\natexlab{b}})\citenamefont {Aroyo}, \citenamefont {Kirov},
  \citenamefont {Capillas}, \citenamefont {Perez-Mato},\ and\ \citenamefont
  {Wondratschek}}]{aroyo2006bilbao2}%
  \BibitemOpen
  \bibfield  {author} {\bibinfo {author} {\bibfnamefont {M.~I.}\ \bibnamefont
  {Aroyo}}, \bibinfo {author} {\bibfnamefont {A.}~\bibnamefont {Kirov}},
  \bibinfo {author} {\bibfnamefont {C.}~\bibnamefont {Capillas}}, \bibinfo
  {author} {\bibfnamefont {J.}~\bibnamefont {Perez-Mato}},\ and\ \bibinfo
  {author} {\bibfnamefont {H.}~\bibnamefont {Wondratschek}},\ }\bibfield
  {title} {\bibinfo {title} {{Bilbao Crystallographic Server. II.
  Representations} of crystallographic point groups and space groups},\ }\href
  {https://doi.org/10.1107/S0108767305040286} {\bibfield  {journal} {\bibinfo
  {journal} {Acta Crystallographica Section A: Foundations of Crystallography}\
  }\textbf {\bibinfo {volume} {62}},\ \bibinfo {pages} {115} (\bibinfo {year}
  {2006}{\natexlab{b}})}\BibitemShut {NoStop}%
\bibitem [{\citenamefont {Aroyo}\ \emph {et~al.}(2011)\citenamefont {Aroyo},
  \citenamefont {Perez-Mato}, \citenamefont {Orobengoa}, \citenamefont {Tasci},
  \citenamefont {de~la Flor},\ and\ \citenamefont
  {Kirov}}]{aroyo2011crystallography}%
  \BibitemOpen
  \bibfield  {author} {\bibinfo {author} {\bibfnamefont {M.~I.}\ \bibnamefont
  {Aroyo}}, \bibinfo {author} {\bibfnamefont {J.}~\bibnamefont {Perez-Mato}},
  \bibinfo {author} {\bibfnamefont {D.}~\bibnamefont {Orobengoa}}, \bibinfo
  {author} {\bibfnamefont {E.}~\bibnamefont {Tasci}}, \bibinfo {author}
  {\bibfnamefont {G.}~\bibnamefont {de~la Flor}},\ and\ \bibinfo {author}
  {\bibfnamefont {A.}~\bibnamefont {Kirov}},\ }\bibfield  {title} {\bibinfo
  {title} {Crystallography online: {Bilbao Crystallographic Server}},\
  }\href@noop {} {\bibfield  {journal} {\bibinfo  {journal} {Bulg. Chem.
  Commun}\ }\textbf {\bibinfo {volume} {43}},\ \bibinfo {pages} {183} (\bibinfo
  {year} {2011})}\BibitemShut {NoStop}%
\bibitem [{\citenamefont {Vergniory}\ \emph {et~al.}(2017)\citenamefont
  {Vergniory}, \citenamefont {Elcoro}, \citenamefont {Wang}, \citenamefont
  {Cano}, \citenamefont {Felser}, \citenamefont {Aroyo}, \citenamefont
  {Bernevig},\ and\ \citenamefont {Bradlyn}}]{vergniory2017graph}%
  \BibitemOpen
  \bibfield  {author} {\bibinfo {author} {\bibfnamefont {M.}~\bibnamefont
  {Vergniory}}, \bibinfo {author} {\bibfnamefont {L.}~\bibnamefont {Elcoro}},
  \bibinfo {author} {\bibfnamefont {Z.}~\bibnamefont {Wang}}, \bibinfo {author}
  {\bibfnamefont {J.}~\bibnamefont {Cano}}, \bibinfo {author} {\bibfnamefont
  {C.}~\bibnamefont {Felser}}, \bibinfo {author} {\bibfnamefont
  {M.}~\bibnamefont {Aroyo}}, \bibinfo {author} {\bibfnamefont {B.~A.}\
  \bibnamefont {Bernevig}},\ and\ \bibinfo {author} {\bibfnamefont
  {B.}~\bibnamefont {Bradlyn}},\ }\bibfield  {title} {\bibinfo {title} {Graph
  theory data for topological quantum chemistry},\ }\href
  {https://doi.org/10.1103/PhysRevE.96.023310} {\bibfield  {journal} {\bibinfo
  {journal} {Physical Review E}\ }\textbf {\bibinfo {volume} {96}},\ \bibinfo
  {pages} {023310} (\bibinfo {year} {2017})}\BibitemShut {NoStop}%
\bibitem [{\citenamefont {Elcoro}\ \emph {et~al.}(2017)\citenamefont {Elcoro},
  \citenamefont {Bradlyn}, \citenamefont {Wang}, \citenamefont {Vergniory},
  \citenamefont {Cano}, \citenamefont {Felser}, \citenamefont {Bernevig},
  \citenamefont {Orobengoa}, \citenamefont {Flor},\ and\ \citenamefont
  {Aroyo}}]{elcoro2017double}%
  \BibitemOpen
  \bibfield  {author} {\bibinfo {author} {\bibfnamefont {L.}~\bibnamefont
  {Elcoro}}, \bibinfo {author} {\bibfnamefont {B.}~\bibnamefont {Bradlyn}},
  \bibinfo {author} {\bibfnamefont {Z.}~\bibnamefont {Wang}}, \bibinfo {author}
  {\bibfnamefont {M.~G.}\ \bibnamefont {Vergniory}}, \bibinfo {author}
  {\bibfnamefont {J.}~\bibnamefont {Cano}}, \bibinfo {author} {\bibfnamefont
  {C.}~\bibnamefont {Felser}}, \bibinfo {author} {\bibfnamefont {B.~A.}\
  \bibnamefont {Bernevig}}, \bibinfo {author} {\bibfnamefont {D.}~\bibnamefont
  {Orobengoa}}, \bibinfo {author} {\bibfnamefont {G.}~\bibnamefont {Flor}},\
  and\ \bibinfo {author} {\bibfnamefont {M.~I.}\ \bibnamefont {Aroyo}},\
  }\bibfield  {title} {\bibinfo {title} {Double crystallographic groups and
  their representations on the {Bilbao Crystallographic Server}},\ }\href
  {https://doi.org/10.1107/S1600576717011712} {\bibfield  {journal} {\bibinfo
  {journal} {Journal of Applied Crystallography}\ }\textbf {\bibinfo {volume}
  {50}},\ \bibinfo {pages} {1457} (\bibinfo {year} {2017})}\BibitemShut
  {NoStop}%
\bibitem [{bil()}]{bilbao_tools}%
  \BibitemOpen
  \href@noop {} {\bibinfo {title} {{Bilbao Crystallographic Server}}},\
  \bibinfo {howpublished} {\url{https://cryst.ehu.es}},\ \bibinfo {note}
  {\texttt{BANDREP}, \texttt{WYCKPOS}, and \texttt{Representations PG}
  tools}\BibitemShut {NoStop}%
\bibitem [{\citenamefont {Alexandradinata}\ and\ \citenamefont
  {H{\"o}ller}(2018)}]{alexandradinata2018no}%
  \BibitemOpen
  \bibfield  {author} {\bibinfo {author} {\bibfnamefont {A.}~\bibnamefont
  {Alexandradinata}}\ and\ \bibinfo {author} {\bibfnamefont {J.}~\bibnamefont
  {H{\"o}ller}},\ }\bibfield  {title} {\bibinfo {title} {No-go theorem for
  topological insulators and high-throughput identification of {Chern}
  insulators},\ }\href {https://doi.org/10.1103/PhysRevB.98.184305} {\bibfield
  {journal} {\bibinfo  {journal} {Physical Review B}\ }\textbf {\bibinfo
  {volume} {98}},\ \bibinfo {pages} {184305} (\bibinfo {year}
  {2018})}\BibitemShut {NoStop}%
\bibitem [{\citenamefont {Cano}\ \emph {et~al.}(2021)\citenamefont {Cano},
  \citenamefont {Fang}, \citenamefont {Pixley},\ and\ \citenamefont
  {Wilson}}]{cano2021moire}%
  \BibitemOpen
  \bibfield  {author} {\bibinfo {author} {\bibfnamefont {J.}~\bibnamefont
  {Cano}}, \bibinfo {author} {\bibfnamefont {S.}~\bibnamefont {Fang}}, \bibinfo
  {author} {\bibfnamefont {J.}~\bibnamefont {Pixley}},\ and\ \bibinfo {author}
  {\bibfnamefont {J.~H.}\ \bibnamefont {Wilson}},\ }\bibfield  {title}
  {\bibinfo {title} {Moir{\'e} superlattice on the surface of a topological
  insulator},\ }\href {https://doi.org/10.1103/PhysRevB.103.155157} {\bibfield
  {journal} {\bibinfo  {journal} {Physical Review B}\ }\textbf {\bibinfo
  {volume} {103}},\ \bibinfo {pages} {155157} (\bibinfo {year}
  {2021})}\BibitemShut {NoStop}%
\bibitem [{\citenamefont {Scheer}\ and\ \citenamefont
  {Lian}(2023)}]{scheer2023kagome}%
  \BibitemOpen
  \bibfield  {author} {\bibinfo {author} {\bibfnamefont {M.~G.}\ \bibnamefont
  {Scheer}}\ and\ \bibinfo {author} {\bibfnamefont {B.}~\bibnamefont {Lian}},\
  }\bibfield  {title} {\bibinfo {title} {Kagome and honeycomb flat bands in
  moir{\'e} graphene},\ }\href {https://doi.org/10.1103/PhysRevB.108.245136}
  {\bibfield  {journal} {\bibinfo  {journal} {Physical Review B}\ }\textbf
  {\bibinfo {volume} {108}},\ \bibinfo {pages} {245136} (\bibinfo {year}
  {2023})}\BibitemShut {NoStop}%
\bibitem [{\citenamefont {Cr{\'e}pel}\ and\ \citenamefont
  {Cano}(2025)}]{crepel2025efficient}%
  \BibitemOpen
  \bibfield  {author} {\bibinfo {author} {\bibfnamefont {V.}~\bibnamefont
  {Cr{\'e}pel}}\ and\ \bibinfo {author} {\bibfnamefont {J.}~\bibnamefont
  {Cano}},\ }\bibfield  {title} {\bibinfo {title} {Efficient prediction of
  superlattice and anomalous miniband topology from quantum geometry},\ }\href
  {https://doi.org/10.1103/PhysRevX.15.011004} {\bibfield  {journal} {\bibinfo
  {journal} {Physical Review X}\ }\textbf {\bibinfo {volume} {15}},\ \bibinfo
  {pages} {011004} (\bibinfo {year} {2025})}\BibitemShut {NoStop}%
\bibitem [{\citenamefont {Yang}\ \emph {et~al.}(2024)\citenamefont {Yang},
  \citenamefont {Xu}, \citenamefont {Feng}, \citenamefont {Schindler},
  \citenamefont {Xu}, \citenamefont {Bi}, \citenamefont {Bernevig},
  \citenamefont {Tang},\ and\ \citenamefont {Liu}}]{yang2024topological}%
  \BibitemOpen
  \bibfield  {author} {\bibinfo {author} {\bibfnamefont {K.}~\bibnamefont
  {Yang}}, \bibinfo {author} {\bibfnamefont {Z.}~\bibnamefont {Xu}}, \bibinfo
  {author} {\bibfnamefont {Y.}~\bibnamefont {Feng}}, \bibinfo {author}
  {\bibfnamefont {F.}~\bibnamefont {Schindler}}, \bibinfo {author}
  {\bibfnamefont {Y.}~\bibnamefont {Xu}}, \bibinfo {author} {\bibfnamefont
  {Z.}~\bibnamefont {Bi}}, \bibinfo {author} {\bibfnamefont {B.~A.}\
  \bibnamefont {Bernevig}}, \bibinfo {author} {\bibfnamefont {P.}~\bibnamefont
  {Tang}},\ and\ \bibinfo {author} {\bibfnamefont {C.-X.}\ \bibnamefont
  {Liu}},\ }\bibfield  {title} {\bibinfo {title} {Topological minibands and
  interaction driven quantum anomalous hall state in topological insulator
  based moir{\'e} heterostructures},\ }\href
  {https://doi.org/10.1038/s41467-024-46717-7} {\bibfield  {journal} {\bibinfo
  {journal} {Nature communications}\ }\textbf {\bibinfo {volume} {15}},\
  \bibinfo {pages} {2670} (\bibinfo {year} {2024})}\BibitemShut {NoStop}%
\bibitem [{\citenamefont {Yang}\ \emph {et~al.}(2025)\citenamefont {Yang},
  \citenamefont {Liu}, \citenamefont {Schindler},\ and\ \citenamefont
  {Liu}}]{yang2025engineering}%
  \BibitemOpen
  \bibfield  {author} {\bibinfo {author} {\bibfnamefont {K.}~\bibnamefont
  {Yang}}, \bibinfo {author} {\bibfnamefont {Y.}~\bibnamefont {Liu}}, \bibinfo
  {author} {\bibfnamefont {F.}~\bibnamefont {Schindler}},\ and\ \bibinfo
  {author} {\bibfnamefont {C.-X.}\ \bibnamefont {Liu}},\ }\bibfield  {title}
  {\bibinfo {title} {Engineering miniband topology via band folding in
  moir{\'e} superlattice materials},\ }\href
  {https://doi.org/10.1103/PhysRevB.111.L241104} {\bibfield  {journal}
  {\bibinfo  {journal} {Physical Review B}\ }\textbf {\bibinfo {volume}
  {111}},\ \bibinfo {pages} {L241104} (\bibinfo {year} {2025})}\BibitemShut
  {NoStop}%
\bibitem [{\citenamefont {Lhachemi}\ \emph {et~al.}(2026)\citenamefont
  {Lhachemi}, \citenamefont {Cr{\'e}pel},\ and\ \citenamefont
  {Cano}}]{lhachemi2026efficient}%
  \BibitemOpen
  \bibfield  {author} {\bibinfo {author} {\bibfnamefont {M.~N.~Y.}\
  \bibnamefont {Lhachemi}}, \bibinfo {author} {\bibfnamefont {V.}~\bibnamefont
  {Cr{\'e}pel}},\ and\ \bibinfo {author} {\bibfnamefont {J.}~\bibnamefont
  {Cano}},\ }\bibfield  {title} {\bibinfo {title} {Efficient prediction of
  topological superlattice bands with spin-orbit coupling},\ }\href
  {https://doi.org/10.1103/fcy2-2py5} {\bibfield  {journal} {\bibinfo
  {journal} {Physical Review B}\ }\textbf {\bibinfo {volume} {113}},\ \bibinfo
  {pages} {165109} (\bibinfo {year} {2026})}\BibitemShut {NoStop}%
\bibitem [{\citenamefont {McMillan}(1976)}]{mcmillan1976theory}%
  \BibitemOpen
  \bibfield  {author} {\bibinfo {author} {\bibfnamefont {W.~L.}\ \bibnamefont
  {McMillan}},\ }\bibfield  {title} {\bibinfo {title} {Theory of
  discommensurations and the commensurate-incommensurate charge-density-wave
  phase transition},\ }\href {https://doi.org/10.1103/PhysRevB.14.1496}
  {\bibfield  {journal} {\bibinfo  {journal} {Physical Review B}\ }\textbf
  {\bibinfo {volume} {14}},\ \bibinfo {pages} {1496} (\bibinfo {year}
  {1976})}\BibitemShut {NoStop}%
\bibitem [{\citenamefont {Gr{\"u}ner}(1988)}]{gruner1988dynamics}%
  \BibitemOpen
  \bibfield  {author} {\bibinfo {author} {\bibfnamefont {G.}~\bibnamefont
  {Gr{\"u}ner}},\ }\bibfield  {title} {\bibinfo {title} {The dynamics of
  charge-density waves},\ }\href {https://doi.org/10.1103/RevModPhys.60.1129}
  {\bibfield  {journal} {\bibinfo  {journal} {Reviews of {M}odern {P}hysics}\
  }\textbf {\bibinfo {volume} {60}},\ \bibinfo {pages} {1129} (\bibinfo {year}
  {1988})}\BibitemShut {NoStop}%
\bibitem [{\citenamefont {Wieder}\ \emph
  {et~al.}(2020{\natexlab{b}})\citenamefont {Wieder}, \citenamefont {Lin},\
  and\ \citenamefont {Bradlyn}}]{wieder2020axionic}%
  \BibitemOpen
  \bibfield  {author} {\bibinfo {author} {\bibfnamefont {B.~J.}\ \bibnamefont
  {Wieder}}, \bibinfo {author} {\bibfnamefont {K.-S.}\ \bibnamefont {Lin}},\
  and\ \bibinfo {author} {\bibfnamefont {B.}~\bibnamefont {Bradlyn}},\
  }\bibfield  {title} {\bibinfo {title} {{Axionic} band topology in
  inversion-symmetric {Weyl}-charge-density waves},\ }\href
  {https://doi.org/10.1103/PhysRevResearch.2.042010} {\bibfield  {journal}
  {\bibinfo  {journal} {Physical Review Research}\ }\textbf {\bibinfo {volume}
  {2}},\ \bibinfo {pages} {042010} (\bibinfo {year}
  {2020}{\natexlab{b}})}\BibitemShut {NoStop}%
\bibitem [{\citenamefont {Devescovi}\ \emph {et~al.}(2024)\citenamefont
  {Devescovi}, \citenamefont {Morales-P{\'e}rez}, \citenamefont {Hwang},
  \citenamefont {Garc{\'\i}a-D{\'\i}ez}, \citenamefont {Robredo}, \citenamefont
  {Luis~Ma{\~n}es}, \citenamefont {Bradlyn}, \citenamefont
  {Garc{\'\i}a-Etxarri},\ and\ \citenamefont {Vergniory}}]{devescovi2024axion}%
  \BibitemOpen
  \bibfield  {author} {\bibinfo {author} {\bibfnamefont {C.}~\bibnamefont
  {Devescovi}}, \bibinfo {author} {\bibfnamefont {A.}~\bibnamefont
  {Morales-P{\'e}rez}}, \bibinfo {author} {\bibfnamefont {Y.}~\bibnamefont
  {Hwang}}, \bibinfo {author} {\bibfnamefont {M.}~\bibnamefont
  {Garc{\'\i}a-D{\'\i}ez}}, \bibinfo {author} {\bibfnamefont {I.}~\bibnamefont
  {Robredo}}, \bibinfo {author} {\bibfnamefont {J.}~\bibnamefont
  {Luis~Ma{\~n}es}}, \bibinfo {author} {\bibfnamefont {B.}~\bibnamefont
  {Bradlyn}}, \bibinfo {author} {\bibfnamefont {A.}~\bibnamefont
  {Garc{\'\i}a-Etxarri}},\ and\ \bibinfo {author} {\bibfnamefont {M.~G.}\
  \bibnamefont {Vergniory}},\ }\bibfield  {title} {\bibinfo {title} {{Axion}
  topology in photonic crystal domain walls},\ }\href
  {https://doi.org/10.1038/s41467-024-50766-3} {\bibfield  {journal} {\bibinfo
  {journal} {Nature {C}ommunications}\ }\textbf {\bibinfo {volume} {15}},\
  \bibinfo {pages} {6814} (\bibinfo {year} {2024})}\BibitemShut {NoStop}%
\bibitem [{\citenamefont {Fang}\ and\ \citenamefont
  {Cano}(2023)}]{fang2023symmetry}%
  \BibitemOpen
  \bibfield  {author} {\bibinfo {author} {\bibfnamefont {Y.}~\bibnamefont
  {Fang}}\ and\ \bibinfo {author} {\bibfnamefont {J.}~\bibnamefont {Cano}},\
  }\bibfield  {title} {\bibinfo {title} {Symmetry indicators in commensurate
  magnetic flux},\ }\href {https://doi.org/10.1103/PhysRevB.107.245108}
  {\bibfield  {journal} {\bibinfo  {journal} {Physical Review B}\ }\textbf
  {\bibinfo {volume} {107}},\ \bibinfo {pages} {245108} (\bibinfo {year}
  {2023})}\BibitemShut {NoStop}%
\bibitem [{\citenamefont {Qi}\ \emph {et~al.}(2006)\citenamefont {Qi},
  \citenamefont {Wu},\ and\ \citenamefont {Zhang}}]{qi2006topological}%
  \BibitemOpen
  \bibfield  {author} {\bibinfo {author} {\bibfnamefont {X.-L.}\ \bibnamefont
  {Qi}}, \bibinfo {author} {\bibfnamefont {Y.-S.}\ \bibnamefont {Wu}},\ and\
  \bibinfo {author} {\bibfnamefont {S.-C.}\ \bibnamefont {Zhang}},\ }\bibfield
  {title} {\bibinfo {title} {Topological quantization of the spin {Hall} effect
  in two-dimensional paramagnetic semiconductors},\ }\href
  {https://doi.org/10.1103/PhysRevB.74.085308} {\bibfield  {journal} {\bibinfo
  {journal} {Physical Review B}\ }\textbf {\bibinfo {volume} {74}},\ \bibinfo
  {pages} {085308} (\bibinfo {year} {2006})}\BibitemShut {NoStop}%
\bibitem [{\citenamefont {Geier}\ \emph {et~al.}(2018)\citenamefont {Geier},
  \citenamefont {Trifunovic}, \citenamefont {Hoskam},\ and\ \citenamefont
  {Brouwer}}]{geier2018second}%
  \BibitemOpen
  \bibfield  {author} {\bibinfo {author} {\bibfnamefont {M.}~\bibnamefont
  {Geier}}, \bibinfo {author} {\bibfnamefont {L.}~\bibnamefont {Trifunovic}},
  \bibinfo {author} {\bibfnamefont {M.}~\bibnamefont {Hoskam}},\ and\ \bibinfo
  {author} {\bibfnamefont {P.~W.}\ \bibnamefont {Brouwer}},\ }\bibfield
  {title} {\bibinfo {title} {Second-order topological insulators and
  superconductors with an order-two crystalline symmetry},\ }\href
  {https://doi.org/10.1103/PhysRevB.97.205135} {\bibfield  {journal} {\bibinfo
  {journal} {Physical Review B}\ }\textbf {\bibinfo {volume} {97}},\ \bibinfo
  {pages} {205135} (\bibinfo {year} {2018})}\BibitemShut {NoStop}%
\bibitem [{\citenamefont {Khalaf}(2018)}]{khalaf2018higher}%
  \BibitemOpen
  \bibfield  {author} {\bibinfo {author} {\bibfnamefont {E.}~\bibnamefont
  {Khalaf}},\ }\bibfield  {title} {\bibinfo {title} {Higher-order topological
  insulators and superconductors protected by inversion symmetry},\ }\href
  {https://doi.org/10.1103/PhysRevB.97.205136} {\bibfield  {journal} {\bibinfo
  {journal} {Physical Review B}\ }\textbf {\bibinfo {volume} {97}},\ \bibinfo
  {pages} {205136} (\bibinfo {year} {2018})}\BibitemShut {NoStop}%
\end{thebibliography}%

\let\addcontentsline\oldaddcontentsline
\clearpage

\onecolumngrid
\begin{center}
\textbf{\large Supplemental Material: \\ \ourtitle}
\end{center}

\setcounter{section}{0}
\setcounter{figure}{0}
\setcounter{equation}{0}
\renewcommand{\thefigure}{S\arabic{figure}}
\renewcommand{\theequation}{S\arabic{equation}}
\renewcommand{\thesection}{S\arabic{section}}

\tableofcontents
\hfill \\

\section{pseudoinverse of the band-representation matrix}
\label{app:pseudoinv}
In this section, we explain how to construct a pseudoinverse of the band-representation matrix $BR$ and use it to solve
\bg
\bb v = BR \cdot \bb m
\label{seq:v_from_m}
\eg
for $\bb m$ up to contributions from the kernel of $BR$.
(Of course, the following discussion applies to any integer-valued matrix.)
Here $BR$ is an integer matrix of size $D_{\bb v} \times D_{\bb m}$, mapping the site-symmetry irrep multiplicity vector $\bb m \in \Z^{D_{\bb m}}$ to the symmetry-data vector $\bb v \in \Z^{D_{\bb v}}$.
Any integer matrix $BR$ admits a Smith normal form $\Lambda$: the matrix $BR$ can be decomposed into
\bg
BR = L \cdot \Lambda \cdot R,
\label{seq:smith}
\eg
where $L$ and $R$ are unimodular integer matrices such that $|{\rm det} L| = |{\rm det} R| =1$, and the normal form $\Lambda$ is a diagonal matrix whose diagonal part is given by
\bg
{\rm Diag} (\Lambda) = (\lambda_1, \lambda_2, \dots, \lambda_{r_{BR}}, 0, \dots , 0)
\label{seq:lambda_diag}
\eg
Here, $r_{BR} = {\rm rank}(BR)$ is the rank of $BR$, $\lambda_i \in \N$ ($i=1, \dots, r_{BR}$) are the elementary divisors (or invariant factors) satisfying $\lambda_1 | \lambda_2 | \cdots | \lambda_{r_{BR}}$.
($a|b$ denotes that $a$ divides $b$.)

We define the pseudoinverse $\Lambda^\ddagger$ of $\Lambda$ as
\bg
(\Lambda^\ddagger)_{ii} = \lambda_i^{-1}
\quad (i=1,\dots,r_{BR})
\eg
and zero otherwise.
It is straightforward to verify that
\bg
\Lambda \cdot \Lambda^\ddagger \cdot \Lambda = \Lambda,
\quad
\Lambda^\ddagger \cdot \Lambda \cdot \Lambda^\ddagger = \Lambda^\ddagger.
\label{seq:lambda_cond}
\eg
A pseudoinverse for a general integer matrix can be constructed via its Smith decomposition so that it satisfies analogous identities.
We define the pseudoinverse of $BR$ by
\bg
BR^\ddagger = R^{-1} \cdot \Lambda^\ddagger \cdot L^{-1}.
\label{seq:BR_inv}
\eg
Using Eq.~\eqref{seq:lambda_cond}, we directly compute
\bg
BR \cdot BR^\ddagger \cdot BR
= L \cdot \Lambda \cdot \Lambda^\ddagger \cdot \Lambda \cdot R
= L \cdot \Lambda \cdot R = BR.
\label{seq:inverse_cond}
\eg
A similar computation shows $BR^\ddagger \cdot BR \cdot BR^\ddagger = BR^\ddagger$.

Now we show that $BR^\ddagger$ provides a solution to Eq.~\eqref{seq:v_from_m}.
Suppose $\bb v$ lies in the image of $BR$, so that Eq.~\eqref{seq:v_from_m} admits solutions.
Acting with $BR^\ddagger$ gives
\bg
BR^\ddagger \cdot \bb v = BR^\ddagger \cdot BR \cdot \bb m.
\eg
Using Eq.~\eqref{seq:inverse_cond}, we find $BR \cdot ( BR^\ddagger \cdot \bb v )
= BR \cdot \bb m = \bb v$.
Therefore, the general solution $\bb m$ can be expressed as
\bg
\bb m = BR^\ddagger \cdot \bb v + \bb m_{\rm ker},
\label{seq:m_general}
\eg
where $\bb m_{\rm ker} \in {\rm ker} (BR)$ such that $BR \cdot \bb m_{\rm ker} = 0$.

The kernel of $BR$ is also determined by the Smith decomposition in Eq.~\eqref{seq:smith}.
In particular, let $(R^{-1})_{:,\alpha}$ denote the $\alpha$th column of $R^{-1}$.
Then the general integer-valued kernel of $BR$ can be expressed as
\bg
\bb m_{\rm ker}
= \sum_{\alpha=1}^{D_{\bb m}-r_{BR}}
(R^{-1})_{:,\alpha+r_{BR}} \, n_\alpha,
\quad n_\alpha \in \Z.
\eg
Equivalently, $\bb m_{\rm ker}
= R^{-1} \cdot ( 0, \dots, 0, n_1, \dots,n_{D_{\bb m}-r_{BR}} )^{\rm T}$, where the first $r_{BR}$ entries vanish.
In other words, any $\bb m_{\rm ker}$ is an integer linear combination of the columns $(R^{-1})_{:,r_{BR}+1, \dots, D_{\bb m}}$.
One immediately verifies that $BR \cdot \bb m_{\rm ker}=0$:
\bg
BR \cdot \bb m_{\rm ker}
=L \cdot \Lambda \cdot R \cdot R^{-1}
\cdot (0, \dots, 0, n_1, \dots, n_{D_{\bb m}-r_{BR}} )^{\rm T}
= L \cdot \Lambda \cdot (0, \dots, 0, n_1, \dots, n_{D_{\bb m}-r_{BR}} )^{\rm T}
= 0,
\eg
since only the first $r_{BR}$ diagonal entries of $\Lambda$ are nonzero.

\section{Symmetry-compatible unit-cell enlargement}
\label{app:supercell}
We begin by fixing notation.
Let $L$ be a two-dimensional Bravais lattice generated by primitive vectors $\bb a_1, \bb a_2$, collected into the $2 \times 2$ matrix
\bg
\mc A = (\bb a_1, \bb a_2).
\eg
The translation subgroup is $\mc T = \{ \mc A \, \bb n | \bb n \in \Z^2 \}$.
A wallpaper-group element is written as
\bg
g = \{ O_g | \bb t_g \},
\eg
where $O_g \in O(2)$ is the point-group part and $\bb t_g \in \mc A \Q^2$ is the translation part, which may be fractional in nonsymmorphic groups.
The action of $g$ on the lattice basis is
\bg
g : \mc A \to O_g \, \mc A = \mc A \, M_g,
\label{seq:Mg_def}
\eg
where $M_g \in GL(2,\Z)$ is the integer matrix representing the induced action on lattice coordinates in the chosen conventional setting.

As a simple example, consider the $pm$ wallpaper group in the conventional rectangular setting $\bb a_1 = (1,0)$ and $\bb a_2 = (0,1)$.
The mirror $m_{10}$ that flips $x$ sends $\bb a_1 \to -\bb a_1$ and $\bb a_2 \to \bb a_2$.
Hence
\bg
O_{m_{10}} = \bpm -1 & 0 \\ 0 & 1 \epm,
\quad
M_{m_{10}} = \bpm -1 & 0 \\ 0 & 1 \epm.
\eg
In contrast, in the $pg$ wallpaper group, one has a glide mirror
\bg
g_{10} = \{ m_{10} | \tfrac12 \bb a_2 \},
\eg
for which the point-group part is identical but the translation part is fractional, $\bb t_g = \frac12 \bb a_2$.

We now describe unit-cell enlargement.
A supercell is defined by an integer matrix $\mc S_{\rm uc}$ with nonzero determinant,
\bg
\mc A' = \mc A \, \mc S_{\rm uc},
\quad
N_{\rm sc} = |\det \mc S_{\rm uc}|,
\label{seq:newlat_def}
\eg
so that the enlarged lattice is $L' = \{ \mc A' \, \bb n | \bb n \in \Z^2 \}$.
We require that the enlargement preserve the wallpaper group $G$ in the same conventional setting.
In particular, we impose symmetry compatibility using the integer matrices $M_g$ of the point group $P=G/\mc T$ defined in the conventional setting.
After enlargement according to Eq.~\eqref{seq:newlat_def}, we require that the enlarged lattice still realizes the same conventional integer representation of the point group.
Concretely, let $M_P = \{ M_g | g \in P \}$ denote the set of representation matrices for $P$ in the conventional setting.
Then, we demand that conjugation by $\mc S_{\rm uc}$ permutes this set:
\bg
\mc S_{\rm uc}^{-1} \, M_P \, \mc S_{\rm uc} = M_P,
\label{seq:MP_conj_invariant1}
\eg
i.e. for every $g \in P$ there exists a (generally different) $g' \in P$ such that
\bg
\mc S_{\rm uc}^{-1} \, M_g \, \mc S_{\rm uc}=M_{g'}.
\label{seq:MP_conj_invariant2}
\eg
Such permutations occur, for example, in $p4mm$, where an enlargement can interchange axial and diagonal mirror matrices within $M_P$ while preserving the full set.

We now turn to the nonsymmorphic part.
Given $g = \{ O_g | \bb t_g \}$, the element may combine with primitive translations.
Hence, in checking symmetry compatibility, one must allow for representatives of the form
\bg
\tilde g = \{ O_g | \bb t_g + \bs \tau \},
\quad \bs \tau \in L,
\eg
since such elements belong to the same coset of the translation subgroup $\mc T$ in the original group $G$.

After enlargement, the point-group part $O_g$ has already been accounted for by the point-group condition in Eq.~\eqref{seq:MP_conj_invariant2}, i.e. we suppose that $g \to g'$ is relabeled accordingly.
The remaining requirement concerns the fractional translation part.
Denote the fractional translation part of $g$ as $\bb t_g = \mc A \, \bb q_g = (\bb q_g)_1 \, \bb a_1 + (\bb q_g)_2 \, \bb a_2$, with $\bb q_g \in \Q$, and similarly for $g'$.
To ensure that $g'$ can be expressed consistently in the enlarged cell, we require that
\bg
\mc A \, \bb q_g + \bs \tau = \mc A' \, \bb q_{g'}
\eg
for some suitable $\bs \tau \in L$.
If such a translation $\bs \tau$ exists, then $g'$ is well defined in the enlarged basis.
Requiring the existence of such a $\bs \tau$ imposes constraints on $\mc S_{\rm uc}$.
We will illustrate this procedure in the example of $pg$ below.

In practice, the symmetry-compatibility conditions derived above may admit a large family of integer matrices $\mc S_{\rm uc}$.
Different choices of $\mc S_{\rm uc}$ can describe the same enlarged lattice or lead to equivalent supercells.
To avoid redundant parameterizations and to simplify the analysis, it is convenient to introduce normal forms for integer matrices.

Given $\mc S_{\rm uc} \in GL(2,\Z)$, one may perform unimodular transformations that correspond to changing the primitive basis of either the original lattice $L$ or the enlarged lattice $L'$.
Such transformations do not change the underlying lattices $L$ and $L'$, but only their coordinate descriptions.
Indeed, for a unimodular matrix $U \in GL(2,\Z)$ with $\det U \in \{+1, -1\}$, we have
\bg
\mc A \, (n_1,n_2)^{\rm T} =
\mc A \, U \, [ U^{-1} \, (n_1,n_2)^{\rm T}],
\eg
and since $U^{-1} \, (n_1,n_2)^{\rm T}$ again ranges over $\Z^2$, the lattice $L = \mc A \, \Z^2$ is unchanged.
Similarly, $\mc A' \, (n'_1,n'_2)^{\rm T} = \mc A' \, U' \, [ U'^{-1} \, (n'_1,n'_2)^{\rm T}]$, so changing the basis of $L'$ by a unimodular matrix $U'$ also leaves the superlattice invariant.
Under such transformations, the matrix $\mc S_{\rm uc}$ is replaced by
\bg
\mc S_{\rm uc} \to U^{-1} \, \mc S_{\rm uc} \, U',
\eg
which corresponds to describing the same pair of lattices in different integer bases.
Therefore, it is often useful to bring $\mc S_{\rm uc}$ into a canonical representative within its equivalence class under left and right unimodular transformations.

Two standard tools are the Hermite and Smith normal forms.
The Hermite normal form provides a canonical lower-triangular representative of an integer matrix under right multiplication by unimodular matrices.
In the present context, this corresponds to fixing the primitive basis of $L$ while changing only the description of the supercell basis of $L'$.
The Smith normal form diagonalizes the matrix under both left and right unimodular transformations, revealing its invariant factors [see also Eq.~\eqref{seq:lambda_diag}].
This corresponds to simultaneously redefining the bases of $L$ and $L'$, and makes the supercell index and its divisibility structure manifest.
However, such unimodular transformations also modify the explicit integer realization of the point-group representation set $M_P$ in the conventional setting.
Since in this work we fix that conventional realization, it is generally preferable to impose the symmetry constraints on $\mc S_{\rm uc}$ first and only then use normal forms to remove remaining redundancies.
In highly symmetric cases, such as $p2$, the symmetry constraints on $\mc S_{\rm uc}$ may leave substantial freedom, and working directly with the Smith normal form can significantly simplify the classification, as we explain below.

\tocless{\subsection*{Examples: $p2$, $pg$, $p4$, and $p4mm$}}{}
We now illustrate the above procedure in several representative cases.
These examples highlight different aspects of the symmetry-compatibility condition and the role of normal forms.

\paragraph{$p2$:} the wallpaper group $p2$ contains only a two-fold rotation $C_2$.
In the conventional setting, the point-group representation set $M_P$ consists of a single nontrivial element, $M_{C_2} = -\mathds{1}_{2 \times 2}$, together with $M_{\rm id} = \mathds{1}_{2 \times 2}$, corresponding to the trivial point-group element ${\rm id}$.
The symmetry condition in Eq.~\eqref{seq:MP_conj_invariant2} is automatically satisfied for any $\mc S_{\rm uc} \in GL(2,\Z)$, since $\mc S_{\rm uc}^{-1} \, (-\mathds{1}_{2 \times 2}) \, \mc S_{\rm uc} = -\mathds{1}_{2 \times 2}$.
Thus the point-group condition imposes no restriction on $\mc S_{\rm uc}$.
As a result, all supercells are allowed from the point-group perspective.
In this situation, the freedom in $\mc S_{\rm uc}$ is large, and it is convenient to use the Smith normal form to classify inequivalent supercells by their invariant factors.
Since we are working in two dimensions, there are only two invariant factors, namely two nonnegative integers that characterize the supercell.
This significantly simplifies the analysis of enlargement in the $p2$ case.

\paragraph{$pg$:} the group $pg$ provides the simplest nonsymmorphic example.
In the conventional rectangular setting, it contains a glide mirror $g = \{ m_{10} | \frac{1}{2} \bb a_2 \}$.
The point-group part $m_{10}$ acts as $M_{m_{10}} = \bsm -1 & 0 \\ 0 & 1 \esm$.
The symmetry condition in Eq.~\eqref{seq:MP_conj_invariant2} restricts $\mc S_{\rm uc}$ to the diagonal form $\bsm s_1 & 0 \\ 0 & s_2 \esm$ with integers $s_1$ and $s_2$.
Beyond this, the fractional translation part must also be compatible with the enlarged basis.
Following the general condition derived above, we check whether there exists $\bs \tau \in L$ such that $\bb t_g + \bs \tau$ can be expressed as $\frac{1}{2} \bb a'_2$ in the enlarged lattice basis.
Writing $\bs \tau = \alpha \bb a_1 + \beta \bb a_2$ with integers $\alpha$ and $\beta$, we obtain $\bb t_g + \bs \tau = \frac{1}{2} \bb a_2 + \alpha \bb a_1 + \beta \bb a_2 = \frac{\alpha}{s_1} \bb a_1 + \frac{1+ 2\beta}{2 s_2} \bb a'_2$.
For $\alpha=0$ and $s_2 = 1+ 2 \beta$, this reduces to $\frac{1}{2} \bb a'_2$, as required.
Therefore, the general form of $\mc S_{\rm uc}$ compatible with $pg$ is $\bsm s_1 & 0 \\ 0 & s_2 \esm$ with $s_1 \in \Z$ and $s_2$ an odd integer.
This case illustrates how nonsymmorphic constraints further restrict $\mc S_{\rm uc}$ beyond the point-group condition.

\paragraph{$p4$:} the group $p4$ contains a four-fold rotation.
In the conventional square setting, the point-group representation set $M_P$ is generated by $M_{C_4} = \bsm 0 & -1 \\ 1 & 0 \esm$.
After imposing the point-group condition [Eq.~\eqref{seq:MP_conj_invariant2}], the allowed supercells take the form, $\mc S_{\rm uc} = \bsm s_1 & s_2 \\ -s_2 & s_1 \esm$ with $s_1, s_2 \in \Z$.
This gives the supercell index $N_{\rm sc} = |\det \mc S_{\rm uc}| = s_1^2 + s_2^2$.
Thus, the allowed enlargement indices are integers representable as a sum of two squares, such as $1,2,4,5,8,9,\dots$.
In generic cases, pairs $(s_1,s_2)$ yielding the same $N_{\rm sc}$ are related by sign changes or by the exchange $s_1 \leftrightarrow s_2$, corresponding to lattice rotations, and hence describe equivalent supercells.
However, exceptional cases can occur.
For example, $N_{\rm sc}=50$ admits both $(1,7)$ and $(5,5)$, which are not related by sign changes or the exchange and have distinct Smith normal forms, hence correspond to inequivalent lattices of $\Z^2$.
Nevertheless, they are related by a rational rotation matrix.
By contrast, for $N_{\rm sc}=65$, the pairs $(1,8)$ and $(4,7)$ share the same Smith normal form and are therefore equivalent under unimodular integer transformations.

\paragraph{$p4mm$:} we illustrate how exchanging symmetries within the conventional set $M_P$ constrains $\mc S_{\rm uc}$.
In the conventional square setting, take $
M_{C_4} = \bsm 0&-1 \\ 1&0 \esm$, $M_{m_{10}} = \bsm -1&0 \\ 0&1 \esm$, and $M_{m_{1 \bar 1}} = \bsm 0&1 \\ 1&0 \esm$, where the mirror $m_{10}$ flips the sign of $\bb a_1$, and $m_{1 \bar 1}$ flips $\bb a_1-\bb a_2$.
As discussed above, the $C_4$ condition already restricts $\mc S_{\rm uc}$ to take the form $\mc S_{\rm uc} = \bsm s_1 & s_2 \\ -s_2 & s_1 \esm$ with $s_1, s_2 \in \Z$.
Imposing the set-level condition in Eq.~\eqref{seq:MP_conj_invariant1}, we have two distinct possibilities for the image of $m_{10}$ under conjugation:
(i) $m_{10}$ is mapped to itself, i.e. $\mc S_{\rm uc}^{-1} \, M_{m_{10}} \, \mc S_{\rm uc} = M_{m_{10}}$.
This yields $\mc S_{\rm uc} = \bsm s_1 & 0 \\ 0 & s_1 \esm$, hence $N_{\rm sc} = |\det \mc S_{\rm uc}| = s_1^2$, i.e. $N_{\rm sc} = 1,4,9,\dots$.
(ii) $m_{10}$ is exchanged with the diagonal mirror $m_{1 \bar 1}$, $\mc S_{\rm uc}^{-1} \, M_{m_{10}} \, \mc S_{\rm uc} = M_{m_{1 \bar 1}}$.
A representative solution is $\mc S_{\rm uc} = \bsm s_1 & -s_1 \\ s_1 & s_1 \esm$, giving $N_{\rm sc} = |\det \mc S_{\rm uc}| = 2s_1^2$, i.e. $N_{\rm sc}=2,8,18,\dots$.
Therefore, symmetry-compatible supercells in $p4mm$ fall into the two families $N_{\rm sc} = s_1^2$ and $N_{\rm sc} = 2s_1^2$, reflecting the allowed exchange $m_{10} \leftrightarrow m_{1 \bar 1}$ within the conventional representation set $M_P$.

\section{Tight-binding model and Wilson-loop analysis in $p2$}
The purpose of this section is to provide details of the tight-binding model used in the main text to demonstrate trivialization of a fragile phase in $p2$.
We consider a rectangular unit cell with primitive lattice vectors $\bb a_1 = (1,0)$ and $\bb a_2 = (0,1)$.
For simplicity, we place four orbitals at the unit-cell center, so $\bb x_{1,2,3,4}=(0,0)$.
Two orbitals are $s$-like and two are $p$-like.
Thus, under the two-fold rotation $C_2$, they transform as $U_{C_2} = {\rm Diag}(+1,+1,-1,-1)$.
Using Pauli matrices, we write $U_{C_2} = \tau_z \sg_0$.

The tight-binding Hamiltonian is
\bg
H_0 (\bk) = \sin k_x \, \tau_x \sg_x + \sin k_y \, \tau_x \sg_y + (1 - \cos k_x - \cos k_y) \tau_z \sg_0 + m (\tau_z \sg_x + \tau_z \sg_y).
\label{seq:h0_p2}
\eg
When $m=0$, this corresponds to two copies of the Qi-Wu-Zhang model~\cite{qi2006topological}.
The occupied subspace consists of the two lowest bands.
They carry Chern numbers $+1$ and $-1$, so the net Chern number vanishes and there is no stable topology.
The $m$ term mixes the two Chern bands and removes accidental surface states that would otherwise appear~\cite{geier2018second,khalaf2018higher,hwang2019fragile}.
Throughout this section we choose $m=0.25$.

\begin{figure}[t!]
\centering
\includegraphics[width=0.95\textwidth]{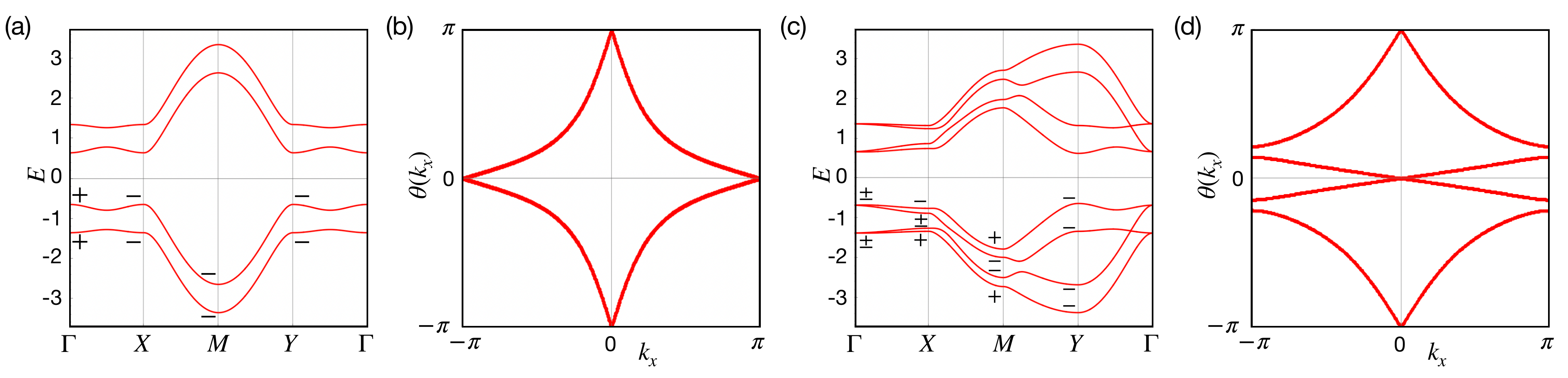}
\caption{(a) Band structure of the tight-binding model $H_0 (\bk)$ in the primitive cell, with $C_2$ eigenvalues indicated at high-symmetry momenta.
(b) Wilson-loop spectrum $\{\theta_a(k_x)\}_{a=1}^2$ for the two occupied bands, showing the characteristic winding of the fragile phase.
(c) Band structure in the $2 \times 1$ supercell after adding the symmetry-preserving perturbation.
(d) Wilson-loop spectrum $\{\theta_a(k_x)\}_{a=1}^4$ in the supercell, where the winding is removed and the spectrum becomes fully gapped.}
\label{sfig:model}
\end{figure}

The occupied bands have $C_2$ eigenvalues at the four high-symmetry momenta
\bg
\bb v =( n_{\Gamma_+}, n_{\Gamma_-}, n_{X_+}, n_{X_-}, n_{Y_+}, n_{Y_-}, n_{M_+}, n_{M_-})^{\rm T}
= (2, 0, 0, 2, 0, 2, 0, 2)^{\rm T},
\eg
where $\Gamma=(0,0)$, $X=(\pi,0)$, $Y=(0,\pi)$, and $M=(\pi,\pi)$, and $n_{\bar \bk_\pm}$ denotes the multiplicity of occupied bands with $C_2$ eigenvalues $\pm 1$ at high-symmetry momentum $\bar \bk$.
The band structure with symmetry eigenvalues is shown in Fig.~\ref{sfig:model}(a).

The fragile topology can be diagnosed from the Wilson-loop spectrum~\cite{yu2011equivalent}.
We compute the $k_y$-directed non-Abelian Berry phase for the two occupied bands while varying $k_x$.
Denoting the Wilson-loop phases by $\theta_a (k_x)$ for $a=1,2$, the spectrum $\{\theta_a (k_x)\}_{a=1}^2$ exhibits a winding structure characteristic of fragile topology protected by $C_2$ symmetry~\cite{alexandradinata2014wilson,wieder2018axion,hwang2019fragile}, as shown in Fig.~\ref{sfig:model}(b).
%

We now construct the $2 \times 1$ supercell.
The enlarged unit cell contains eight orbitals: four located at $(0,0)$ and four at $(1/2,0)$, where the latter position is expressed in the rescaled supercell coordinate system in which the new lattice vector along $x$ has unit length.
Accordingly, we denote their positions as $\bb x_{1,\dots,4}=(0,0)$ and $\bb x_{5,\dots,8}=(1/2,0)$.
The supercell Hamiltonian can be constructed systematically from $H_0(\bk)$.
We first express $H_0(\bk)$ in real-space hopping form:
\bg
H_0 (\bk)_{\alpha \beta} = \sum_{\Delta \bR} \, t(\Delta \bR)_{\alpha \beta} \, e^{-i \bk \cdot (\Delta \bR + \bx_\alpha - \bx_\beta)},
\eg
where $\alpha, \beta=1,\dots,4$ label orbitals,
$\Delta \bR \in L = \{n_1 \bb a_1 + n_2 \bb a_2 | (n_1,n_2) \in \Z^2\}$, and $t(\Delta \bR)_{\alpha \beta}$ denotes the hopping amplitude from orbital $\alpha$ at $\bR'+\Delta \bR + \bx_\alpha$ to orbital $\beta$ at $\bR' + \bx_\beta$.
By translation symmetry, the choice of lattice vector $\bR' \in L$ is arbitrary and $t (\Delta \bR)_{\alpha \beta}$ depends only on $\Delta \bR$.
The nonzero hopping matrices are
\ba
& t [(0,0)] = \bpm +1 & (1-i)m & 0 & 0 \\ (1+i)m & +1 & 0 & 0 \\ 0 & 0 & -1 & -(1-i)m \\ 0 & 0 & -(1+i)m & -1 \epm,
\nn
& t [(+1,0)] = \frac{1}{2} \bpm -1 & 0 & 0 & i \\ 0 & -1 & i & 0 \\ 0 & i & +1 & 0 \\ i & 0 & 0 & +1 \epm,
\quad
t [(0,+1)] = \frac{1}{2} \bpm -1 & 0 & 0 & +1 \\ 0 & -1 & -1 & 0 \\ 0 & +1 & +1 & 0 \\ -1 & 0 & 0 & +1 \epm,
\ea
with $t[(-1,0)] = t[(+1,0)]^\dg$ and
$t[(0,-1)] = t[(0,+1)]^\dg$ by hermiticity of the Hamiltonian.

From these matrices, the supercell hopping matrices are constructed as
\bg
t' [(0,0)] = \bpm t[(0,0)] & t[(-1,0)] \\ t[(+1,0)] & t[(0,0)] \epm,
\quad
t' [(+1,0)] = \bpm 0 & t[(+1,0)] \\ 0 & 0 \epm,
\quad
t' [(0,+1)] = \bpm t[(0,+1)] & 0 \\ 0 & t[(0,+1)] \epm,
\eg
where $0$ denotes the $4 \times 4$ zero matrix.
The remaining hoppings are obtained by hermitian conjugation, $t'[(-1,0)] = t'[(+1,0)]^\dg$ and $t'[(0,-1)] = t'[(0,+1)]^\dg$.
The supercell Hamiltonian is then
\bg
H_{\rm sc} (\bk)_{AB} = \sum_{\Delta \bR} \, t'(\Delta \bR)_{AB} \, e^{-i \bk \cdot (\Delta \bR + \bx_A - \bx_B)},
\eg
with $A,B =1, \dots, 8$ and $\bR \in \{(0,0), (\pm 1,0), (0,\pm 1)\}$.
The unit-cell doubling produces band folding, and the Wilson-loop winding persists.
To remove this residual, accidental winding, we introduce a $C_2$-preserving perturbation
\bg
H_{\rm pert} (\bk) = 0.2 \bpm \tau_0 \sg_x & 0 \\ 0 & -\tau_0 \sg_x \epm.
\eg
With this perturbation, the Wilson-loop spectrum $\{\theta_a (k_x)\}_{a=1}^4$ for the four occupied bands in the supercell becomes fully gapped.
The corresponding perturbed band structure and Wilson-loop spectrum are shown in Figs.~\ref{sfig:model}(c) and (d), respectively.
Also, the symmetry data of the occupied bands become $\bb v = (2,2,2,2,0,4,2,2)^{\rm T}$, which is atomic, as discussed in the main text.
This confirms that the $2 \times 1$ enlargement with supercell index $N_{\rm sc}=2$ trivializes the fragile topology in this model.

\section{A nonminimal enlargement in spinless $p4mm$}
\label{app:example}
In this section, we present a representative example in spinless $p4mm$ for which the smallest nontrivial symmetry-compatible enlargement does not trivialize all fragile roots.
Instead, some fragile roots remain non-atomic at $N_{\rm sc}=2$, the smallest nontrivial supercell index, while all of them trivialize at the next enlargement $N_{\rm sc}=4$.
Since time-reversal symmetry does not play a special role in this example, the same enlargement behavior occurs both with and without it.
Our purpose here is to illustrate concretely how the richer point-group and Wyckoff-position structure in $p4mm$ can obstruct trivialization at the smallest enlargement scale.

\tocless{\subsection*{Group-theoretical data and fragile roots}}{}
We first summarize the symmetry setting of spinless $p4mm$ in the conventional square unit cell.
Note that we follow the notation of the Bilbao Crystallographic Server~\cite{aroyo2006bilbao1,aroyo2006bilbao2,aroyo2011crystallography,vergniory2017graph,elcoro2017double}.
The primitive lattice vectors are $\bb a_1 = (1,0)$ and $\bb a_2 = (0,1)$, and the point group is generated by the four-fold rotation $C_4$ together with mirror reflections.
In the conventional setting, we use the following symmetry operations:
\bg
C_4 : (x,y) \to (-y,x),
\quad
m_{10} : (x,y) \to (-x,y),
\quad
m_{01} : (x,y) \to (x,-y),
\nn
m_{11} : (x,y) \to (-y,-x),
\quad
m_{1 \bar 1} : (x,y) \to (y,x).
\eg
Here $m_{10}$ and $m_{01}$ are mirrors about the coordinate axes, while $m_{11}$ and $m_{1\bar 1}$ are diagonal mirrors.

\begin{figure}[t!]
\centering
\includegraphics[width=0.95\textwidth]{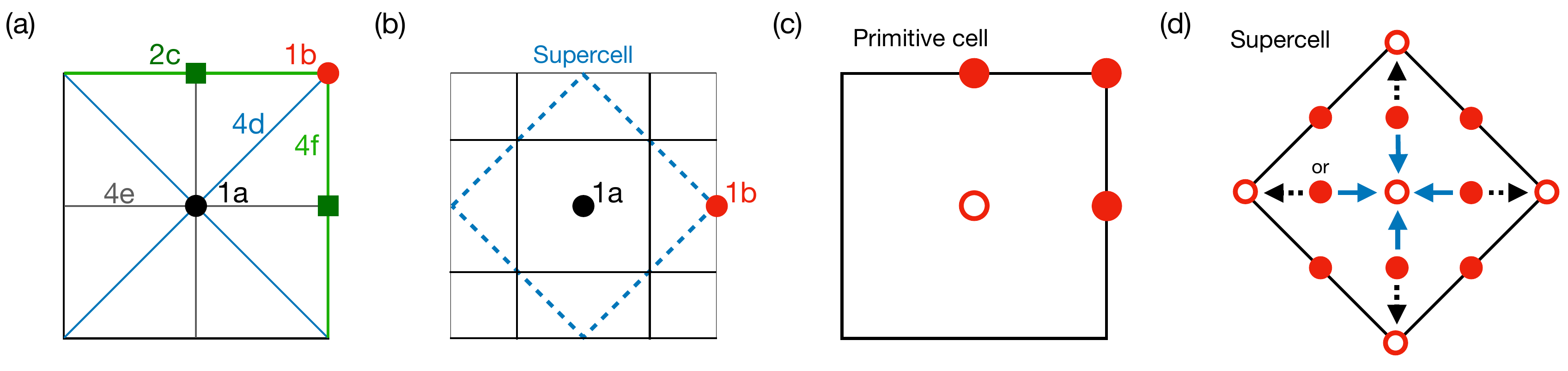}
\caption{
(a) Wyckoff positions (WPs) of $p4mm$.
The WPs $4d$, $4e$, and $4f$ lie on mirror-invariant lines.
(b) $\sqrt 2 \times \sqrt 2$ supercell.
The dashed blue diamond indicates the supercell, with maximal WPs $1a$ and $1b$.
(c) Real-space electron-positron configuration corresponding to $\rho = (A_1)_{1b} \oplus (B_1)_{2c} \ominus (A_1)_{1a}$ [Eq.~\eqref{seq:p4mm_rho_example}].
For clarity, filled and open circles denote electrons and positrons, respectively, without distinguishing their site-symmetry irreps.
(d) Real-space configuration corresponding to $\rho' = (A_1)_{2c} \oplus (A')_{4d} \ominus (A_1)_{1a} \ominus (A_1)_{1b}$ [Eq.~\eqref{seq:p4mm_rho2_example}] in the supercell.
The $(A')_{4d}$ electrons can be adiabatically deformed either into the $1a$ WPs (blue arrows) or into the $1b$ WPs (black dashed arrows).
In either case, one positron necessarily remains, so complete pair annihilation is impossible.
Hence, fragile topology remains in this unit-cell enlargement with index $N_{\rm sc}=2$.}
\label{sfig:p4mm}
\end{figure}

The relevant Wyckoff positions (WPs) are shown in Fig.~\ref{sfig:p4mm}(a).
In particular, the maximal WPs are $1a:(0,0)$, $1b:(1/2,1/2)$, and $2c:\{(1/2,0),(0,1/2)\}$.
The nonmaximal WPs $4d$, $4e$, $4f$, and $8g$ are given by
\ba
4d:& \{ (x,x), (-x,-x), (-x,x), (x,-x) \},
\nn
4e:& \{ (x,0), (-x,0), (0,x), (0,-x) \},
\nn
4f:& \{ (x,1/2), (-x,1/2), (1/2,x), (1/2,-x)\},
\nn
8g:& \{ (x,y), (-x,-y), (-y,x), (y,-x), (x,-y), (-x,y), (-y,-x), (y,x) \},
\label{seq:p4mm_wps}
\ea
where $8g$ denotes the general position with free real parameters $x$ and $y$.
The site-symmetry groups at $1a$ and $1b$ are both $4mm$, generated by fourfold rotation and mirrors, whereas the site-symmetry group at $2c$ is $2mm$, generated by twofold rotation and mirrors.
The WPs $4d$, $4e$, and $4f$ have site-symmetry group $m$, generated by a single mirror.

At the $1a$ WP, the relevant irreducible representations (irreps) of the site-symmetry group $4mm$ are $A_1$, $A_2$, $B_1$, $B_2$, and $E$, with character table
\begin{center}
\begin{tabular}{c|ccccc}
& ${\rm id}$ & $C_4$ & $C_2$ & $m_{10}$ & $m_{1 \bar 1}$ \\
\hline
$A_1$ & $+1$ & $+1$ & $+1$ & $+1$ & $+1$ \\
$A_2$ & $+1$ & $+1$ & $+1$ & $-1$ & $-1$ \\
$B_1$ & $+1$ & $-1$ & $+1$ & $+1$ & $-1$ \\
$B_2$ & $+1$ & $-1$ & $+1$ & $-1$ & $+1$ \\
$E$ & $+2$ & 0 & $-2$ & 0 & 0
\end{tabular}
\end{center}
Here, ${\rm id}$ in the first row denotes the identity.

The $1b$ WP also has a site-symmetry group isomorphic to $4mm$.
However, the symmetry operations leaving $(1/2,1/2)$ invariant are accompanied by lattice translations.
For example, the fourfold rotation is realized as $\{C_4|\bb a_1\}$, meaning that $C_4$ is applied first and then a translation by $\bb a_1$.
Explicitly, $\{C_4|\bb a_1\}: (x,y) \to (-y+1,x)$, which leaves $(1/2,1/2)$ invariant.
Similarly, mirrors are represented by operators such as $\{m_{10}|\bb a_1\}$.
Since the group is isomorphic to $4mm$, the irreps at $1b$ are labeled by the same symbols $A_1$, $A_2$, $B_1$, $B_2$, and $E$.

At the $2c$ WP, whose site-symmetry group is $2mm$, we first define the irreps at the position $(1/2,0)$.
The character table is
\begin{center}
\begin{tabular}{c|ccccc}
& ${\rm id}$ & $C_2$ & $\{m_{10}|\bb a_1\}$ & $m_{01}$ \\
\hline
$A_1$ & $+1$ & $+1$ & $+1$ & $+1$ \\
$A_2$ & $+1$ & $+1$ & $-1$ & $-1$ \\
$B_1$ & $+1$ & $-1$ & $-1$ & $+1$ \\
$B_2$ & $+1$ & $-1$ & $+1$ & $-1$
\end{tabular}
\end{center}
The irreps at the other position $(0,1/2)$ of the WP $2c$ are obtained by conjugation under $C_4$.
Since $(0,1/2) = C_4 (1/2,0)$, the site-symmetry group at $(0,1/2)$ is the conjugate of that at $(1/2,0)$.
Accordingly, it is generated by $\{m_{01}|\bb a_2\}$ and $m_{10}$.
Indeed, the conjugation relations, $C_4 \, m_{10} \, C_4^{-1} = \{m_{01}|\bb a_2\}$ and $C_4 \, m_{01} \, C_4^{-1} = m_{10}$, relate the generators of the two site-symmetry groups at $(1/2,0)$ and $(0,1/2)$.
This relation must be taken into account when constructing supercells.
For example, the $1a$ or $1b$ WP in the primitive cell can map to either $(1/2,0)$ or $(0,1/2)$ of the WP $2c$ in the supercell.
To determine how irreps of $4mm$ at $1a$ or $1b$ reduce to irreps (or their linear combinations) of $2mm$ at $2c$, the above conjugation structure must be considered.
%

We now define the real-space multiplicity vector $\bb m$ by listing the site-symmetry irreps at the above WPs in a fixed order:
\ba
\bb m =\big[ & n_{(A_1)_{1a}}, n_{(A_2)_{1a}}, n_{(B_1)_{1a}}, n_{(B_2)_{1a}}, n_{(E)_{1a}}, n_{(A_1)_{1b}}, n_{(A_2)_{1b}}, n_{(B_1)_{1b}}, n_{(B_2)_{1b}}, n_{(E)_{1b}},
\nn
& n_{(A_1)_{2c}}, n_{(A_2)_{2c}}, n_{(B_1)_{2c}}, n_{(B_2)_{2c}}, n_{(A')_{4d}}, n_{(A'')_{4d}}, n_{(A')_{4e}}, n_{(A'')_{4e}}, n_{(A')_{4f}}, n_{(A'')_{4f}}, n_{(A)_{8g}} \big]^{\rm T},
\label{seq:p4mm_m_def}
\ea

The symmetry-data vector $\bb v$ is defined by the multiplicities of irreps at the high-symmetry momenta $\Gamma=(0,0)$, $M=(\pi,\pi)$, and $X=(0,\pi)$.
The corresponding little-group irreps are $\Gamma_{1,\dots,5}$, $M_{1,\dots,5}$, and $X_{1,\dots,4}$.
Hence we define
\bg
\bb v = ( n_{\Gamma_1}, n_{\Gamma_2}, n_{\Gamma_3}, n_{\Gamma_4}, n_{\Gamma_5}, n_{M_1}, n_{M_2}, n_{M_3}, n_{M_4}, n_{M_5}, n_{X_1}, n_{X_2}, n_{X_3}, n_{X_4} )^{\rm T}.
\label{seq:p4mm_v_def}
\eg
With these conventions, the band-representation matrix $BR$ and its pseudoinverse $BR^\ddagger$ [Eq.~\eqref{seq:BR_inv}] are defined by
\bg
BR = \bsm
1 & 0 & 0 & 0 & 0 & 1 & 0 & 0 & 0 & 0 & 1 & 0 & 0 & 0 & 1 & 0 & 1 & 0 & 1 & 0 & 1 \\
0 & 0 & 1 & 0 & 0 & 0 & 0 & 1 & 0 & 0 & 1 & 0 & 0 & 0 & 0 & 1 & 1 & 0 & 1 & 0 & 1 \\
0 & 0 & 0 & 1 & 0 & 0 & 0 & 0 & 1 & 0 & 0 & 1 & 0 & 0 & 1 & 0 & 0 & 1 & 0 & 1 & 1 \\
0 & 1 & 0 & 0 & 0 & 0 & 1 & 0 & 0 & 0 & 0 & 1 & 0 & 0 & 0 & 1 & 0 & 1 & 0 & 1 & 1 \\
0 & 0 & 0 & 0 & 1 & 0 & 0 & 0 & 0 & 1 & 0 & 0 & 1 & 1 & 1 & 1 & 1 & 1 & 1 & 1 & 2 \\
1 & 0 & 0 & 0 & 0 & 0 & 0 & 0 & 1 & 0 & 0 & 0 & 1 & 0 & 1 & 0 & 1 & 0 & 0 & 1 & 1 \\
0 & 0 & 1 & 0 & 0 & 0 & 1 & 0 & 0 & 0 & 0 & 0 & 1 & 0 & 0 & 1 & 1 & 0 & 0 & 1 & 1 \\
0 & 0 & 0 & 1 & 0 & 1 & 0 & 0 & 0 & 0 & 0 & 0 & 0 & 1 & 1 & 0 & 0 & 1 & 1 & 0 & 1 \\
0 & 1 & 0 & 0 & 0 & 0 & 0 & 1 & 0 & 0 & 0 & 0 & 0 & 1 & 0 & 1 & 0 & 1 & 1 & 0 & 1 \\
0 & 0 & 0 & 0 & 1 & 0 & 0 & 0 & 0 & 1 & 1 & 1 & 0 & 0 & 1 & 1 & 1 & 1 & 1 & 1 & 2 \\
1 & 0 & 1 & 0 & 0 & 0 & 0 & 0 & 0 & 1 & 1 & 0 & 1 & 0 & 1 & 1 & 2 & 0 & 1 & 1 & 2 \\
0 & 1 & 0 & 1 & 0 & 0 & 0 & 0 & 0 & 1 & 0 & 1 & 0 & 1 & 1 & 1 & 0 & 2 & 1 & 1 & 2 \\
0 & 0 & 0 & 0 & 1 & 1 & 0 & 1 & 0 & 0 & 1 & 0 & 0 & 1 & 1 & 1 & 1 & 1 & 2 & 0 & 2 \\
0 & 0 & 0 & 0 & 1 & 0 & 1 & 0 & 1 & 0 & 0 & 1 & 1 & 0 & 1 & 1 & 1 & 1 & 0 & 2 & 2
\esm,
\quad
BR^\ddagger = \bsm
0 & 0 & 0 & 0 & 0 & 1 & 0 & 0 & 0 & 0 & 0 & 0 & 0 & 0 \\
0 & -1 & 0 & 0 & 0 & -1 & 0 & 0 & 1 & 0 & 1 & 0 & 0 & 0 \\
0 & -1 & 0 & -1 & 0 & -1 & 1 & 0 & 1 & 0 & 1 & 0 & 0 & 0 \\
0 & 0 & 1 & 0 & 0 & 0 & 0 & 0 & 0 & 0 & 0 & 0 & 0 & 0 \\
0 & 0 & 0 & 0 & 1 & 0 & 0 & 0 & 0 & 0 & 0 & 0 & 0 & 0 \\
1 & -1 & 0 & -1 & 0 & -1 & 1 & 0 & 1 & 0 & 0 & 0 & 0 & 0 \\
0 & 1 & 0 & 1 & 0 & 1 & 0 & 0 & -1 & 0 & -1 & 0 & 0 & 0 \\
0 & 1 & 0 & 0 & 0 & 1 & 0 & 0 & 0 & 0 & -1 & 0 & 0 & 0 \\
0 & 0 & 0 & 0 & 0 & 0 & 0 & 0 & 0 & 0 & 0 & 0 & 0 & 0 \\
0 & 0 & 0 & 0 & 0 & 0 & 0 & 0 & 0 & 0 & 0 & 0 & 0 & 0 \\
0 & 1 & 0 & 1 & 0 & 0 & -1 & 0 & -1 & 0 & 0 & 0 & 0 & 0 \\
0 & 0 & 0 & 0 & 0 & 0 & 0 & 0 & 0 & 0 & 0 & 0 & 0 & 0 \\
0 & 0 & 0 & 0 & 0 & 0 & 0 & 0 & 0 & 0 & 0 & 0 & 0 & 0 \\
0 & 0 & 0 & 0 & 0 & 0 & 0 & 0 & 0 & 0 & 0 & 0 & 0 & 0 \\
0 & 0 & 0 & 0 & 0 & 0 & 0 & 0 & 0 & 0 & 0 & 0 & 0 & 0 \\
0 & 0 & 0 & 0 & 0 & 0 & 0 & 0 & 0 & 0 & 0 & 0 & 0 & 0 \\
0 & 0 & 0 & 0 & 0 & 0 & 0 & 0 & 0 & 0 & 0 & 0 & 0 & 0 \\
0 & 0 & 0 & 0 & 0 & 0 & 0 & 0 & 0 & 0 & 0 & 0 & 0 & 0 \\
0 & 0 & 0 & 0 & 0 & 0 & 0 & 0 & 0 & 0 & 0 & 0 & 0 & 0 \\
0 & 0 & 0 & 0 & 0 & 0 & 0 & 0 & 0 & 0 & 0 & 0 & 0 & 0 \\
0 & 0 & 0 & 0 & 0 & 0 & 0 & 0 & 0 & 0 & 0 & 0 & 0 & 0
\esm.
\eg
The matrix $BR$ has dimension $D_{\bb v} \times D_{\bb m}$ with $D_{\bb v}=14$ and $D_{\bb m}=21$.
The symmetry data are obtained from the real-space multiplicities through $\bb v = BR \cdot \bb m$.

Following the method discussed in the main text, we obtain the Hilbert basis of symmetry-data vectors for $\aiset$, the set of all atomic phases.
The $14$ atomic generators are
\bg
\bb h^{(\rm A)}_1 = (0, 0, 0, 1, 0, 0, 0, 0, 1, 0, 0, 1, 0, 0)^{\rm T},
\quad
\bb h^{(\rm A)}_2 = (0, 0, 0, 1, 0, 0, 1, 0, 0, 0, 0, 0, 0, 1)^{\rm T},
\nn
\bb h^{(\rm A)}_3 = (0, 0, 1, 0, 0, 0, 0, 1, 0, 0, 0, 1, 0, 0)^{\rm T},
\quad
\bb h^{(\rm A)}_4 = (0, 0, 1, 0, 0, 1, 0, 0, 0, 0, 0, 0, 0, 1)^{\rm T},
\nn
\bb h^{(\rm A)}_5 = (0, 1, 0, 0, 0, 0, 0, 0, 1, 0, 0, 0, 1, 0)^{\rm T},
\quad
\bb h^{(\rm A)}_6 = (0, 1, 0, 0, 0, 0, 1, 0, 0, 0, 1, 0, 0, 0)^{\rm T},
\nn
\bb h^{(\rm A)}_7 = (1, 0, 0, 0, 0, 0, 0, 1, 0, 0, 0, 0, 1, 0)^{\rm T},
\quad
\bb h^{(\rm A)}_8 = (1, 0, 0, 0, 0, 1, 0, 0, 0, 0, 1, 0, 0, 0)^{\rm T},
\nn
\bb h^{(\rm A)}_9 = (0, 0, 0, 0, 1, 0, 0, 0, 0, 1, 0, 0, 1, 1)^{\rm T},
\quad
\bb h^{(\rm A)}_{10} = (0, 0, 0, 0, 1, 0, 0, 0, 0, 1, 1, 1, 0, 0)^{\rm T},
\nn
\bb h^{(\rm A)}_{11} = (0, 0, 0, 0, 1, 0, 0, 1, 1, 0, 0, 1, 1, 0)^{\rm T},
\quad
\bb h^{(\rm A)}_{12} = (0, 0, 0, 0, 1, 1, 1, 0, 0, 0, 1, 0, 0, 1)^{\rm T},
\nn
\bb h^{(\rm A)}_{13} = (0, 0, 1, 1, 0, 0, 0, 0, 0, 1, 0, 1, 0, 1)^{\rm T},
\quad
\bb h^{(\rm A)}_{14} = (1, 1, 0, 0, 0, 0, 0, 0, 0, 1, 1, 0, 1, 0)^{\rm T}.
\label{seq:p4mm_gen_a}
\eg
For the set of all atomic and fragile phases, $\afset$, the corresponding symmetry-data vectors are generated by 26 generators.
Among them, $14$ coincide with the atomic generators $\{\bb h^{(\rm A)}_a\}_{a=1}^{14}$, while the remaining $12$ generators correspond to fragile roots $\{\bb v^{(\rm FR)}_c\}_{c=1}^{12}$:
\bg
\bb v^{(FR)}_1 = (0, 0, 0, 0, 1, 0, 1, 1, 0, 0, 0, 0, 1, 1)^{\rm T},
\quad
\bb v^{(FR)}_2 = (0, 0, 0, 0, 1, 0, 1, 1, 0, 0, 1, 1, 0, 0)^{\rm T},
\nn
\bb v^{(FR)}_3 = (0, 0, 0, 0, 1, 1, 0, 0, 1, 0, 0, 0, 1, 1)^{\rm T},
\quad
\bb v^{(FR)}_4 = (0, 0, 0, 0, 1, 1, 0, 0, 1, 0, 1, 1, 0, 0)^{\rm T},
\nn
\bb v^{(FR)}_5 = (0, 1, 1, 0, 0, 0, 0, 0, 0, 1, 0, 0, 1, 1)^{\rm T},
\quad
\bb v^{(FR)}_6 = (0, 1, 1, 0, 0, 0, 0, 0, 0, 1, 1, 1, 0, 0)^{\rm T},
\nn
\bb v^{(FR)}_7 = (0, 1, 1, 0, 0, 0, 1, 1, 0, 0, 0, 0, 1, 1)^{\rm T},
\quad
\bb v^{(FR)}_8 = (0, 1, 1, 0, 0, 1, 0, 0, 1, 0, 1, 1, 0, 0)^{\rm T},
\nn
\bb v^{(FR)}_9 = (1, 0, 0, 1, 0, 0, 0, 0, 0, 1, 0, 0, 1, 1)^{\rm T},
\quad
\bb v^{(FR)}_{10} = (1, 0, 0, 1, 0, 0, 0, 0, 0, 1, 1, 1, 0, 0)^{\rm T},
\nn
\bb v^{(FR)}_{11} = (1, 0, 0, 1, 0, 0, 1, 1, 0, 0, 1, 1, 0, 0)^{\rm T},
\quad
\bb v^{(FR)}_{12} = (1, 0, 0, 1, 0, 1, 0, 0, 1, 0, 0, 0, 1, 1)^{\rm T}.
\label{seq:p4mm_gen_fr}
\eg
Note that, in general, the set of generators $\{\bb h^{(\rm A)}_a\}_{a=1}^{d_{\rm A}}$ for $\aiset$ does not have to coincide with the subset of generators $\{\bb h^{(\rm AF)}_b\}_{b=1}^{d_{\rm AF}}$ for $\afset$ that correspond to atomic phases.
This is because some atomic generators in $\{\bb h^{(\rm A)}_a\}_{a=1}^{d_{\rm A}}$ can be expressed as integer combinations of generators in $\{\bb h^{(\rm AF)}_b\}_{b=1}^{d_{\rm AF}}$ that include fragile roots.

\tocless{\subsection*{Smallest nontrivial enlargement with $N_{\rm sc}=2$}}{}
As explained in Sec.~\ref{app:supercell}, the smallest nontrivial symmetry-compatible enlargement in $p4mm$ has supercell index $N_{\rm sc}=2$.
In the conventional setting, this corresponds to the supercell transformation $\mc S_{\rm uc} = \bsm 1 & 1 \\ -1 & 1 \esm$, which is equivalent to the representative supercell shown in Fig.~\ref{sfig:p4mm}(b).
Site-symmetry irreps and WPs in the primitive lattice are reorganized under this enlargement.
For example, two $1a$ WPs in the primitive cells appear in the supercell description.
One of them remains a $1a$ WP in the supercell, while the other becomes the $1b$ WP.
Under this enlargement, the mirrors $m_{10}$ and $m_{1 \bar 1}$ in the primitive cell are mapped to the mirrors $m_{11}$ and $m_{10}$ in the supercell.
Because of this exchange, the irreps $A_1$, $A_2$, $B_1$, $B_2$, $E$ defined at the $1a$ WP are mapped to $A_1$, $A_2$, $B_2$, $B_1$, $E$, respectively, at the $1a$ and $1b$ WPs in the supercell.
For example, under this enlargement, $(A_1)_{1a} \to (A_1)_{1a} \oplus (A_1)_{1b}$, where the left- and right-hand sides are written in terms of the site-symmetry irreps of the primitive cell and the supercell, respectively.
(This is encoded in the first column of $\mc M_{\rm orb, 2}$ below.)
Repeating this procedure for all site-symmetry irreps determines the orbital map $\mc M_{\rm orb,2}$ and the corresponding symmetry-data transformation matrix $\mc M_{\bb v, 2}
= BR \cdot \mc M_{\rm orb, 2} \cdot BR^\ddagger$:
\bg
\mc M_{\rm orb, 2} = \bsm
1 & 0 & 0 & 0 & 0 & 0 & 0 & 0 & 0 & 0 & 0 & 0 & 0 & 0 & 0 & 0 & 0 & 0 & 0 & 0 & 0 \\
0 & 1 & 0 & 0 & 0 & 0 & 0 & 0 & 0 & 0 & 0 & 0 & 0 & 0 & 0 & 0 & 0 & 0 & 0 & 0 & 0 \\
0 & 0 & 0 & 1 & 0 & 0 & 0 & 0 & 0 & 0 & 0 & 0 & 0 & 0 & 0 & 0 & 0 & 0 & 0 & 0 & 0 \\
0 & 0 & 1 & 0 & 0 & 0 & 0 & 0 & 0 & 0 & 0 & 0 & 0 & 0 & 0 & 0 & 0 & 0 & 0 & 0 & 0 \\
0 & 0 & 0 & 0 & 1 & 0 & 0 & 0 & 0 & 0 & 0 & 0 & 0 & 0 & 0 & 0 & 0 & 0 & 0 & 0 & 0 \\
1 & 0 & 0 & 0 & 0 & 0 & 0 & 0 & 0 & 0 & 0 & 0 & 0 & 0 & 0 & 0 & 0 & 0 & 0 & 0 & 0 \\
0 & 1 & 0 & 0 & 0 & 0 & 0 & 0 & 0 & 0 & 0 & 0 & 0 & 0 & 0 & 0 & 0 & 0 & 0 & 0 & 0 \\
0 & 0 & 0 & 1 & 0 & 0 & 0 & 0 & 0 & 0 & 0 & 0 & 0 & 0 & 0 & 0 & 0 & 0 & 0 & 0 & 0 \\
0 & 0 & 1 & 0 & 0 & 0 & 0 & 0 & 0 & 0 & 0 & 0 & 0 & 0 & 0 & 0 & 0 & 0 & 0 & 0 & 0 \\
0 & 0 & 0 & 0 & 1 & 0 & 0 & 0 & 0 & 0 & 0 & 0 & 0 & 0 & 0 & 0 & 0 & 0 & 0 & 0 & 0 \\
0 & 0 & 0 & 0 & 0 & 1 & 0 & 0 & 1 & 0 & 0 & 0 & 0 & 0 & 0 & 0 & 0 & 0 & 0 & 0 & 0 \\
0 & 0 & 0 & 0 & 0 & 0 & 1 & 1 & 0 & 0 & 0 & 0 & 0 & 0 & 0 & 0 & 0 & 0 & 0 & 0 & 0 \\
0 & 0 & 0 & 0 & 0 & 0 & 0 & 0 & 0 & 1 & 0 & 0 & 0 & 0 & 0 & 0 & 0 & 0 & 0 & 0 & 0 \\
0 & 0 & 0 & 0 & 0 & 0 & 0 & 0 & 0 & 1 & 0 & 0 & 0 & 0 & 0 & 0 & 0 & 0 & 0 & 0 & 0 \\
0 & 0 & 0 & 0 & 0 & 0 & 0 & 0 & 0 & 0 & 1 & 0 & 1 & 0 & 0 & 0 & 2 & 0 & 0 & 0 & 0 \\
0 & 0 & 0 & 0 & 0 & 0 & 0 & 0 & 0 & 0 & 0 & 1 & 0 & 1 & 0 & 0 & 0 & 2 & 0 & 0 & 0 \\
0 & 0 & 0 & 0 & 0 & 0 & 0 & 0 & 0 & 0 & 0 & 0 & 0 & 0 & 1 & 0 & 0 & 0 & 0 & 0 & 0 \\
0 & 0 & 0 & 0 & 0 & 0 & 0 & 0 & 0 & 0 & 0 & 0 & 0 & 0 & 0 & 1 & 0 & 0 & 0 & 0 & 0 \\
0 & 0 & 0 & 0 & 0 & 0 & 0 & 0 & 0 & 0 & 0 & 0 & 0 & 0 & 1 & 0 & 0 & 0 & 0 & 0 & 0 \\
0 & 0 & 0 & 0 & 0 & 0 & 0 & 0 & 0 & 0 & 0 & 0 & 0 & 0 & 0 & 1 & 0 & 0 & 0 & 0 & 0 \\
0 & 0 & 0 & 0 & 0 & 0 & 0 & 0 & 0 & 0 & 0 & 0 & 0 & 0 & 0 & 0 & 0 & 0 & 1 & 1 & 2
\esm,
\quad
\mc M_{\bb v, 2}
= \bsm
1 & 0 & 0 & 0 & 0 & 1 & 0 & 0 & 0 & 0 & 0 & 0 & 0 & 0 \\
1 & -1 & 2 & -1 & 0 & -1 & 1 & 0 & 1 & 0 & 0 & 0 & 0 & 0 \\
0 & 1 & 0 & 0 & 0 & 0 & 1 & 0 & 0 & 0 & 0 & 0 & 0 & 0 \\
0 & 0 & 0 & 1 & 0 & 0 & 0 & 0 & 1 & 0 & 0 & 0 & 0 & 0 \\
0 & 1 & 0 & 1 & 2 & 0 & -1 & 0 & -1 & 0 & 0 & 0 & 0 & 0 \\
0 & 0 & 0 & 0 & 0 & 0 & 0 & 0 & 0 & 0 & 1 & 0 & 0 & 0 \\
0 & -1 & 1 & 0 & 0 & -1 & 0 & 0 & 1 & 0 & 1 & 0 & 0 & 0 \\
0 & 0 & 0 & 0 & 0 & 0 & 0 & 0 & 0 & 0 & 1 & 0 & 0 & 0 \\
0 & -1 & 1 & 0 & 0 & -1 & 0 & 0 & 1 & 0 & 1 & 0 & 0 & 0 \\
1 & 2 & 0 & 1 & 2 & 1 & 0 & 0 & -1 & 0 & -2 & 0 & 0 & 0 \\
1 & 0 & 1 & 0 & 1 & 0 & 0 & 0 & 0 & 0 & 0 & 0 & 0 & 0 \\
0 & 1 & 0 & 1 & 1 & 0 & 0 & 0 & 0 & 0 & 0 & 0 & 0 & 0 \\
1 & 0 & 1 & 0 & 1 & 0 & 0 & 0 & 0 & 0 & 0 & 0 & 0 & 0 \\
0 & 1 & 0 & 1 & 1 & 0 & 0 & 0 & 0 & 0 & 0 & 0 & 0 & 0
\esm.
\label{seq:p4mm_Morb_2}
\eg

Applying $\mc M_{\bb v,2}$ to the fragile roots, we find that not all of them become atomic.
Six fragile roots, $\bb v^{(\rm FR)}_{1,3,5,7,9,12}$, remain nontrivial.
To gain intuition, let us focus on $\bb v^{(\rm FR)}_1$ in Eq.~\eqref{seq:p4mm_gen_fr} as a representative example.
Other fragile roots can be understood in a similar way.
Using the pseudoinverse method in Eq.~\eqref{seq:m_general}, one can obtain a representative $\bb m$ up to the kernel of $BR$, $\bb m_{\rm ker}$, for $\bb v^{(\rm FR)}_1$:
\bg
\bb m = (-1, 0, 0, 0, 0, 1, 0, 0, 0, 0, 0, 0, 1, 0, 0, 0, 0, 0, 0, 0, 0)^{\rm T},
\label{seq:p4mm_m_example}
\eg
corresponding to
\bg
\rho = (A_1)_{1b} \oplus (B_1)_{2c} \ominus (A_1)_{1a},
\label{seq:p4mm_rho_example}
\eg
as illustrated in Fig.~\ref{sfig:p4mm}(c).
Because of the freedom associated with $\bb m_{\rm ker}$, other representatives of $\bb m$ are also possible.
For example, $\td \rho = (A_2)_{1b} \oplus (B_2)_{2c} \ominus (A_2)_{1a}$.
Note that $\rho$ and $\td \rho$ are stably equivalent in the sense defined in Ref.~\cite{hwang2026stable}: they can be deformed into each other in the presence of suitable auxiliary orbitals $\rho_{\rm aux}$, i.e. $\rho \oplus \rho_{\rm aux} \Leftrightarrow \td \rho \oplus \rho_{\rm aux}$.
Whether this fragile root can be trivialized under the enlargement is determined by the symmetry-data vector $\bb v$, not by the particular representative $\bb m$.
Nevertheless, by choosing a convenient representative $\bb m$ and using the electron–positron picture, we can gain intuition for why trivialization is obstructed.

After the enlargement, $\bb m$ in Eq.~\eqref{seq:p4mm_m_example} transforms to
\bg
\bb m' = \mc M_{\rm orb, 2} \cdot \bb m
= (-1, 0, 0, 0, 0, -1, 0, 0, 0, 0, 1, 0, 0, 0, 1, 0, 0, 0, 0, 0, 0)^{\rm T},
\eg
corresponding to the real-space representation,
\bg
\rho' = (A_1)_{2c} \oplus (A')_{4d} \ominus (A_1)_{1a} \ominus (A_1)_{1b}.
\label{seq:p4mm_rho2_example}
\eg
according to Eq.~\eqref{seq:p4mm_Morb_2}.
Since $(A_1)_{2c}$ is located at the maximal WP $2c$, it alone cannot adiabatically move to other WPs.
By contrast, $4d$ is a nonmaximal WP connected to $1a$ and $1b$.
Because of this connectivity, $(A')_{4d}$ can be adiabatically transformed either to $(E)_{1a} \oplus (A_1)_{1a} \oplus (B_2)_{1a}$ or to $(E)_{1b} \oplus (A_1)_{1b} \oplus (B_2)_{1b}$, as shown in Fig.~\ref{sfig:p4mm}(d).
This can be seen from the character table above.
Each of these possibilities leads to $\rho' = (E)_{1a} \oplus (B_2)_{1a} \oplus (A_1)_{2c} \ominus (A_1)_{1b}$ or $(E)_{1b} \oplus (B_2)_{1b} \oplus (A_1)_{2c} \ominus (A_1)_{1a}$.
In both cases, one positron remains.
This situation can be compared with the $p2$ example discussed in the main text:
There are two positrons, one at $1a$ and one at $1b$.
To annihilate these positrons, four orbitals transforming as $(A')_{4d}$ irreps must move to those positions.
These four orbitals must respect $C_4$, and their motion is therefore restricted to the diagonal mirror lines.
As a result, all four orbitals can move only to one of $1a$ or $1b$.
This contrasts with the $p2$ case, where two of them can move to $1a$ and the other two to $1b$.
Thus, although some orbitals can move along such symmetry-preserving lines, the available deformation processes are not sufficient to annihilate all positrons because of the richer group structure.
In conclusion, $N_{\rm sc}=2$ does not suffice to trivialize all fragile roots in spinless $p4mm$.

\tocless{\subsection*{Next enlargement with $N_{\rm sc}=4$}}{}
We now consider the next relevant symmetry-compatible enlargement with supercell index $N_{\rm sc}=4$.
Proceeding as above, one obtains the orbital map
\bg
\mc M_{\rm orb, 4} = \bsm
1 & 0 & 0 & 0 & 0 & 0 & 0 & 0 & 0 & 0 & 0 & 0 & 0 & 0 & 0 & 0 & 0 & 0 & 0 & 0 & 0 \\
0 & 1 & 0 & 0 & 0 & 0 & 0 & 0 & 0 & 0 & 0 & 0 & 0 & 0 & 0 & 0 & 0 & 0 & 0 & 0 & 0 \\
0 & 0 & 1 & 0 & 0 & 0 & 0 & 0 & 0 & 0 & 0 & 0 & 0 & 0 & 0 & 0 & 0 & 0 & 0 & 0 & 0 \\
0 & 0 & 0 & 1 & 0 & 0 & 0 & 0 & 0 & 0 & 0 & 0 & 0 & 0 & 0 & 0 & 0 & 0 & 0 & 0 & 0 \\
0 & 0 & 0 & 0 & 1 & 0 & 0 & 0 & 0 & 0 & 0 & 0 & 0 & 0 & 0 & 0 & 0 & 0 & 0 & 0 & 0 \\
1 & 0 & 0 & 0 & 0 & 0 & 0 & 0 & 0 & 0 & 0 & 0 & 0 & 0 & 0 & 0 & 0 & 0 & 0 & 0 & 0 \\
0 & 1 & 0 & 0 & 0 & 0 & 0 & 0 & 0 & 0 & 0 & 0 & 0 & 0 & 0 & 0 & 0 & 0 & 0 & 0 & 0 \\
0 & 0 & 1 & 0 & 0 & 0 & 0 & 0 & 0 & 0 & 0 & 0 & 0 & 0 & 0 & 0 & 0 & 0 & 0 & 0 & 0 \\
0 & 0 & 0 & 1 & 0 & 0 & 0 & 0 & 0 & 0 & 0 & 0 & 0 & 0 & 0 & 0 & 0 & 0 & 0 & 0 & 0 \\
0 & 0 & 0 & 0 & 1 & 0 & 0 & 0 & 0 & 0 & 0 & 0 & 0 & 0 & 0 & 0 & 0 & 0 & 0 & 0 & 0 \\
1 & 0 & 1 & 0 & 0 & 0 & 0 & 0 & 0 & 0 & 0 & 0 & 0 & 0 & 0 & 0 & 0 & 0 & 0 & 0 & 0 \\
0 & 1 & 0 & 1 & 0 & 0 & 0 & 0 & 0 & 0 & 0 & 0 & 0 & 0 & 0 & 0 & 0 & 0 & 0 & 0 & 0 \\
0 & 0 & 0 & 0 & 1 & 0 & 0 & 0 & 0 & 0 & 0 & 0 & 0 & 0 & 0 & 0 & 0 & 0 & 0 & 0 & 0 \\
0 & 0 & 0 & 0 & 1 & 0 & 0 & 0 & 0 & 0 & 0 & 0 & 0 & 0 & 0 & 0 & 0 & 0 & 0 & 0 & 0 \\
0 & 0 & 0 & 0 & 0 & 1 & 0 & 0 & 1 & 1 & 0 & 0 & 0 & 0 & 2 & 0 & 0 & 0 & 0 & 0 & 0 \\
0 & 0 & 0 & 0 & 0 & 0 & 1 & 1 & 0 & 1 & 0 & 0 & 0 & 0 & 0 & 2 & 0 & 0 & 0 & 0 & 0 \\
0 & 0 & 0 & 0 & 0 & 0 & 0 & 0 & 0 & 0 & 1 & 0 & 1 & 0 & 0 & 0 & 2 & 0 & 0 & 0 & 0 \\
0 & 0 & 0 & 0 & 0 & 0 & 0 & 0 & 0 & 0 & 0 & 1 & 0 & 1 & 0 & 0 & 0 & 2 & 0 & 0 & 0 \\
0 & 0 & 0 & 0 & 0 & 0 & 0 & 0 & 0 & 0 & 1 & 0 & 1 & 0 & 0 & 0 & 2 & 0 & 0 & 0 & 0 \\
0 & 0 & 0 & 0 & 0 & 0 & 0 & 0 & 0 & 0 & 0 & 1 & 0 & 1 & 0 & 0 & 0 & 2 & 0 & 0 & 0 \\
0 & 0 & 0 & 0 & 0 & 0 & 0 & 0 & 0 & 0 & 0 & 0 & 0 & 0 & 1 & 1 & 0 & 0 & 2 & 2 & 4
\esm.
\label{seq:p4mm_Morb_4}
\eg
Note that $\mc M_{\rm orb, 4} = \mc M_{\rm orb, 2}^2$, since the $N_{\rm sc}=4$ supercell can be obtained by applying the $N_{\rm sc}=2$ enlargement twice.
(The orientation may differ, but it is irrelevant in the present case.)
The corresponding symmetry-data transformation is
\bg
\mc M_{\bb v, 4}
= \bsm
1 & 0 & 0 & 0 & 0 & 1 & 0 & 0 & 0 & 0 & 1 & 0 & 0 & 0 \\
0 & 1 & 0 & 0 & 0 & 0 & 1 & 0 & 0 & 0 & 1 & 0 & 0 & 0 \\
1 & -2 & 3 & -1 & 0 & -2 & 1 & 0 & 2 & 0 & 1 & 0 & 0 & 0 \\
0 & -1 & 1 & 1 & 0 & -1 & 0 & 0 & 2 & 0 & 1 & 0 & 0 & 0 \\
1 & 3 & 0 & 2 & 4 & 1 & -1 & 0 & -2 & 0 & -2 & 0 & 0 & 0 \\
1 & 0 & 1 & 0 & 1 & 0 & 0 & 0 & 0 & 0 & 0 & 0 & 0 & 0 \\
0 & 1 & 0 & 1 & 1 & 0 & 0 & 0 & 0 & 0 & 0 & 0 & 0 & 0 \\
1 & 0 & 1 & 0 & 1 & 0 & 0 & 0 & 0 & 0 & 0 & 0 & 0 & 0 \\
0 & 1 & 0 & 1 & 1 & 0 & 0 & 0 & 0 & 0 & 0 & 0 & 0 & 0 \\
1 & 1 & 1 & 1 & 2 & 0 & 0 & 0 & 0 & 0 & 0 & 0 & 0 & 0 \\
1 & 2 & 0 & 1 & 2 & 1 & 0 & 0 & -1 & 0 & 0 & 0 & 0 & 0 \\
1 & 0 & 2 & 1 & 2 & -1 & 0 & 0 & 1 & 0 & 0 & 0 & 0 & 0 \\
1 & 2 & 0 & 1 & 2 & 1 & 0 & 0 & -1 & 0 & 0 & 0 & 0 & 0 \\
1 & 0 & 2 & 1 & 2 & -1 & 0 & 0 & 1 & 0 & 0 & 0 & 0 & 0
\esm.
\label{seq:p4mm_Mv_4}
\eg

For this enlargement, every fragile root maps to an atomic symmetry-data vector.
As in the previous subsection, we illustrate the trivialization using the representative fragile root $\bb v^{(\rm FR)}_1$.
The orbital multiplicity in Eq.~\eqref{seq:p4mm_m_example} for this fragile root becomes
\bg
\bb m'' = (-1, 0, 0, 0, 0, -1, 0, 0, 0, 0, -1, 0, 0, 0, 1, 0, 1, 0, 1, 0, 0)^{\rm T}.
\eg
This corresponds to
\bg
\rho'' = (A')_{4d} \oplus (A')_{4e} \oplus (A')_{4f} \ominus (A_1)_{1a} \ominus (A_1)_{1b} \ominus (A_1)_{2c}.
\eg
All positrons can now be annihilated through suitable adiabatic deformation processes.
The WPs $4d$, $4e$, and $4f$ are connected to $1a$, $2c$, and $1b$, respectively, as can be seen from the definitions of the WPs in Eq.~\eqref{seq:p4mm_wps}, leading to the following deformations:
\bg
(A')_{4d} \Leftrightarrow (E)_{1a} \oplus (A_1)_{1a} \oplus (B_2)_{1a},
\quad
(A')_{4e} \Leftrightarrow (A_1)_{2c} \oplus (B_1)_{2c},
\quad
(A')_{4f} \Leftrightarrow (E)_{1b} \oplus (A_1)_{1b} \oplus (B_1)_{1b}.
\eg
After these processes, $\rho''$ becomes
\bg
(E)_{1a} \oplus (B_2)_{1a} \oplus (E)_{1b} \oplus (B_1)_{1b} \oplus (B_1)_{2c},
\eg
which contains no positrons and therefore corresponds to an atomic configuration.
The other fragile roots can be understood in a similar manner, consistent with the symmetry-data analysis.
This confirms that the $N_{\rm sc}=4$ enlargement suffices to trivialize all fragile roots in spinless $p4mm$.

\end{document}